\documentclass[10pt,preprint]{aastex}
\usepackage{subfigure}

\begin{document}


\title{Ring Star Formation Rates in Barred and Nonbarred Galaxies}


\author{R. D. Grouchy\altaffilmark{1,2,3,5}, R. J. Buta\altaffilmark{1,2},
H. Salo\altaffilmark{3}, and E. Laurikainen\altaffilmark{3,4}}
\altaffiltext{1}{Visiting Astronomer, Cerro Tololo Inter-American 
Observatory, La Serena, Chile} \altaffiltext{2}{Department of
Physics and Astronomy, University of Alabama, Tuscaloosa, AL,
35487-0324, USA}\altaffiltext{3}{Department of Physical Sciences 
Astronomy Division, University of Oulu, FI-90014 Oulu, Finland} 
\altaffiltext{4}{Finnish Centre for Astronomy with ESO (FINCA), 
University of Turku, Turku, Finland}
\altaffiltext{5}{Now at L'Observatoire de Paris; 61 Av. de l'Observatoire, 
F-75014 Paris, France}

%%\email{aastex-help@aas.org}

%% Mark off your abstract in the "abstract" environment. In the manuscript

\begin{abstract}
Nonbarred ringed galaxies are relatively normal galaxies showing bright
rings of star formation in spite of lacking a strong bar.  This
morphology is interesting because it is generally accepted that a
typical galactic disk ring forms when material collects near a resonance,
set up by the pattern speed of a bar or bar-like perturbation. Our goal
in this paper is to examine whether the star formation properties of
rings are related to the strength of a bar or, in the absence of a bar,
to the non-axisymmetric gravity potential in general. For this purpose,
we obtained H$\alpha$ emission line images and calculated the line fluxes
and star formation rates (SFRs) for 16 nonbarred SA galaxies and four weakly
barred SAB galaxies with rings. For comparison, we combine our new
observations with a re-analysis of previously published data on five SA, 
seven SAB, and 15 SB galaxies with rings, three of which are duplicates from
our sample.  With these data, we examine what role a bar may play
in the {\it star formation process} in rings. Compared to
barred ringed galaxies, we find that the inner ring SFRs and H$\alpha$+[NII] 
equivalent widths in nonbarred ringed galaxies show a similar range and 
trend with absolute blue magnitude,
revised Hubble type, and other parameters.  On the whole, the star
formation properties of inner rings, excluding the distribution of 
HII regions, are independent of the ring shapes and the bar strength 
in our small samples. We confirm that the deprojected axis ratios of 
{\it inner rings} correlate with maximum relative gravitational 
force $Q_g$; however, if we consider all rings, a better correlation 
is found when local bar forcing at the radius of the ring, $Q_r$, is 
used.  Individual cases are described and other correlations are 
discussed. By studying the physical properties of these galaxies, we 
hope to gain a better understanding of their placement in the scheme 
of the Hubble sequence and how they formed rings without the driving 
force of a bar.
\end{abstract}




\section{Introduction} 

A nonbarred ringed galaxy is a normal galaxy that shows one or
more large-scale ring-shaped patterns in the absence of a conspicuous
bar. These galaxies are interesting because they pose a small dilemma
for existing dynamical theories: based on numerical simulations, a bar
is believed to be an {\it essential element} in ring formation (e.g.,
Schwarz 1981, 1984a; see also reviews by Sellwood \& Wilkinson 1993;
see also review by Buta \& Combes 1996). These simulations have 
indicated that rings 
form easily in galaxies through bar-driven gravity torques.  These 
torques can lead to the secular accumulation of interstellar 
gas into well-defined regions, usually the main low-order resonances 
associated with the bar pattern speed. Invariant manifolds of orbits 
near unstable Lagrangian points have also been proposed to explain 
specific characteristics of galaxy rings (Romero-G\'{o}mez et al. 
2006; Athanassoula et al. 2009).

Alternate theories have been proposed to explain the presence of
rings in nonbarred galaxies. One such theory suggests that nonbarred 
ringed galaxies were once more strongly barred, but the bar has mostly 
or completely dissolved, leaving behind the rings that formed when the 
bar was stronger. Bar dissolution might occur if a bar drives enough gas
to the center to build up a central mass concentration (CMC; e.g., Shen \&
Sellwood 2004), or if gas exerts sufficient torques on the bar
(Bournaud et al. 2005). The idea of bar dissolution and reformation is
an active topic at this time (Bournaud \& Combes 2002; Bournaud et al.
2005), but has not been unambiguously confirmed by observations.  In 
2002, Block et al. found some evidence for bar dissolution, but a 
similar analysis done by Buta et al. (2004) was not able to confirm the 
results.  Buta (1991) suggested that NGC 7702 could be an example of a 
ringed galaxy whose main bar has 
dissolved because of the presence of a nuclear bar, a possible relic of 
a past major bar episode (Friedli \& Pfenniger 1991). Athanassoula (1996) 
examined the effects of bar dissolution on a simulated outer ring, and 
found that the ring survives for a long time after the bar dissolves.

It is possible that the rings of some nonbarred galaxies formed in
response to a strong spiral density wave (e.g., the SA(r)bc spiral NGC
5364). Mark (1974) suggested that a ring could form at the inner
Lindblad resonance (ILR) of the spiral wave pattern speed.  Buta \& 
Combes (1996) argued that such rings would be infrequent due
to the inefficiency of the ring formation process in the presence of
what might be an unsteady spiral.  However, Rautiainen \& Salo (2000) 
concluded that in models with a hot disk that never formed a bar (Toomre
$Q$-parameter=2.5), a spiral potential can still effectively form a
ring at the spiral ILR.

Another possible explanation for ring formation in nonbarred galaxies 
is that some of these rings are in fact still bar-driven features, 
but the bar is only detectable in the near-infrared where the galaxy 
is more transparent to dust. Most galaxies are selected for given 
properties from catalogs that only give blue-light classifications 
(e.g., de Vaucouleurs et al. 1991; Sandage \& Tammann 1981). 
Blue light is a very good passband for detecting rings, dust, and young stars, 
but bars tend to be made of older stellar populations that are 
less prominent in such a waveband.  Infrared light penetrates the 
dust more effectively and is more sensitive to the stellar population 
typically found in bars. A recently identified example of a ringed 
galaxy classified as nonbarred in blue light but which shows a bar in 
the near-IR is the SA(rs)bc spiral NGC 3147 (Casasola et al. 2008). 
In related cases, a bar might be missed even if potentially detectable 
in blue light because it is viewed end-on at a high inclination angle 
(e.g., the Sb(r) galaxy NGC 7184; Sandage \& Tammann 1981).

It is also possible that a ring of a nonbarred galaxy can form 
through the minor merger of a companion.  Sil'chenko 
\& Moiseev (2006) suggested that the small rings seen in NGC 7217 and 
NGC 7742, both known to have counter-rotating components, 
are due to tidally induced distortion of the stellar disk due to the 
(now mostly-merged) companion. Some rings, as in Hoag's object and 
IC 2006, have been interpreted in terms of accretion of a gas-rich 
companion (e.g., Schweizer et al. 1987, 1989). These ideas have also 
been considered to explain the large star-forming UV ring seen in 
the nonbarred galaxy ESO 381$-$47 (Donovan et al. 2009).

In addition to the mode of ring formation, another intriguing aspect 
of some nonbarred ringed galaxies is the presence of counter-winding 
spiral patterns in which two sets of spiral structure appear to open 
opposite to one another.  The implication, assuming both features 
are in the same plane and are not counter-rotating, is that one set 
of spiral structure is trailing while the other is leading.  
The best-known example of counter-winding spiral structure is the 
SA(r)a galaxy NGC 4622 (Buta et al. 1992, 2003), and we have recently 
identified a new case in the SA(rs)bc spiral ESO 297$-$27 (Grouchy et 
al. 2008, hereafter Paper 1).  In NGC 4622, the inner ring is made 
partly of leading and trailing arm segments, while in ESO 297$-$27, 
the inner pseudoring is made mainly of a single inner spiral arm.  
Although it has been suggested in these papers that such patterns 
could indicate that an interaction has occurred, no definitive theoretical 
studies have yet been made that can explain the special characteristics 
of each case.  

In this paper, we examine the star formation properties in both barred and
nonbarred ringed galaxies with several objectives in mind. First,
concentrated star formation is an aspect that both barred and nonbarred
galaxy rings share (Sandage 1961; Kormendy 1979; Buta 1988 and
references therein; Pogge 1989; Buta et al. 2004). This would seem to
point to a galaxy-wide mechanism for collecting gas into rings. We
wish to examine whether any aspect of the ring-shaped star formation,
such as the star formation rate (SFR), might connect to the strength of the
non-axisymmetric perturbation. We are also interested in the triggering
mechanism of star formation in rings. If bars actually trigger the star
formation in rings, we might expect that barred galaxies will on
average show higher ring SFRs. Previous studies have suggested that this 
is not the case. Some nonbarred or weakly barred galaxies are exceptional 
sites of star formation (e.g., NGC 4736 and
7742), and these make it unclear whether a bar is an essential element
to the {\it star formation process} occurring in rings. The problem is
akin to spiral structure: do density waves trigger star formation in
galaxies, or do they act merely as pattern-organizing structures with
little or no role in triggering the collapse of clouds?  Elmegreen \&
Elmegreen (1986) and McCall (1986) independently concluded that density
waves are not likely to be the main triggering mechanism for star
formation in galaxies.

Our second objective is to re-examine how intrinsic ring shape connects 
to perturbation strength and ring star formation. Buta (2002) had
suggested that ring shape does not depend significantly on bar strength, 
defined as the maximum gravitational torque per unit mass per
unit square of the circular speed.  This finding is at odds with numerical
simulation studies (e.g., Salo et al. 1999), and we consider an
alternative approach that uses the strength of the perturbation at
the position of the ring as the controlling parameter for ring shape.

Our final goal is to bring more attention to the diversity of properties
of the rings in nonbarred galaxies. This diversity is greater
than what is seen in barred galaxy rings, whose statistical properties,
such as relative sizes, shapes, and orientations relative to the
bar have been studied previously (Buta 1995).

Our study is based on an analysis of new H$\alpha$+[NII] images of 
20 nonbarred ringed galaxies (Grouchy 2008) complemented by optical 
$BVI$ images that will be more fully presented in a separate paper.  
In order to cover a wide range of apparent bar strengths, our 
sample is combined with a re-analysis of previously published data 
for 20 strongly barred and 12 weakly barred galaxies initially analyzed
by Crocker et al. (1996, hereafter CBB96).  The main finding 
of CBB96 was a correlation between the way HII regions are distributed 
around inner rings and the intrinsic axis ratio of the rings. Circular 
inner rings have HII regions distributed more uniformly in azimuth than 
do elliptical inner rings. In the latter, HII regions concentrate around 
the ring major axis. We also further explore this issue here.

\enlargethispage{\baselineskip}
H$\alpha$+[NII] images are used to derive SFRs for both the global HII 
region distribution and for the HII regions confined to specific rings, 
using the flux conversion technique of Kennicutt (1983, 1998a). Previous 
studies (Finn et al. 2004; Kennicutt et al. 1994) have found
connections between a galaxy's type and its SFR (see Finn's Figure 9).
This paper complements these previous studies with more comparisons of
the general properties of nonbarred ringed galaxies (color, absolute
magnitude, equivalent width, and type) to their SFRs.

This paper is arranged as follows. In Section 2, we describe 
the galaxies in our sample as well as the CBB96 sample.  We 
discuss the observations and the process of data analysis, 
including a re-analysis of the CBB96 data. We also describe 
the process of calculating the galaxy's SFR as well as our 
estimates in the accuracy of the calibration.  Section 3 
focusses on the derivation of ring parameters and includes a 
description of the H$\alpha$ distribution for each galaxy 
observed. In Section 4, we explain the process of deriving 
the non-axisymmetric perturbations.  In Section 5, we discuss 
the analysis of the galaxy and ring properties addressing possible 
correlations. The discussion of our findings and conclusions can 
be found in Sections 6 and 7, respectively.  

\section{The Data}

\subsection{Sample Selection}

The sample of nonbarred ringed galaxies observed for this paper was
selected from the Third Reference Catalogue of Bright Galaxies (RC3; de
Vaucouleurs et al. 1991) and the Catalogue of Southern Ringed Galaxies
(CSRG; Buta 1995). The main requirement was that the galaxy has a
well-defined inner or outer ring, or both types of features.  We not
only chose our galaxies by morphological type, but we also constrained
the isophotal galaxy diameter to $D_{25}\geq 2^{\prime}$ and the
isophotal axis ratio to $R_{25}^{-1} \leq 0.5$ (implying an inclination
$i \leq 60\degr$) so that we would be able to resolve the spiral and
ring structures within the galaxy's disk. The sample cannot be regarded
as statistical, but merely as a set of well-defined cases.  The type
range is S0$^+$ to Sbc, where rings are most typically found in
nonbarred as well as barred galaxies.

Our data set of 20 galaxies cannot stand alone because it needs a
point of comparison. We want to know how the properties of star 
formation in nonbarred galaxy rings might compare to what is 
normally seen in barred galaxies. For this purpose, the sample of 
CBB96 is especially useful. CBB96 observed 32 ringed galaxies and 
derived HII region luminosity functions, galaxy-wide SFRs, and the 
distribution of HII regions around inner rings, covering a wide 
range of apparent bar strengths, but mostly emphasizing barred 
galaxies. 

The CBB96 sample was selected in a similar manner as our new sample,
and the average properties of the two are similar.  For example, the 
type range for the two data sets is similar, and both samples 
emphasize high luminosity galaxies.  The latter can 
be seen by comparing the mean absolute blue magnitudes for the new 
sample ($\langle$M$_B\rangle = -20.4 \pm 0.8$ and the CBB96 sample 
($\langle$M$_B\rangle = -20.4 \pm 0.6$).  Combining our sample with 
CBB96, we have 27 nonbarred (SA and S\underline{{\rm A}}B) and 20 
barred (SB) or weakly barred (SAB and SA\underline{{\rm B}}) galaxies. 
Since three galaxies are in common between the two data sets, the 
combined sample includes 44 different galaxies. Our newly observed 
galaxies and their general properties are listed in Table 1, while 
the properties of the CBB96 sample are listed in Table 2.

\subsection{Observations and Reduction}

Our new H$\alpha$ observations were obtained in 2002 August with a Tek
2K CCD camera attached to the 1.5 m telescope of Cerro-Tololo
Inter-American Observatory (CTIO). The CCD was operated with a gain of
1.5 $e^{-}$ per ADU and has a pixel scale of $0\arcsec.434$, giving a
field of view of 14\rlap{.}$^{\prime}$8 x 14\rlap{.}$^{\prime}$8. We
imaged each galaxy twice, once in the redshifted H$\alpha$ emission
filter followed by an image offset in wavelength for continuum
subtraction.  

The images were reduced using standard IRAF procedures and calibrated
to a standard scale by following the steps detailed in Jacoby et al. 
(1987). Each night, we observed spectrophotometric standard
stars (Hamuy et al. 1992) in the same filters as the galaxy images. The
standard stars were used to calculate the sensitivity, $S(\lambda)$, of
our system, which allows us to convert the instrumental H$\alpha$
emission flux to a flux measured in erg s$^{-1}$ cm$^{-2}$. Table 3
lists the calculated sensitivity of our system for each star observed.
Using a 3$\sigma$ rejection, we adopted a mean value of $S(\lambda$)
for each filter.  We observed the spectrophotometric standard stars
through all filters available; however, seven observations were omitted
either due to tracking problems or the fact that the filter was not
used that night.

\begin{figure}
\epsscale{0.65}\plotone{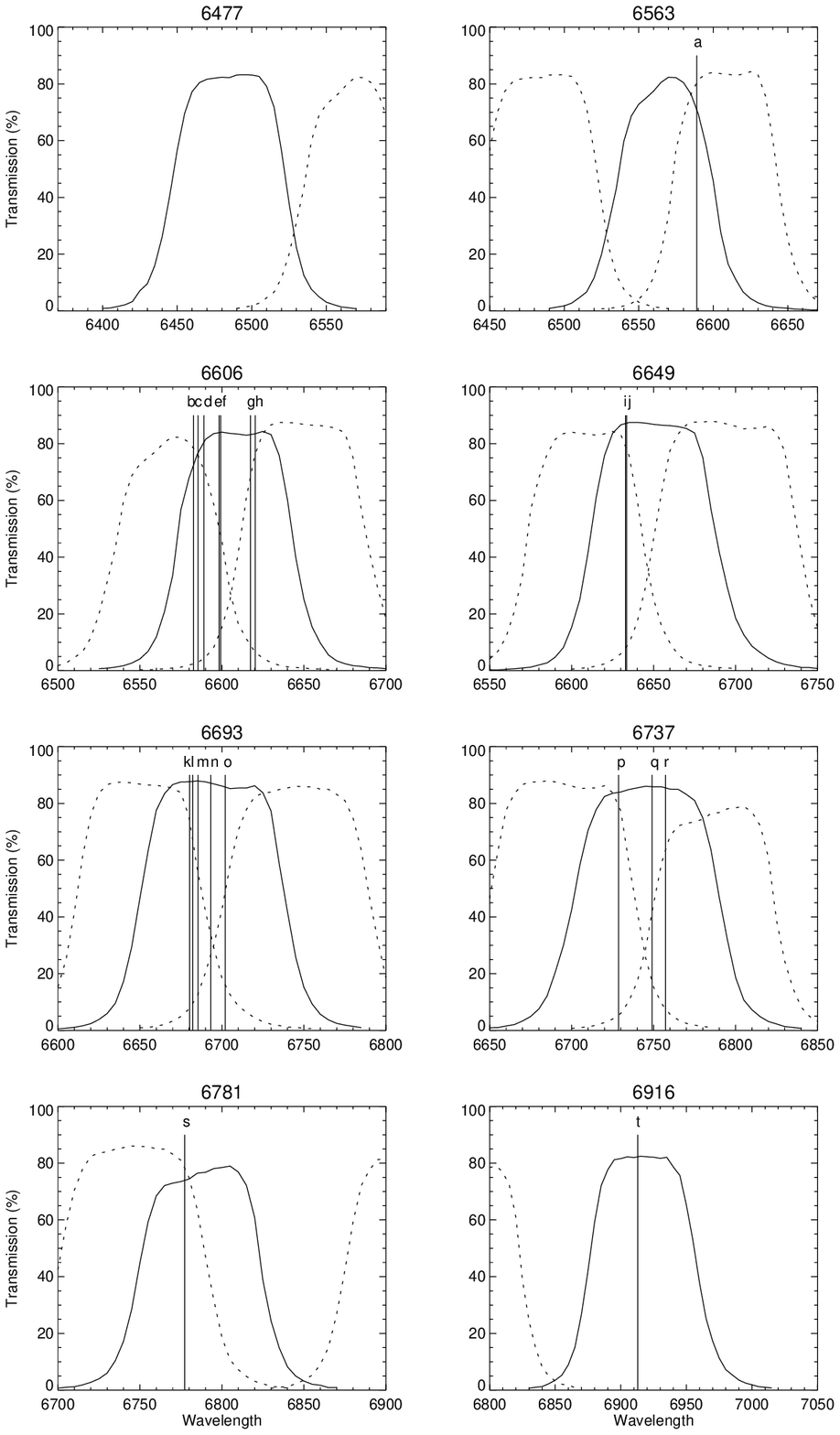}
\caption{Transmission
properties of the narrowband filters used to observe our sample.
Each panel is centered on a filter (solid line) whose central
wavelength in Angstroms is labeled above. Nearby filters (dashed
lines) are included for comparison. For each galaxy,
we included a vertical line representing the wavelength of its
redshifted H$\alpha$ emission.  The lines are labeled alphabetically
and correspond to the sample in the following way:
a) NGC 5530,  
b) NGC 7217, c) IC 1993, d) NGC 5364, e) NGC 7742, f) IC 5267, g) ESO 399-25, 
h) NGC 7187, 
i) NGC 7020, j) NGC 7702,
k) ESO 286$-$10, l) ESO 198$-$13, m) ESO 234$-$11, n) ESO 526$-$7, o) ESO 297$-$27
p) ESO 231$-$1, q) ESO 409$-$3, r) ESO 576$-$57,
s) ESO 111$-$22, and t) ESO 236$-$29 } \label{filters}
\end{figure}

To measure the amount of H$\alpha$ emission from each galaxy, we first 
subtracted the continuum from the redshifted H$\alpha$ image.  Each 
filter is approximately 75 \AA\ wide and has a unique transmission 
profile, which depends on wavelength (see Figure~\ref{filters}).  Due
to this variation in transmission properties for each filter, the nearby
continuum may not be an accurate model of the continuum underlying the
H$\alpha$ emission.  Since stars are not strong sources of H$\alpha$
emission, their fluxes are expected to be the same in both images, and
any difference in the foreground stellar flux measurements is attributed 
to differences in filter properties. After matching the point-spread 
function of the foreground stars, we modified the nearby continuum 
to approximate the underlying continuum in the H$\alpha$ image by using 
the ratio of the flux from these stars as a scale factor.  In many 
cases, we find a residual flux after subtracting the scaled continuum.  
Using these residuals, we find a range of 0.1\%-9.1\% uncertainty in 
the continuum subtraction from the H$\alpha$ image (see Table 4).  We 
must keep these uncertainties in mind for galaxies in our samples with 
low levels of emission which may be artifacts from this subtraction process.  

After subtracting the continuum, the sky, and the residual foreground 
star fluxes from the line images, the total H$\alpha$ fluxes were 
integrated within ellipses having the shape and orientation listed in 
Table 1. The sizes of these ellipses were determined by the visible 
extent of the observed H$\alpha$ emission.

According to a survey done by Kennicutt \& Kent (1983) on $\approx170$
galaxies, the average equivalent width of H$\alpha$ emission from a
typical galaxy ranges from 20 to 30 \AA. The narrowband filters we used are
wide enough so that the additional emission lines, [NII] $\lambda$6548
and $\lambda$6584, contribute to the measured H$\alpha$ flux.  This
contamination is removed from each measurement by using an estimated
average ratio of [NII] to H$\alpha$ of 0.5 for normal spiral galaxies
(Kennicutt et al. 1994).  Table 4 summarizes the
H$\alpha$+[NII] fluxes for our sample and the CBB96 sample. We then 
converted these fluxes (erg s$^{-1}$ cm$^{-2}$) to a luminosity 
(erg s$^{-1}$) by using NASA/IPAC Extragalactic Database (NED) 
Galactic standard of rest (GSR) distances to the galaxies (see Table 1). 
NED distances assume a cosmological expansion rate of 
$H_{0}$ = 73$\pm$5 km s$^{-1}$ Mpc$^{-1}$.

\subsection{Calculation of Star Formation Rate}
We corrected our H$\alpha$ luminosity for extinction within the galaxy
(internal extinction) as well as absorption of H$\alpha$ emission
between the galaxy and our detector (Galactic extinction). Kennicutt 
et al. (1994) found, on average, that the internal extinction in a 
typical HII region of a normal spiral galaxy is approximately 1.1 mag. 
The Galactic extinction, on the other hand, was calculated for each 
galaxy in Table 1 using Seaton's (1979) relation of 
\begin{equation}A(\lambda)=E(B-V)\times X(\lambda),\end{equation} 
where $E(B-V)$ is the Galactic extinction taken from NED and $X(6563)$, 
the interstellar extinction at 6563 \AA, is taken to be 2.46 (Howarth 1983). 
Both sources of extinction were 
subtracted from the H$\alpha$ magnitude to get the corrected luminosities 
listed in Column 6 of Table 4.

The H$\alpha$ luminosity was used as a direct measure of the SFR within 
a galaxy.  Kennicutt, Tamblyn, \& Congdon (1994) tested the rate of 
star formation for galaxies with varying initial mass functions. They 
found the relation of $1M_{\odot}$ yr$^{-1} = 1.26$$\times$10$^{41}$ erg s$^{-1}$ 
based on a Salpeter (1955) initial mass function with a mass range from 
0.1 M$_{\odot}$ to 100 M$_{\odot}$.  This relation assumes solar metallicity 
and a constant SFR for the galaxy. Our SFRs for the 20 galaxies are listed 
in Table 4.

We estimated the equivalent width of H$\alpha$ emission for each
galaxy by comparing the H$\alpha$ flux to that of the scaled
continuum image. This is a distance-independent parameter (owing to
normalization by the continuum flux) that has been interpreted as a
ratio of current to past star formation (e.g., Hameed \& Devereux
1999). We used the following relationship to calculate the
equivalent width (Romanishin 1990):
\begin{equation}W_{\lambda} = \int\frac{(F_{obs} - \alpha
F_{cont})}{\alpha F_{cont}}\times\frac{T_{\lambda}}{T_{H\alpha}}d\lambda,\end{equation}
where $F_{obs}$ is the flux from the H$\alpha$ image, $F_{cont}$ is
the flux from the continuum image, $\alpha$ is the scale factor
determined from foreground stars, $T_{H\alpha}$ is the transmission
coefficient of the filter at the galaxy's redshifted H$\alpha$
emission, and $T_{\lambda}$ is the fraction of light with 
wavelength $\lambda$, which is transmitted through the filter.
All values calculated from the narrowband images such as
$L_{H\alpha}$, SFR, and $W_{\lambda}$ are listed in Table 4.

For the CBB96 sample, the observations, image processing techniques,
and calibration are fully described in CBB96 and will not be repeated
here.  The original analysis of these data is in any case similar to 
what we have done in this paper with the exception that we have used 
the York Extinction Solver (McCall 2004) to correct the fluxes for 
Galactic extinction.  

\subsection{Accuracy of Calibrations and Flux Measurements}
To test the accuracy of the calibration of the new sample, we
observed the planetary nebulae NGC 6818 and NGC 6326 through the 
H$\alpha$ and the nearby continuum filters. We applied the 
system's sensitivity to the planetary nebulae using 
the following equation from Jacoby et al. (1987):

\begin{equation}S(\lambda) =
\frac{F(\lambda)}{C\times10^{0.4K(\lambda)A}},\end{equation} 

\noindent
where $F$($\lambda$) is the known flux of the planetary nebulae at
wavelength $\lambda$, $C$ is the photon count rate, $K(\lambda)$ is the
CTIO atmospheric extinction, and $A$ is the airmass. These nebulae were
the subject of a previous study by Kohoutek \& Martin (1981, hereafter
KM81); therefore, the H$\alpha$ fluxes ($F$($\lambda$) in Equation (3))
for these nebulae are known. We used the IRAF routine PHOT to measure a
count rate within an aperture large enough to enclose the optical
extent of the nebula.  By comparing our measured fluxes to those of
KM81, we found that, on average, our calibration was too bright by
$\approx 0.15$ mag. This discrepancy may indicate that our 
aperture sizes were larger than those used by KM81; however, we 
were unable to confirm this as the source of the error since the 
aperture sizes were not listed in KM81. Our observations compared with 
measurements from KM81 are listed in Table 5. The KM81 values refer to 
the sum of the H$\alpha$ and [NII] fluxes listed in their paper.

\begin{figure}
\centering
\includegraphics[width=0.5\textwidth,angle=0,clip=]{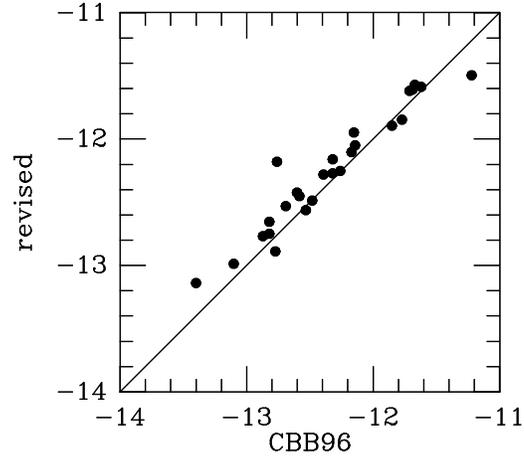}
\vspace{0cm} \caption{Comparison between H$\alpha$+[NII] fluxes
estimated for 30 galaxies by CBB96 and those estimated in the
re-analysis. The solid line has a unit slope.} \label{comparison}
\end{figure}

For our purposes, we need more information from the CBB96 sample 
than was given by CBB96. First, we wanted to estimate ring
SFRs for the CBB96 galaxies in a similar manner as we have derived them
for our present sample. Second, we wanted to derive quantitative
estimates of bar strength for this sample to more firmly examine the
possible significance of bars in the ring star formation process. These
parameters necessitated that we re-analyze the original CBB96 images.
The main difference with CBB96 will be in the selection of foreground 
stars for the continuum subtraction and minor details of the sky 
subtraction. Therefore, we can evaluate the {\it internal} uncertainties 
in the continuum subtraction simply by the action of two independent 
analyses of the same data set.  Figure~\ref{comparison} shows a comparison 
between the revised total H$\alpha$ fluxes (see Table 6) for the CBB96 
sample and those listed in Column 2 of Table 7 of CBB96. The line is a 
perfect correlation for comparison.  The re-analysis differs from the 
original analysis in the sense 
$\langle log(CBB96_{original}/CBB96_{revised})\rangle=0.08\pm0.03$ with a standard
deviation of 0.15 dex for 26 objects. The scatter of $\pm$0.15 dex is
larger than the 20\% uncertainties Kennicutt (1983) quoted as typical
in H$\alpha$ flux measurements, but this is not surprising given the
early Hubble types and the faintness of the emission in many cases.
Table 6 summarizes the results of our re-analysis of the CBB96 data.

\begin{figure}
\centering
\includegraphics[width=0.5\textwidth,angle=0,bb=26 20 463 444,clip=]{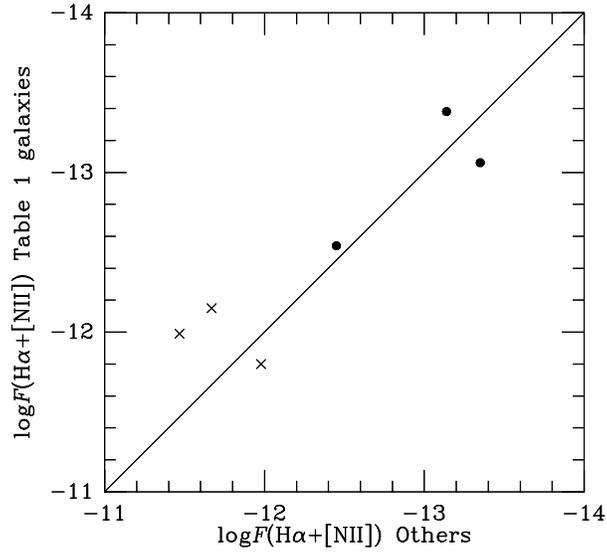}
\caption{Comparison between H$\alpha$+[NII] fluxes for six
galaxies in Table 1 compared with values for the same galaxies from
Kennicutt \& Kent (1983, crosses: NGC 5364, 7217, and NGC 7742) and 
from Table 2 of this paper (filled circles: NGC 7020, NGC 7187, and NGC 7702). 
The solid line has a unit slope.} \label{comparison2}
\end{figure}

\begin{figure}
\centering
\includegraphics[width=0.5\textwidth,angle=0,bb=26 20 463 444,clip=]{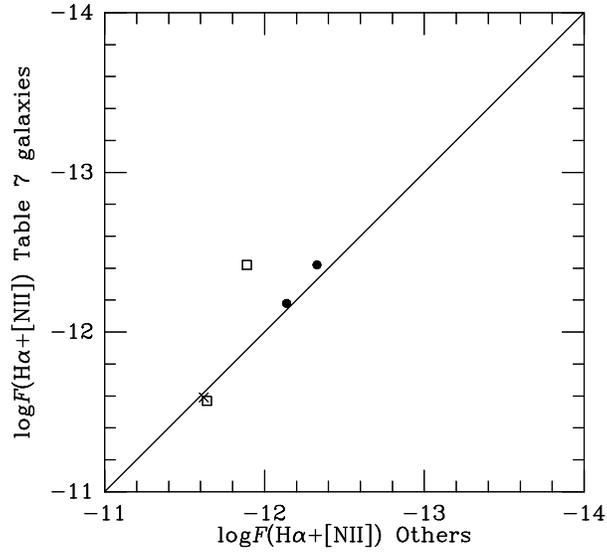}
\caption{Comparison between H$\alpha$+[NII] fluxes for three
galaxies in Table 2 compared with values for the same galaxies from
Kennicutt \& Kent (1983, crosses: NGC 1832), Hameed \& Devereux (1999, open
squares: NGC 1350 and NGC 1433), and Koopmann \& Kenney (2006, 
filled circles: IC 5240 and NGC 7098). The solid line has a unit slope.} 
\label{comparison3}
\end{figure}

Our sample shares three galaxies with CBB96, and it shares three 
galaxies with a study done by Kennicutt \& Kent (1983, hereafter KK83).  
Figure~\ref{comparison2} compares the H$\alpha$ + [NII] fluxes of 
our galaxies with the measurements from CBB96 and KK83.  The mean 
difference between the log of our flux and that from CBB96 or KK83 
(i.e. $\langle$log(F/F$_{other}$)$\rangle$) for all six galaxies is $-$0.14 
with a standard error in the mean of $\pm$0.14.  
The standard deviation of each flux estimate is 0.33 which is about 
twice as high as the standard deviation of the internal comparison 
in Figure~\ref{comparison}. The scatter is larger than for the typical 
late-type galaxies studied by KK83 owing partly to the relatively early 
types of the galaxies and to the faintness of the emission. Similarly, 
the CBB96 sample shares galaxies with previous studies.  Figure~\ref{comparison3} 
compares the revised measurements from the CBB96 sample with the 
measurements from one galaxy in the KK83 sample, two galaxies studied 
by Hameed \& Devereux (1999), and two galaxies studied by Koopmann 
\& Kenney (2006). The agreement is very good for four of the galaxies. 
The discrepant point is NGC 1350, which suffers from serious scattered 
light on the CBB96 image.

\section{Derivation of Ring Parameters}

\subsection{H$\alpha$ and Star Formation in the Rings} 
In addition to {\it global} H$\alpha$ fluxes and SFRs, we measured 
these values for the rings in the sample. For this purpose, we 
fitted an ellipse to the projected H$\alpha$ ring to determine its 
size and ellipticity.  Using IRAF routine TVMARK, we visually selected 
prominent HII regions and regions of diffused emission assumed to be 
associated with the ring feature.  This method isolated the ring along 
the line of its maximum H$\alpha$ emission. Next, a least-squares fit 
was used to derive the center coordinates, major and minor axes dimensions, 
and major axis position angle (P.A.).  If a ring was not visible in the 
H$\alpha$ image, then its parameters were determined from the 
broadband image.  We acknowledge the potential for variations in 
the ring parameters based on the shape defined by the observer. To 
estimate this error, two individuals measured the ring properties 
independently. The size of the semimajor axis radius was affected less 
than the ring's axis ratio with errors that were, on average, 0.6\% 
and 1.8\% of the fitted values, respectively. To approximate a total 
error in our ring sizes and shapes, we added, in quadrature, the 
human error with the error resulting from fitting the points to 
an ellipse.  We calculated the parameters for inner and outer
rings as noted in the galaxy's Hubble$-$de Vaucouleurs type.  Here we
establish the convention of defining the ''main" ring of the galaxy as
the inner ring unless the galaxy has no inner ring (e.g., NGC 7020,
7217, and IC 1993) in which case the main ring refers to the outer
ring. We then used the IRAF routine ELLIPSE to measure the flux within
an elliptical annulus encompassing the projected ring.  We found 
that the shape and P.A. of the rings rarely coincided with 
the same values derived from the outer isophotes of the broadband images 
(Grouchy 2008). Table 7 lists the diameter, width, shape, P.A., and 
SFR of the projected rings while Table 8 shows the deprojected size 
and shape of the main ring.  

For the CBB96 data, a similar procedure was followed to define the ring
properties.   We used a cursor on a TV display to map out the size, shape, 
and major axis P.A. of the inner rings, and then fitted these 
points with an ellipse. Then an elliptical annulus encompassing the entire 
inner ring was defined by visual inspection and the H$\alpha$ flux within 
that annulus was recorded.  These newly derived H$\alpha$ luminosities and 
SFRs will differ somewhat from CBB96 who used slightly 
different distances.  Also, we used the York Extinction Solver (McCall 2004) 
to correct the revised CBB96 fluxes for the small amounts of Galactic 
extinction in these cases.  Table 6 includes our new estimates of the 
ring to total flux ratios of the CBB96 sample, together with the ellipse 
parameters used. These ratios range from essentially zero in NGC 7702 
to 91\% for the barred galaxy NGC 53.

\subsection{Description of H$\alpha$ Distributions}

Grouchy (2008) showed that most rings in the Table 1 sample are
zones of enhanced blue colors, indicating active star formation.
Thus, we expect that some rings will also be zones of ionized gas
enhancement. In this section, we show the H$\alpha$ images and
describe the characteristics of the distribution of HII regions.

\begin{figure}
\includegraphics[width=0.95\textwidth,angle=90]{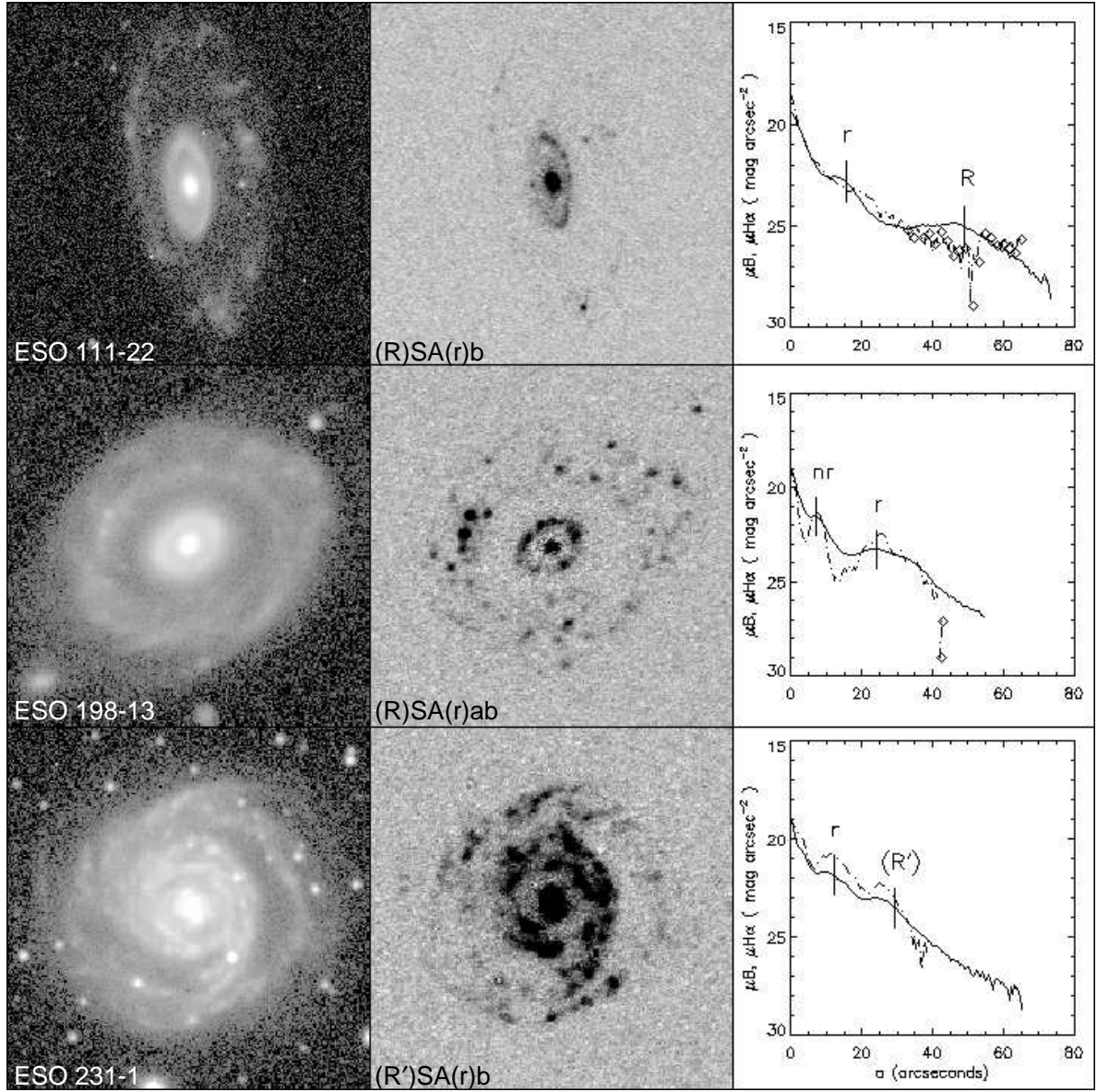}
\hfill \caption{$B$-band (left column) and H$\alpha$ (middle column)
images, and $B$-band (solid line) and H$\alpha$ (dashed line)
surface brightness profiles (right column) of ESO 111$-$22 (top
row), ESO 198$-$13 (middle row), and ESO 231$-$1 (bottom row). The
ESO 198$-$13 and ESO 231$-$1 images have a width of 1.$\arcmin$5.
The ESO 111$-$22 images are 2.$\arcmin$2 across. The vertical lines 
on the surface brightness profiles indicate the semimajor axis of 
the rings (see Table 7).  For ESO 198$-$13, the outer ring was not 
detected in the H$\alpha$ image and is not, therefore, labeled. 
The diamonds on the H$\alpha$ profile indicate the radii at which 
the azimuthally averaged flux drops below the uncertainty in the sky.  
Each image is oriented with north up and east to the left.} \label{img1}
\end{figure}

\begin{figure}
\includegraphics[width=0.95\textwidth,angle=90]{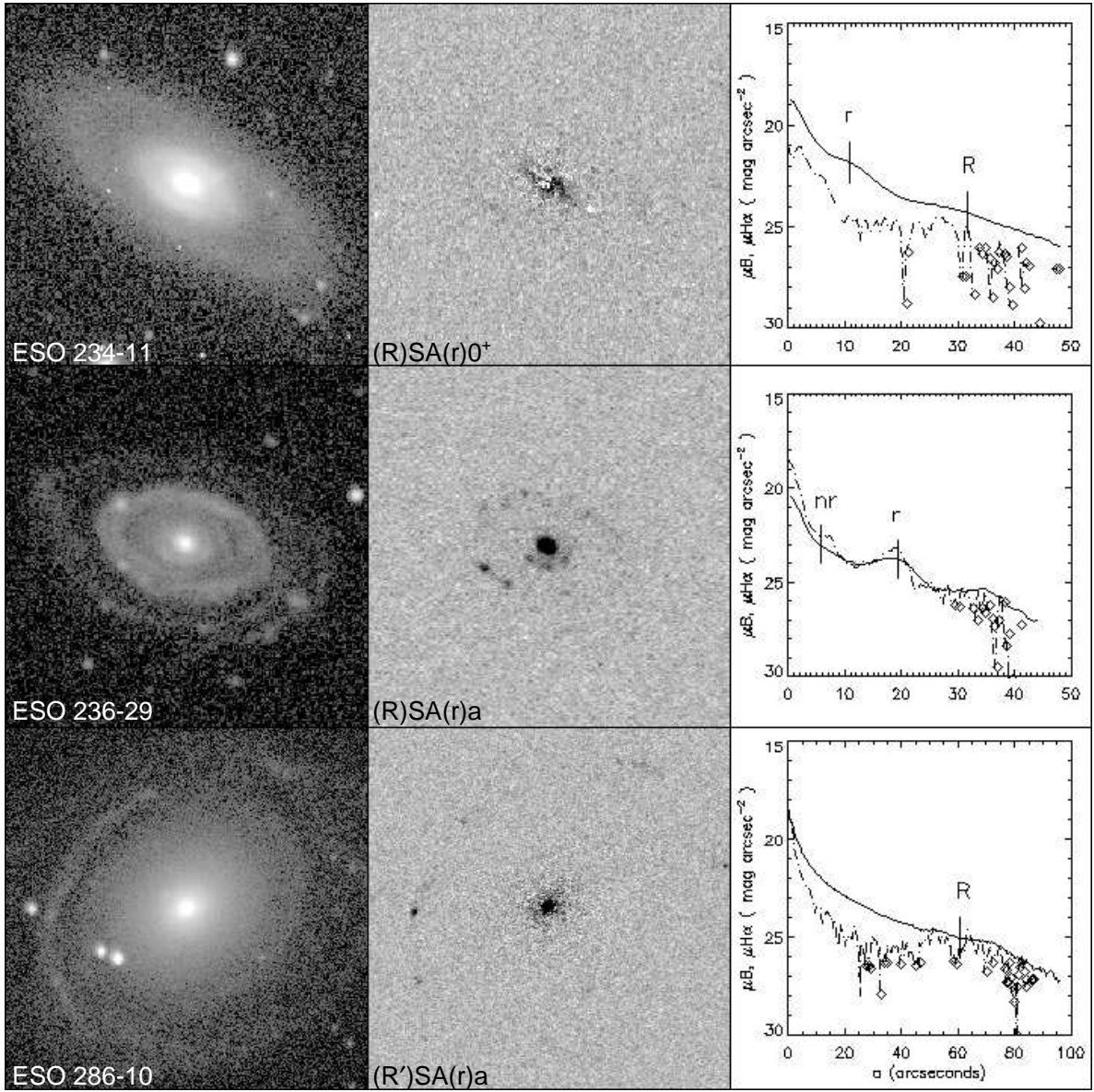}
\hfill \caption{$B$-band (left column) and H$\alpha$ (middle column)
images, and $B$-band (solid line) and H$\alpha$ (dashed line)
surface brightness profiles (right column) of ESO 234$-$11 (top
row), ESO 236$-$29 (middle row), and ESO 286$-$10 (bottom row). The
ESO 234$-$11 and ESO 236$-$29  images have a width of 1.$\arcmin$5.
The ESO 286$-$10 images are 2.$\arcmin$2 across. The vertical lines 
on the surface brightness profiles indicate the semimajor axis of 
each ring (see Table 7).  For ESO 234$-$11, the ring sizes are based 
on the broadband image since neither ring was detected in H$\alpha$.  
For ESO 236$-$29, we were able to detect the nuclear and inner rings, 
but not the outer ring in H$\alpha$.  For ESO 286$-$10, we were able 
to detect the outer ring, but not the inner ring in H$\alpha$.  
The diamonds on the H$\alpha$ profile indicate the radii at which 
the azimuthally averaged flux drops below the uncertainty in the sky. 
Each image is oriented with north up and east to the left.} \label{img2}
\end{figure}

\begin{figure}
\includegraphics[width=0.95\textwidth,angle=90]{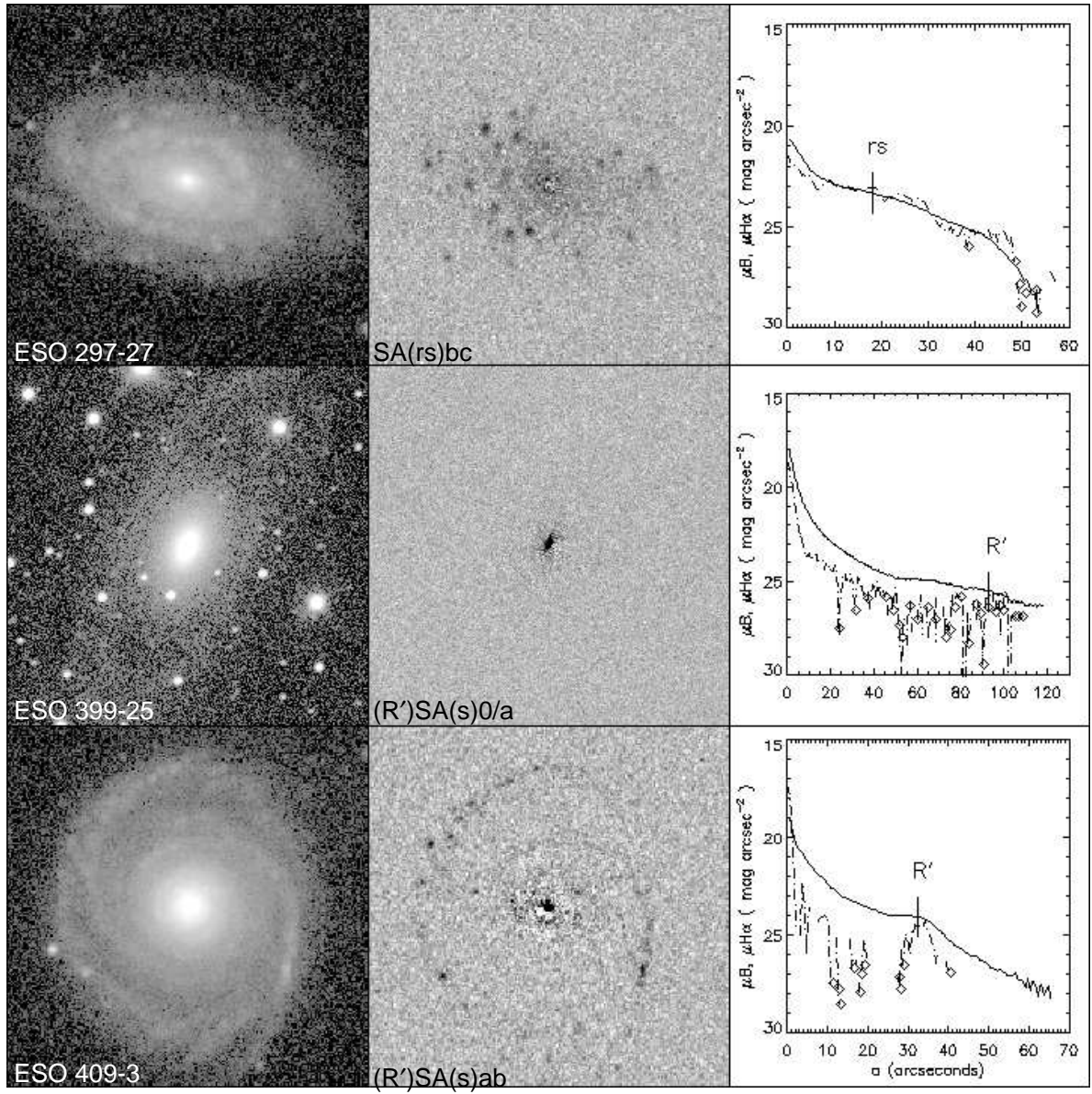}
\hfill \caption{$B$-band (left column) and H$\alpha$ (middle column)
images, and $B$-band (solid line) and H$\alpha$ (dashed line)
surface brightness profiles (right column) of ESO 297$-$27 (top
row), ESO 399$-$25 (middle row), and ESO 409$-$3 (bottom row). The
ESO 297$-$27 and ESO 409$-$3 images are 1.$\arcmin$5 across. The ESO
399$-$25 images are 2.$\arcmin$9 across. The vertical lines on the 
surface brightness profiles indicate the ring's semimajor axis 
(see Table 7).  For ESO 399$-$25, the size of the pseudoring 
is based on the broadband image since the ring was not detected in 
H$\alpha$.  The diamonds on the H$\alpha$ profile indicate the 
radii at which the azimuthally averaged flux drops below the 
uncertainty in the sky. Each image is oriented with north up and east 
to the left.} \label{img3}
\end{figure}

\begin{figure}
\includegraphics[width=0.95\textwidth,angle=90]{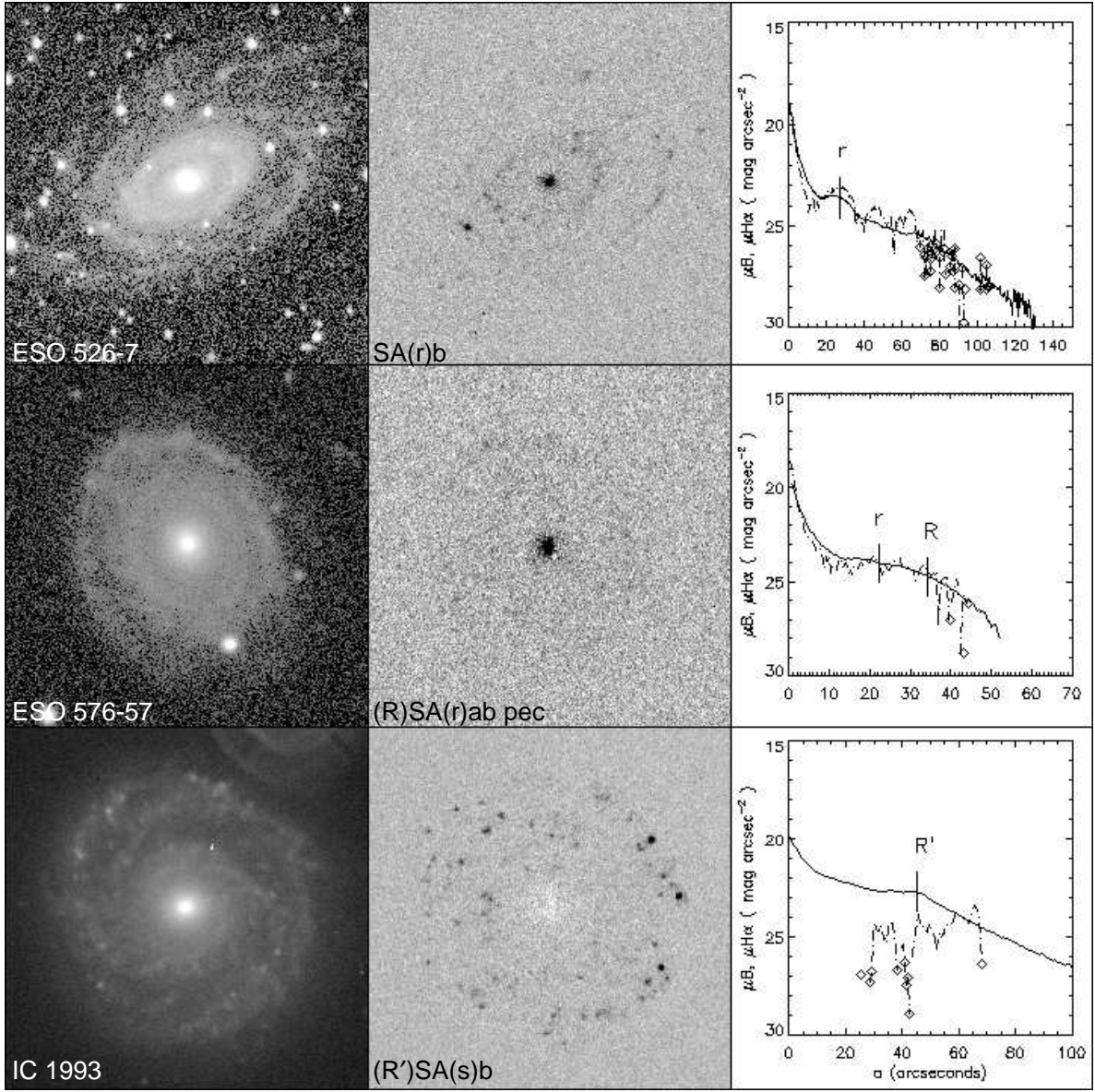}
\hfill \caption{$B$-band (left column) and H$\alpha$ (middle column)
images, and $B$-band (solid line) and H$\alpha$ (dashed line)
surface brightness profiles (right column) of ESO 526$-$7 (top row),
ESO 576$-$57 (middle row), and IC 1993 (bottom row). The ESO 526$-$7
images have a scale of 2.$\arcmin$9, and the ESO 576$-$57 images
have a scale of 1.$\arcmin$8.  The IC 1993 images have a scale of
2.$\arcmin$2. The vertical lines on the surface brightness profiles 
indicate the ring's semimajor axis (see Table 7).  The diamonds on 
the H$\alpha$ profile indicate the radii at which the azimuthally 
averaged flux drops below the uncertainty in the sky. Each image is 
oriented with north up and east to the left.} \label{img4}
\end{figure}

\begin{figure}
\includegraphics[width=0.95\textwidth,angle=90]{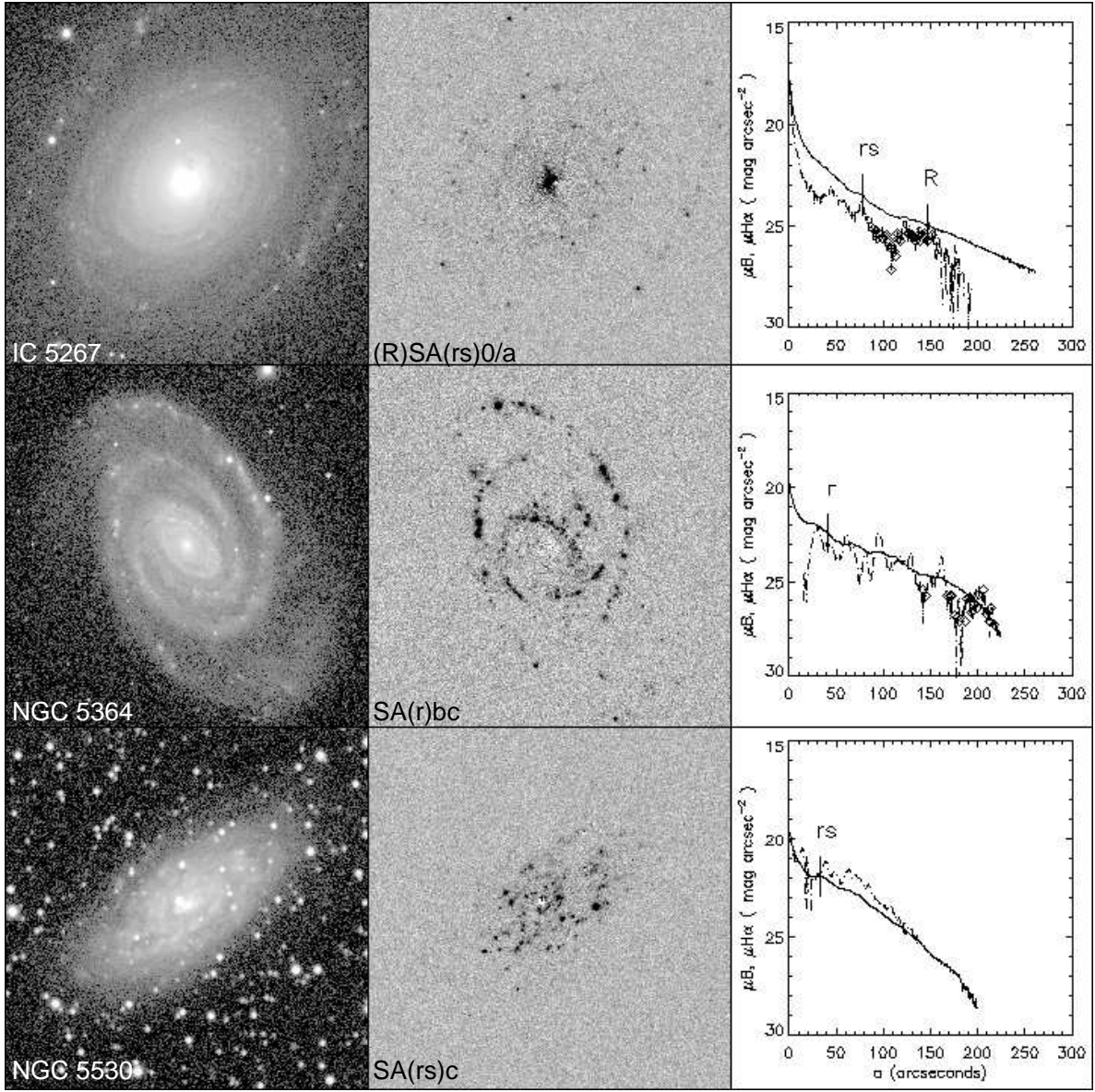}
\hfill \caption{$B$-band (left column) and H$\alpha$ (middle column)
images, and $B$-band (solid line) and H$\alpha$ (dashed line)
surface brightness profiles (right column) of IC 5267 (top row), NGC
5364 (middle row), and NGC 5530 (bottom row). The IC 5267 images
have a scale of 4.$\arcmin$3.  The NGC 5364 images have a scale of
6.$\arcmin$5. The NGC 5530 images have a scale of 5.$\arcmin$8. 
The vertical lines on the surface brightness profiles indicate the 
ring's semimajor axis (see Table 7).  The diamonds on the H$\alpha$ 
profile indicate the radii at which the azimuthally averaged flux drops 
below the uncertainty in the sky.  Each 
image is oriented with north up and east to the left.} \label{img5}
\end{figure}

\begin{figure}
\includegraphics[width=0.95\textwidth,angle=90]{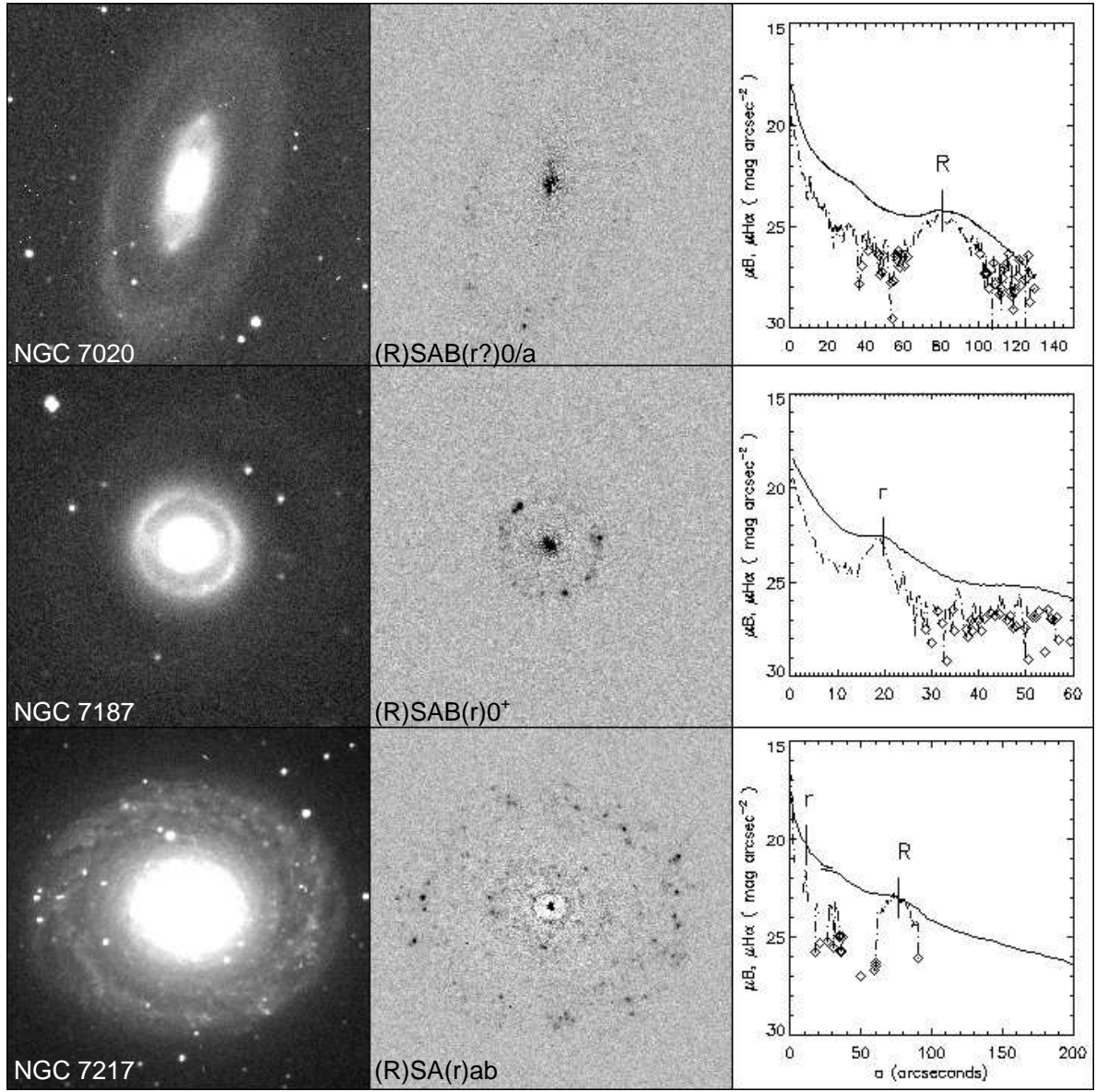}
\hfill \caption{$B$-band (left column) and H$\alpha$ (middle column)
images, and $B$-band (solid line) and H$\alpha$ (dashed line)
surface brightness profiles (right column) of NGC 7020 (top row),
NGC 7187 (middle row), and NGC 7217 (bottom row). The NGC 7020
images have a scale of 3.$\arcmin$3.  The NGC 7187 images have a
scale of 2.$\arcmin$2. The NGC 7217 images have a scale of
3.$\arcmin$5. The vertical lines on the surface brightness profiles 
indicate the ring's semimajor axis (see Table 7).  For NGC 7020, 
the outer ring was faintly detectable, but the inner ring was not
detected in H$\alpha$.  For NGC 7187, the inner ring was present 
while the outer ring was not detected in H$\alpha$.  The diamonds 
on the H$\alpha$ profile indicate the radii at which the azimuthally 
averaged flux drops below the uncertainty in the sky.  Each image is 
oriented with north up and east to the left.} \label{img6}
\end{figure}

\begin{figure}
\includegraphics[width=0.8\textwidth,angle=90]{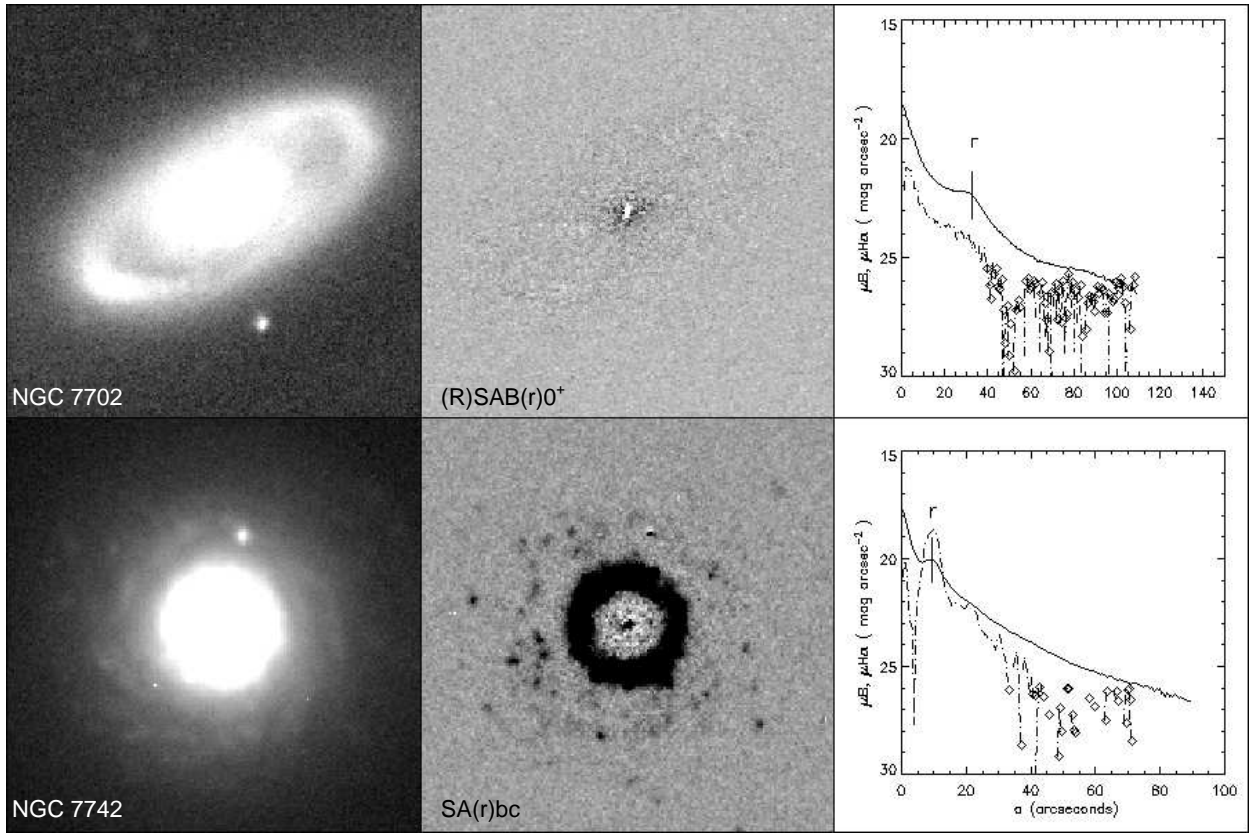}
\hfill \caption{$B$-band (left column) and H$\alpha$ (middle column)
images, and $B$-band (solid line) and H$\alpha$ (dashed line)
surface brightness profiles (right column) of NGC 7702 (top row) and
NGC 7742 (bottom row). The NGC 7702 images a scale of 1.$\arcmin$5,
and the NGC 7742 images have a scale of 1.$\arcmin$3. The vertical lines 
on the surface brightness profiles indicate the ring's semimajor axis 
(see Table 7).  For NGC 7702, the size of the inner ring is based on 
the broadband image of the galaxy since neither ring was detected in 
H$\alpha$.  It should also be noted that the faint outer ring of NGC 7702 
is not shown in the $B$-band image.  The diamonds on the H$\alpha$ 
profile indicate the radii at which the azimuthally averaged flux 
drops below the uncertainty in the sky. Each image is oriented with 
north up and east to 
the left.} \label{img7}
\end{figure}

The images are presented in Figures~\ref{img1} -~\ref{img7} with 
each row showing data for a single galaxy. The first column
is a log scale $B$-band image with the galaxy name labeled in the
lower left corner. The middle column is an H$\alpha$ emission image
with the galaxy type in the lower left corner. The $B$-band and 
H$\alpha$ images for a single galaxy are shown in the same scale 
which is listed in the figure description. The third column compares 
the galaxy's $B$-band (Grouchy 2008) and H$\alpha$ surface brightness 
profiles, which are azimuthally averaged radial profiles.  
The profiles were measured in the plane of the galaxy with elliptical 
orientation parameters based on the broadband images as listed in 
Grouchy (2008). In order to compare the two profiles, the H$\alpha$ 
surface brightness was assigned an arbitrary zero point of 18.945 
mag just as Koopmann \& Kenney (2006) did in their H$\alpha$ survey.

\textbf{ESO 111$-$22} (Figure~\ref{img1}, top row). A strong example of
a double-ringed nonbarred galaxy, type (R)SA(r)b, with a prominent,
smooth inner ring and a large, patchy outer ring seen in the
$B$-band. In spite of the obvious $B$-band morphological differences, 
the two rings of ESO 111$-$22 both show discrete HII regions, the 
inner ring being especially prominent.  Near-side extinction may be 
causing the inner ring to be weak on the east side. The rings appear 
asymmetric; the outer ring, in particular, appears distorted from a 
smooth oval shape.  

Of particular interest are the sizes of the two rings in this galaxy.
Using measured dimensions in arcseconds and the GSR distance scale of
0.645 kpc arcsec$^{-1}$, we derive diameters of 20.1 and 63.1 kpc for
the inner and outer rings, respectively.  The ring diameters are much
larger than the typical rings in nonbarred galaxies studied by de
Vaucouleurs \& Buta (1980). This is discussed further in Section 5.2.
The ratio of projected ring diameters is 3.1 (see Table 8), larger than
is typically seen in double-ringed barred galaxies (see review Buta \& Combes
1996).  We estimate that the main ring of ESO 111$-$22, the inner 
ring, contributes 26\% of the galaxy's total H$\alpha$ emission while 
the outer ring contributes 13\%.
  
Comparing the surface brightness profiles for ESO 111$-$22 shows what
appears to be an offset in the positions of the azimuthally averaged 
peak of emission for the inner ring from the $B$-band to the H$\alpha$. 
The $B$-band ring peaks around 12.$\arcsec$5 while the H$\alpha$ ring 
peaks around 20$\arcsec$.

\textbf{ESO 198$-$13} (Figure~\ref{img1}, middle row). This galaxy is
classified as type (R)SA(r)ab.  ESO 198$-$13 actually has three 
ring features, but only the inner two are displayed in the image.  
The small, bright ring close to the nucleus was interpreted by 
Buta \& Crocker (1993) and Buta (1995) as a {\it nuclear} ring 
since it is very much the analog of a nuclear ring in a barred 
galaxy. However, this feature has a linear diameter of 4.76 kpc, 
considerably larger than the average nuclear ring diameter of 
1.5 kpc ($H_o$ = 75 km s$^{-1}$ kpc$^{-1}$; Buta \& Crocker 1993).  
For the purpose of clarity, we will continue to refer to this 
feature as a nuclear ring even though it may be analogous to an 
inner ring.

In H$\alpha$, the nuclear ring is a prominent well-defined feature 
with discrete HII regions and some central emission. This ring appears
normal and symmetric around the center of the galaxy. The larger, more
flocculent pseudoring appears in the $B$-band image as a single
spiral arm wound $\approx$520$\degr$, but it also could be a
superposition of two rings offset from each other. This region appears
in the H$\alpha$ image as scattered discrete HII regions.  The 
outer ring refers to a larger, faint ring not displayed in the image 
because the ring did not have any H$\alpha$ emission to show.   

The nuclear ring, inner pseudoring, and outer ring have measured
diameters of 4.76kpc, 16.8kpc, and 49.9kpc with the outer ring
measurement taken from the $B$-band image since the ring does not have
any ionized gas. The ratio of the outer to inner pseudoring
diameter is 3. However, if the nuclear ring is interpreted as an inner
ring instead, then the ratio increases to 10.5. The ratio of the
pseudoring to the nuclear ring diameters is 3.5, which is still larger
than the mean barred galaxy outer to inner ring ratio of only 2.2 (Buta
1995).  We have calculated that the nuclear ring emits 12\% of the total 
H$\alpha$ while the main inner pseudoring contributes 55\% of the emission.

From the surface brightness profiles for ESO 198$-$13, we see that
the location and width of the stellar and gaseous rings are similar.
The galaxy appears not to have diffuse ionized gas outside these
ring features which is indicated by the sharp dropoff in H$\alpha$
surface brightness outside the rings. Therefore, we can say that the
ionized gas is confined to the inner and nuclear rings, and it 
follows the spiral pattern.

\textbf{ESO 231$-$1} (Figure~\ref{img1}, bottom row).  This galaxy is
classified as an (R$^{\prime}$)SA(r)b. There are two rings, an inner
ring and an outer pseudoring. Both features are strongly
represented in the narrowband H$\alpha$ image. The inner ring
appears to be doubled or has a double spiral feature tightly wound
into the appearance of a ring. The H$\alpha$ emission is strong,
well ordered, and follows the stellar spiral structure seen in the
$B$-band image.

The ring sizes are not as extreme in this case as the previous
cases. The inner ring has a diameter of 12.2 kpc, still large for an
SA galaxy according to de Vaucouleurs \& Buta (1980) and emits 43\% 
of the galaxy's H$\alpha$ emission.  The outer pseudoring is 29.5 
kpc in diameter, 2.4 times larger, and emits only 18\% of the H$\alpha$.

The surface brightness profiles appear normal and follow each other
fairly well with the two rings seen in both profiles.  The placement
and width of the rings are in agreement.

\textbf{ESO 234$-$11} (Figure~\ref{img2}, top row).  This galaxy is
classified as an (R)SA(r)0$^{+}$.  Since ESO 234$-$11 is an 
early-type galaxy and does not have a lot of gas present, we 
would not expect to see strong H$\alpha$ emission from it.  While 
the $B$-band image shows a faint hexagonal outer ring, the narrowband 
image shows very little ionized gas mainly from the center of the 
galaxy.  The rings are not visible in the H$\alpha$ image, implying 
that they are made up of an older stellar population. The lack of 
ionized gas in the rings is also reflected in the H$\alpha$ surface 
brightness profile. The ring diameters based on the $B$-band image 
were measured to be 8.0 kpc and 23.5 kpc giving a ratio of 2.9.

\textbf{ESO 236$-$29} (Figure~\ref{img2}, middle row).  ESO 236$-$29 
appears to be a relatively normal (R)SA(r)a galaxy with three 
ring features.  Two of the three ring features are partly seen in 
the H$\alpha$ image: the nuclear and the inner rings.  The inner 
ring appears mostly complete and consists of discrete HII regions.  
The strongest feature in H$\alpha$ is from the center of the galaxy 
where ionized gas follows a more elongated distribution than the 
apparent stellar bulge isophotes. The similarity between the shape 
of the inner ionized gas distribution and the outer disk suggests 
that ESO 236$-$29 has an inner ionized gas disk.  In blue light, 
the inner ring is a well-defined feature possibly made of tightly 
wrapped spiral structure. The outer ring could also be interpreted 
as being made of a single spiral arm.

Interestingly, the high radial velocity of ESO 236$-$29 places it 
at a distance of 218.6 Mpc, leading to enormous sizes for all of 
its apparent ring features: 12.1 kpc for the nuclear ring, 40.9 kpc 
for the inner ring, and 73.8 kpc for the outer ring.  The galaxy 
has been interpreted as a collisional ring galaxy by Madore et al. 
(2009), which could in principle explain these large 
sizes.  The ratio of outer to inner ring is only 1.8 while the 
ratio of inner to nuclear ring is 3.4.  The nuclear ring contributes 
18\% of the H$\alpha$ emission, while the main inner ring contributes 
only half as much with 9\% of the galaxy's H$\alpha$.  

The $B$-band and H$\alpha$ profiles appear similar to each other 
except for a more pronounced H$\alpha$ peak occurring around 
7$\arcsec$. The analogous feature in the $B$-band profile appears 
flatter and may be a lens-like feature instead of as a true ring.

\textbf{ESO 286$-$10} (Figure~\ref{img2}, bottom row). This galaxy is
classified as an (R$^{\prime}$)SA(r)a.  In the $B$-band image, we
see a single, faint, off-center ring which is the outer pseudoring. 
It appears to be made mostly of a single spiral arm, although a 
second outer arm is definitely present on the east side. There is 
very little H$\alpha$ emission in the ring-arm or in the galaxy as 
a whole. The nucleus has some ionized gas, but we only see small 
patches of very faint emission from the outer ring of the galaxy. 
The inner ring is a partial feature only weakly evident in $B$ and 
completely invisible in H$\alpha$.  

We estimate the size of the outer ring to be 43 kpc and find that 
it contributes approximately 10\% of the galaxy's total H$\alpha$ 
emission.   

Both the $B$-band and H$\alpha$ profiles are featureless.  The H$\alpha$ 
profile shows no emission in the inner 30$\arcsec$ after a drop off 
from the nucleus.  Even though the outer ring is not visible in the 
H$\alpha$ image, its presence can be discerned by an increase in 
the H$\alpha$ profile beginning around 45$\arcsec$.

\textbf{ESO 297$-$27} (Figure~\ref{img3}, top row).  This galaxy is
classified as an SA(rs)bc and was studied in greater detail in Paper 1.
It appears, at first glance, to be a normal intermediate to late-type
spiral galaxy. However, a closer inspection of the $B$-band images
shows a single inner arm winding counter to faint outer arms. There is
no trace of the counter-winding structure in the H$\alpha$ emission of
the galaxy which appears patchy and disorganized.

Using the $B$-band image, the inner pseudoring is approximately 15.0
kpc in diameter and a corresponding ring in the H$\alpha$ image is
responsible for 7\% of the emission.

The H$\alpha$ surface brightness profile follows the broadband profile 
fairly well including a sharp dropoff beginning around 45$\arcsec$.  
The H$\alpha$ image as well as the surface brightness profile implies that
there is a disk of diffuse ionized gas present throughout the
galaxy, but it is not organized into a ring. 

\textbf{ESO 399$-$25} (Figure~\ref{img3}, middle row). This galaxy is
classified as an (R$^{\prime}$)SA(s)0/a. The $B$-band image shows a
faint outer ring, which is not seen in the H$\alpha$ image. As an 
early-type galaxy, there is little or no ionized gas seen in the 
narrowband image.   

Using the broadband ring derived parameters, we estimate the outer 
ring to be 32 kpc in diameter; however, since it has no measurable 
ionized gas, it is an older stellar component.  The lack of dust and 
ionized gas is reflected in the featureless profiles of the broadband and narrowband
images.  The H$\alpha$ profile shows a sharp drop off of emission beyond
the center into what appears to be a very faint ionized disk.  The
majority of H$\alpha$ emission detected from this galaxy comes from the
center.

\textbf{ESO 409$-$3} (Figure~\ref{img3}, bottom row). This galaxy is
classified as an (R$^{\prime}$)SA(s)ab.  The galaxy has a faint
pseudoring of H$\alpha$ emission, which follows the pseudoring seen 
in the $B$-band image. Most of the ionized gas is concentrated near 
the center of the galaxy.  We estimate that the pseudoring has a 
diameter of 36.5 kpc and emits 53\% of the galaxy's H$\alpha$.

The H$\alpha$ surface brightness profile of this galaxy shows a drop
in emission from the center to the ring.  The ring is seen in the
H$\alpha$ profile and image, but it is not as broad as the stellar
component.  

\textbf{ESO 526$-$7} (Figure~\ref{img4}, top row). This galaxy is
classified as an SA(r)b. In the $B$-band image, we note the presence of
a broad inner ring with patchy, multi-armed spiral structure continuing
beyond the ring.  This inner ring is also seen as a patchy narrow ring
of emission in the H$\alpha$ image.  The inner ring has a diameter of 
21.6 kpc and emits 15\% of the total H$\alpha$ flux from the galaxy. 
The H$\alpha$ and the $B$-band profiles are similar to the ionized gas 
tracing the inner ring as well as the spiral structure beyond.

\textbf{ESO 576$-$57} (Figure~\ref{img4}, middle row). This galaxy is
classified as an (R)SA(r)ab. Two rings are seen in the $B$-band
image: a thin outer ring and a more diffuse inner ring. Neither
feature is exceptionally strong, and the outer ring is only
partially present. The H$\alpha$ emission does not reflect this 
structure however. A small amount of ionized gas was detected 
in the center of this early-type galaxy, and the stellar rings show 
very little sign of H$\alpha$ emission.  While the low level of 
H$\alpha$ emission throughout the galaxy may be real, it may also 
indicate that our observations were made  through the incorrect filter. 
At the time of the observations, the recessional velocity was listed 
on NED as 8883 $\pm$ 53 km s$^{-1}$. 
However, it is now listed at a much lower value of 4627 $\pm$ 10 km s$^{-1}$. 

In linear dimensions, the inner and outer rings are 13.2 and 20.6 kpc 
in diameter, respectively.  Therefore, the ratio of outer to inner ring 
diameter is 1.6.  The main inner ring was estimated to contribute 11\% of 
what little H$\alpha$ was detected from the galaxy.    

The $B$-band and H$\alpha$ surface brightness profiles appear smooth and 
follow each other closely in shape.  Even though the image does not 
show much H$\alpha$ emission, the disk of ESO 576$-$57 does appear to 
be filled with diffuse ionized gas.

\textbf{IC 1993} (Figure~\ref{img4}, bottom row).  This galaxy is a
low-luminosity member of the Fornax Cluster and is classified as an
(R$^{\prime}$)SA(s)b. The broadband image is taken from Buta \&
Crocker (1991) and shows the flocculent but distinctive outer
pseudoring, which appears as multiple spiral arms winding into each
other.  The outer pseudoring appears in the H$\alpha$ image mainly as
discrete HII regions, although many also lie inside the feature. The
brightest of the HII regions is on the western side of the galaxy,
which corresponds to the side with relatively stronger and less
flocculent spiral structure in the $B$-band. There appears to be no
real presence of ionized gas in the center of the galaxy.
We find that the outer pseudoring has a diameter of 5.7 kpc and 
emits 80\% of the galaxy's H$\alpha$.

The profile of H$\alpha$ emission for IC 1993 shows that no ionized 
gas was detected in the center of the galaxy.  The profile does show 
a few patchy peaks of emission corresponding to the flocculent 
pseudoring seen in the broadband image.

\textbf{IC 5267} (Figure~\ref{img5}, top row).  This galaxy is
classified as an (R)SA(rs)0/a.  The broadband image shows two thin 
rings of spiral structure, which are intermittently traced by HII 
regions in the narrowband image.  Overall, the galaxy has very little 
H$\alpha$ emission; however, the emission it does have appears to 
follow the spiral arms. The $B$-band image shows a minor axis dust 
lane, suggesting that IC 5267 has suffered a recent minor merger 
that may account for its complex structure.

We find that the inner and outer ring features have diameters of 
18 kpc and 33 kpc, and output 4\% and 6\% of the galaxy's H$\alpha$ 
flux, respectively.  The ratio of outer to inner ring diameter is 1.8.

The H$\alpha$ emission profile shows areas of emission corresponding 
to the inner pseudoring and the outer ring of the galaxy.  The profile 
drops rapidly from the center, but emission increases at around 30$\arcsec$.  
A similar profile is seen in the $B$-band image, but the change in the 
profile corresponding to the rings is more gradual in this filter.  The 
features may be more difficult to discern in the broadband image due to 
the disk being dominated by light from the bulge (see Grouchy 2008 for 
bulge-disk properties of this galaxy).

\textbf{NGC 5364} (Figure~\ref{img5}, middle row).  This galaxy is
classified as an SA(r)bc. It is an intermediate to late-type galaxy
with bright grand design spiral structure. An off-center inner ring
prominently stands out against the underlying disk and the surrounding 
spiral arms in the $B$-band image. The arms are traced strongly by HII
regions of ionized gas. The ring itself appears to have more H$\alpha$ 
emission on the northern side of the galaxy when compared to the 
southern side. This trend appears true for all of the spiral
structure; there is a northwestern side dominance in the presence of
ionized gas. NGC 5364 has a major companion, NGC 5363, to the north
that may be influencing its apparent morphology.  The inner ring 
diameter is measured to be 6.7 kpc and emits 15\% of total H$\alpha$ 
flux. 

The H$\alpha$ emission profile of NGC 5364 is patchy, showing many 
peaks of emission corresponding to the spiral structure.  The 
center of the galaxy shows no emission, and the inner ring appears to 
be the point of first H$\alpha$ detection.  The emission 
from the inner ring peaks around 30$\arcsec$ and is much narrower than 
the corresponding feature seen in the broadband profile.  In general, 
the $B$-band profile shows similar features as the H$\alpha$ profile, 
but each feature appears narrower in the narrowband.

\textbf{NGC 5530} (Figure~\ref{img5}, bottom row). This galaxy is
classified as an SA($\underline{\rm r}$s)c. The spiral structure is
patchy and disorganized in both the $B$-band and H$\alpha$ images.  We
do get the sense of an inner pseudoring in the $B$-band, which is not
seen outlined by the HII regions. These HII regions, instead, appear
unorganized and randomly distributed.  The inner pseudoring is 4.6 kpc 
in diameter and emits 14\% of the galaxy's total H$\alpha$ flux.

The broadband surface brightness profile shows an increase in 
flux around 35$\arcsec$ followed by an increase in H$\alpha$ flux 
around 40$\arcsec$.  This increased emission is likely due to HII 
regions associated with the inner pseudoring.  There is a second 
feature of increased emission occurring approximately at 80$\arcsec$, 
which is likely due to the flocculent spiral structure outside the 
ring.  Also interesting to note is that the optical disk extends 
well beyond any significant H$\alpha$ emission. 

\textbf{NGC 7020} (Figure~\ref{img6}, top row). This galaxy is
classified as type (R)SAB(r?)0/a. Although exceedingly regular and
well-defined, the interpretation of its bright inner hexagonal zone
as an inner ring and/or a bar is uncertain (Buta 1990b; Buta et al. 2007).

The $B$-band image shows that the large outer ring is completely
detached from the inner hexagonal zone of the galaxy. The possible
inner ring shows ansae at the ends of its major axis. This is a very
striking galaxy for ring structure, which is not reflected in the
H$\alpha$ emission.  Although the center is bright in H$\alpha$, 
the inner hexagonal zone and the two bright ansae are not visible 
in H$\alpha$, confirming their stellar dynamical nature. The galaxy 
appears to be fairly free of ionized gas, which is not surprising 
for an early-type galaxy. The outer ring has a measured diameter 
of 33.6 kpc and is responsible for 31\% of the galaxy's H$\alpha$ 
emission.

The surface brightness profiles of NGC 7020 appear smooth and show
emission from the outer ring in H$\alpha$ and in the $B$-band.
Within the outer ring, the H$\alpha$ profile does not show much 
emission except at the very center. While the H$\alpha$ image does 
not show the ansae features, the profile does indicate a slight 
increase in flux at their location of approximately 35$\arcsec$ 
radius.

\textbf{NGC 7187} (Figure~\ref{img6}, middle row).  This galaxy is
classified as (R)S$\underline{\rm A}$B(r)0$^+$. The morphology 
consists of a very faint outer ring and a striking inner ring, 
which appears normal and symmetric.  There appears to be no 
strong bar, but the bulge shows isophote twists, and there may 
be up to three bar-like features (Buta 1990a; Wozniak et al. 1995).  
While the $B$-band image shows both rings, only the inner ring 
appears in the H$\alpha$ emission.  Even though this is an 
early-type galaxy, gas ionized from a possible recent event of star 
formation exists in the inner ring feature.  Perhaps this is a 
relatively young, new feature of the galaxy while the outer ring 
is an older stellar component.  The inner ring has a diameter of 
7.0 kpc and emits 54\% of the galaxy's total H$\alpha$ flux.  
%The outer ring has a diameter of 17.5 kpc and does not emit 
%any measurable H$\alpha$.

The $B$-profile shows the underlying light due to the large 
bulge of this early-type galaxy, which broadens the detectable 
ring features.  The H$\alpha$ emission from the inner ring is 
seen well in the narrowband surface brightness profile as well 
as the image.  This ring does coincide with a stellar feature, 
but the narrowband emission is much narrower.  The H$\alpha$ 
profile also shows a slight increase in the emission from 
35$\arcsec$ to 60$\arcsec$ which follows a similar stellar 
feature and coincides with the position of the outer ring at 
51.$\arcsec$5.   

\textbf{NGC 7217} (Figure~\ref{img6}, bottom row). This galaxy is
classified as type (R)SA(r)ab. It has a large, flocculent outer 
ring with a sharp, well-defined inner ring inside the bright 
bulge.  The outer ring appears to be the main source of 
H$\alpha$ flux from this galaxy.  There is a wide zone of HII 
regions surrounding the inner ring, and this zone has a sharp, 
ring-like boundary on its outer edge.  Thus, NGC 7217 appears to 
be an example of a three-ring nonbarred galaxy. The illustrated 
broadband image is from Buta et al. (1995).

The inner ring has a diameter of 1.7 kpc and contributes 4\% of 
the total H$\alpha$ emission.  This ring has a zone of HII 
regions surrounding it which extends out to a diameter of 5.1 kpc.  
The outer ring is approximately 12.0 kpc in diameter and 1.4 kpc 
wide.  

The $B$-band profile is smooth and shows the broadened stellar 
outer ring feature.  On the other hand, the H$\alpha$ profile 
shows little emission, and the emission is sporatic.  Of the 
emission that was detected, the majority of it comes from the 
outer ring.

\textbf{NGC 7702} (Figure~\ref{img7}, top row).  This galaxy is
classified as an (R)S$\underline{\rm A}$B(r)0$^{+}$.  It has a 
faint outer ring, 
not displayed in the $B$-band image, which was not visible 
in the H$\alpha$ image.  NGC 7702 also has an inner ring which 
is very striking in the broadband image, but mostly disappears 
in the narrowband image.  

The inner ring of NGC 7702 has a diameter of 13.7 kpc and emits 
22\% of the total H$\alpha$ flux coming from this galaxy.  These 
measurements are based on the inner ring orientation parameters 
derived from the $B$-band image.

While the H$\alpha$ image does not show any significant emission, 
the surface brightness profile does show an increase in flux 
around 30$\arcsec$ which coincides with the inner ring in the 
broadband.  Although we were unable to detect any significant 
H$\alpha$ emission from the outer ring, there does seem to be an 
increase in H$\alpha$ flux beginning around 70$\arcsec$ which 
corresponds to the approximate location of the outer ring as seen 
in the broadband.   

\begin{figure}
\centering
\includegraphics[width=0.4\textwidth,angle=90]{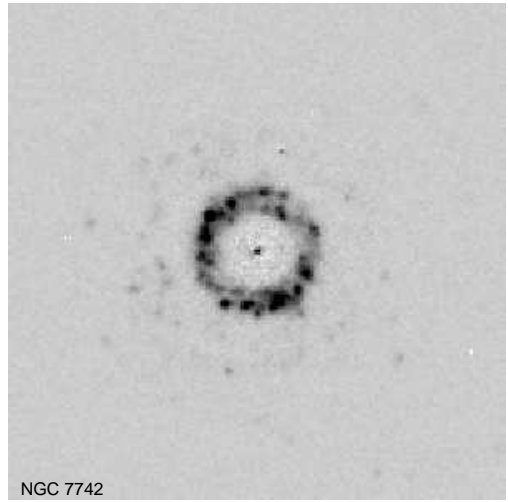}
\hfill \caption{Close up of NGC 7742 showing the braided morphology
of the inner ring.  It appears to be made up of two spiral arms
wound around each other.  The panel is $1.\arcmin45$ wide, and 
the image is oriented with north up and east to the left.}
\label{closeup}
\end{figure}

\textbf{NGC 7742} (Figure~\ref{img6}, bottom row). This galaxy is
classified as type S$\underline{\rm A}$B(r)ab, where the weak 
bar classification recognizes a small, broad oval located inside 
the inner ring. It has an inner ring that shows up strongly in 
the broadband and the narrowband images. The ionized gas appears 
to be distributed in a disk outside of this ring. The ring itself, 
in the narrowband image, when viewed closely, takes on the shape of 
two extremely well-defined and tightly wound spiral arms 
(see Figure~\ref{closeup}).  de Zeeuw et al. (2002) showed that the 
gas in NGC 7742 is rotating counter to the stellar background disk. 
Thus, the extreme appearance of this galaxy may be due to a merger 
of two galaxies.  The inner ring has a diameter of 2.26 kpc and emits 
87\% of the galaxy's H$\alpha$ flux.  

The intense H$\alpha$ emission from the inner ring appears as a narrow, 
sharp peak in the surface brightness profile.  The $B$-band profile 
shows that the ring coincides with the H$\alpha$ ring, but the stellar 
ring feature is not as dramatic.

\subsection{Ring Diameters}

Comparing absolute blue magnitudes from Grouchy (2008) for the 
galaxies in Table 1 to the magnitudes from CBB96 for the Table 2 
sample, we find the same mean, $\langle$M$_B^o\rangle =-20.4\pm0.7$, 
suggesting it is reasonable to compare the two samples. Nevertheless, 
we have noted in Section 3.2 the large diameters of some of the rings in 
the Table 1 sample.  This suggests a selection bias toward larger 
than average rings. To get an idea of how much larger the rings might 
be, we use the analyses of de Vaucouleurs \& Buta (1980) and Buta \& 
de Vaucouleurs (1982).  De Vaucouleurs \& Buta analyzed ratios of 
ring diameters to galaxy diameters for a sample of nearby galaxies, 
and they found that relative ring sizes depend on family (i.e., barred 
or nonbarred) and stage (i.e., early, intermediate, or late type).  
They found that nonbarred galaxy rings have a smaller relative size 
than barred galaxy rings and that early- and late-type galaxies have 
rings smaller in relative size than intermediate types.

\begin{figure}
\centering
\includegraphics[width=0.6\textwidth, angle=0, clip=]{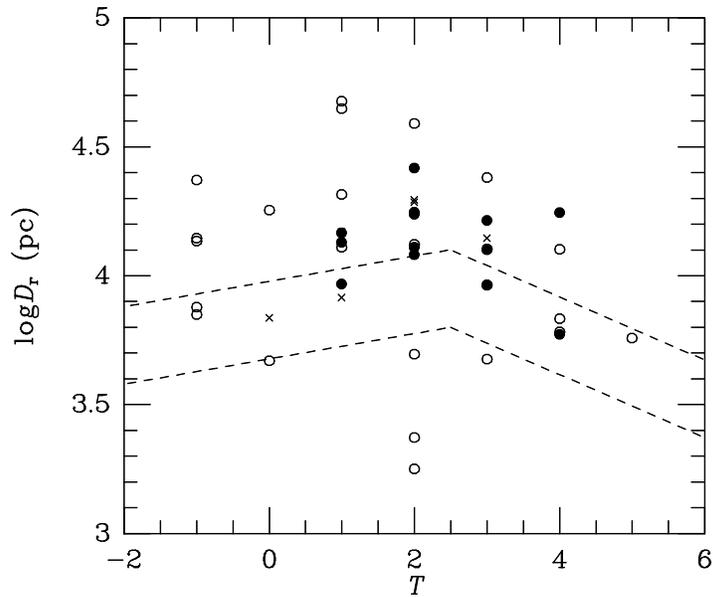}
\caption{Plot of the log of the deprojected inner ring diameter
vs. de Vaucouleurs type index. The dashed lines are for SB
galaxies (upper) and SA galaxies (lower) as derived for a nearby
galaxy sample by Buta \& de Vaucouleurs (1982), adjusted for Hubble
constant. The points plotted are for the combined barred/nonbarred
galaxy sample, with filled circles for SB types, crosses for SAB
types, and open circles for SA types.} \label{ringdiameters}
\end{figure}

Buta \& de Vaucouleurs (1982) then used a subset of galaxies having
known distances from various methods to calibrate the linear
diameters of the rings. Again, these diameters were found to depend 
on family and stage. After an adjustment for the distance scale 
used by Buta \& de Vaucouleurs (1982), which was consistent with 
a Hubble constant of $100\pm10$ km s$^{-1}$ Mpc$^{-1}$ (de
Vaucouleurs \& Bollinger 1979), Figure~\ref{ringdiameters} shows the
stage and family dependence for linear ring diameters. The upper
dashed lines are for SB(r) and SB(rs) galaxies, while the lower
dashed lines are for SA(r) and SA(rs) galaxies. SAB galaxies would
be intermediate between these lines. The points plotted on this
graph are for our combined barred/nonbarred galaxy sample, with
filled circles for SB types, crosses for SAB types, and open circles
for SA types. The diameters are from Tables 8 and 9. The plot shows
that indeed our rings are larger than the typical features seen in
nearby galaxies, for both barred and nonbarred but especially
for the nonbarred galaxies. In fact, the largest and smallest inner
rings are found among the SA galaxies. The larger scatter in SA ring
diameters compared to SB ring diameters was also found by de
Vaucouleurs \& Buta (1980) and Buta \& de Vaucouleurs (1982).

In a substantial revisiting of the general ring size issue, Wu (2008,
2009) has completely reconsidered the dependence of ring radii on
galaxy parameters using a sample from the CSRG which is statistically 
more complete.  Wu found little evidence for the stage and family
dependences highlighted by Buta \& de Vaucouleurs, at least for galaxies 
earlier than Sc. Wu develops a relation between absolute near-infrared
magnitudes, corrected to have a Gaussian error distribution, and the
log of the linear ring diameter $D$ in kiloparsecs, and finds that the
near-IR luminosity $L_{IR} \propto$ D$^{1.2}$, significantly different
from the $D^2$ relation found by Kormendy (1979) for blue-light
absolute magnitudes. Wu argues that the Buta \& de Vaucouleurs and
Kormendy samples suffered biases or selection effects, and indeed for
nonbarred galaxies the issue of ring distinctions (i.e., nuclear versus
inner) can contribute to an apparent family dependence in ring sizes.
Figure~\ref{ringdiameters} shows that the barred and nonbarred galaxies
in our combined samples have similar average diameters, likely because
the CSRG would not have detected the nuclear ring analogs in SA
galaxies.

\section{Derivation of Non-Axisymmetric Perturbations}

\begin{figure}
\includegraphics[width=1.0\textwidth]{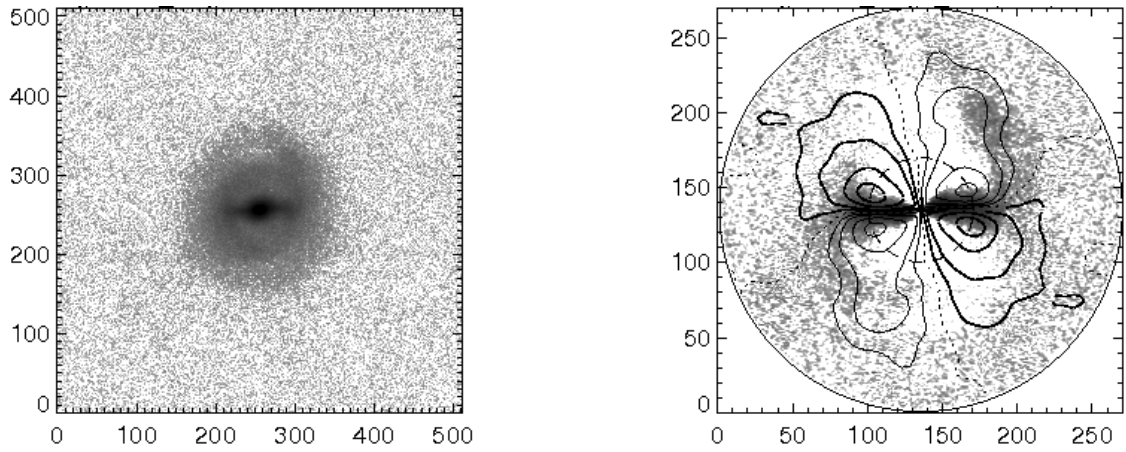}
\caption{Example images associated with measuring the bar
strength for the barred galaxy NGC 6761 from the CBB96 data.
Left: the red continuum image of the galaxy used to
estimate the galaxy's potential. Right: the force ratio
map overlaid onto the red continuum image minus $m=0$ Fourier
component.} \label{force}
\end{figure}

The gravitational torque method (Combes \& Sanders 1981; Buta \&
Block 2001; Laurikainen \& Salo 2002) was used to derive quantitative
estimates of the maximum non-axisymmetric perturbation in the combined
barred/nonbarred ringed galaxy sample.  The method, described in
Laurikainen \& Salo (2002), is based on transforming a red or 
near-infrared image into a gravitational potential under the assumption
of a constant mass-to-light ratio.  We use a polar method to convert
two-dimensional images into two-dimensional potential maps (see Salo et
al. 1999 for details).  The polar method reduces the noise which is
helpful in the fainter, outer regions of the galaxy. The potential is
written as the convolution of the density with the function 1/$\Delta
r$, modified to allow for finite disk thickness. From these potentials,
the radial, $F_R$, and tangential, $F_T$, force components are derived.  
Also, from the $m$=0 Fourier component of the potential, the mean radial 
force, $F_{0R}$, is derived. For barred galaxies, a measure of the bar 
strength is derived from the ratio map $F_T/F_{0R}$. In the presence of 
a bar, this kind of map shows a ``butterfly" pattern with islands of 
strong relative tangential forcing (see Figure~\ref{force}). The sign of 
the tangential force is dependent on a quadrant because this force 
component always points toward the bar ends. From the ratio map, we 
interpolate the following function:

\begin{equation}Q_T(r)=\frac{|F_T(r,\phi)|_{max}}{|F_{0R}(r)|}.\end{equation}

\begin{figure}
\includegraphics[width=0.8\textwidth, angle=90,clip=]{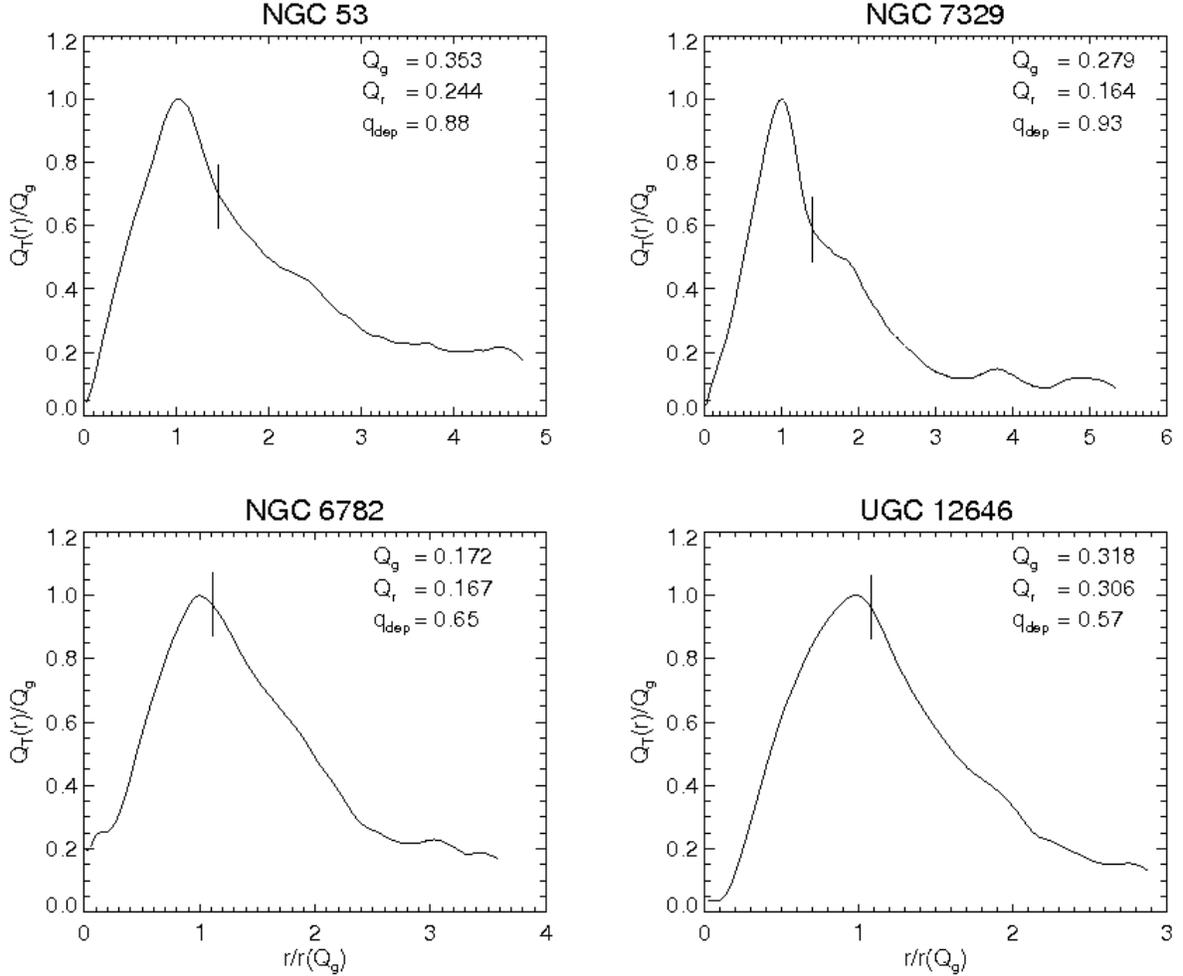}
\caption{Plot demonstrating the relationship between $Q_{g}$, $Q_{b}$, 
and $Q_{r}$ as defined for this study.  Each panel shows the 
strength of the non-axisymmetric perturbation as a function of radius 
for a strongly barred galaxy: NGC 53 (upper left), NGC 7329 (upper right), 
NGC 6782 (lower left), and UGC 12646 (lower right).  Each galaxy has a 
number associated with it which represents the maximum value in this profile.  
We call this maximum value $Q_{g}$.  If $Q_{g}$ occurs in the region 
of the bar, as it does for each of these galaxies, then $Q_{g}$ becomes 
$Q_{b}$ and is a measure of bar strength.  For each galaxy the value of 
$Q_{g}$ is listed in the upper right-hand corner of the panel.  Also for 
each galaxy, we have placed a line representing the position of the ring.  
The measure of the non-axisymmetric perturbation in the location of the ring 
is called $Q_{r}$.  For NGC 6782 and UGC 12646, the position of the ring 
is very close to the position where $Q_{g}$ is measured.  For NGC 53 and 
NGC 7329, the ring position is outside of the bar region.} \label{qring}
\end{figure}

\noindent
The maximum of this function, $Q_g$, gives a single measure of the 
total non-axisymmetry strength (see Figure~\ref{qring}). If the 
galaxy has a significant bar, $Q_g$ is the same as the bar strength, 
$Q_b$, while if a galaxy has a strong spiral but a weak or no bar, 
then $Q_g$ is a measure of the spiral strength, $Q_s$.  For a galaxy 
with an elongated ring, the maximum value of $Q_T$, at the 
{\it deprojected major axis radius of the ring}, is a measure of the 
non-axisymmetric force felt by the ring, which we will call $Q_{r}$ 
(see Figure~\ref{qring}).

For the 20 galaxies in our present sample, we used the $I$-band images
from Grouchy (2008) to derive $Q_{g}$ values, while for the CBB96
sample, we used narrowband red continuum images. In spite of being 
narrowband, the CBB96 images were deep enough to be useful for this 
purpose. The derivation of $Q_{g}$ requires an estimate of the 
scale height, $h_z$, of the vertical density distribution. We assumed 
an exponential vertical density distribution, although this choice has 
little impact on the results. The scale height $h_z$ is not directly 
observable for any of our galaxies (CBB96 or this paper). Instead, we 
must infer a value from the radial scale length $h_r$. For the combined 
barred/nonbarred sample, radial scale lengths were derived using the 
bulge/disk/bar (``bdbar") two-dimensional decomposition code described 
by Laurikainen et al. (2005). Parameters and descriptions of 
the decompositions for the Table 1 galaxies are discussed in a separate 
paper (Grouchy 2008). For the CBB96 sample, we only summarize our fitted 
radial scale lengths. De Grijs (1998) showed that vertical scale heights 
scale in a type-dependent manner with radial scale length, the ratio 
$h_z/h_r$ decreasing with later Hubble types. For types Sa and earlier, 
$h_z/h_r$ $\approx$ 1/4; types Sab to Sbc, $h_z/h_r$ $\approx$ 1/5; and 
types Sc and later, $h_z/h_r$ $\approx$ 1/9.  Laurikainen et al. (2005) 
show that the largest source of error in the measurement of $Q_{g}$ 
comes from the assumption of the disk scale height, which can be as 
high as $\pm_{20\%}^{8\%}$ and easily overshadows errors from 
orientation parameters (see Laurikainen et al. 2005 for details 
of error analysis). Figure~\ref{qring} shows the ratio map and $Q_{T}$(r) 
profile for four barred galaxies from  the CBB96 sample.

Table 8 summarizes our force calculation results. In addition to $Q_g$
and $Q_{r}$, the $m=2$ Fourier component at the ring radius (denoted
as $A_{2r}$), and the radius where $Q_{g}$ occurs (denoted as $r(Q_{g})$), 
are also compiled. Table 9 summarizes the same parameters for 27 of 
the CBB96 galaxies. The radial scale lengths listed in both tables 
come from one-dimensional and two-dimensional decompositions. The full 
details of the two-dimensional decompositions will be presented in a 
separate paper.  For our present purposes, we compared the two-dimensional 
derived disk scalelengths with those derived from a one-dimensional fit and 
chose the best fit. The galaxies which have one-dimensional fitted scale 
lengths are noted in the tables. The value of $Q_g$ for the combined sample 
ranges from $\approx0.04$ to 0.55.

\section{Rings and Galaxy Properties}

\subsection{Ring Star Formation Rates}

\begin{figure}
\includegraphics[width=0.9\textwidth]{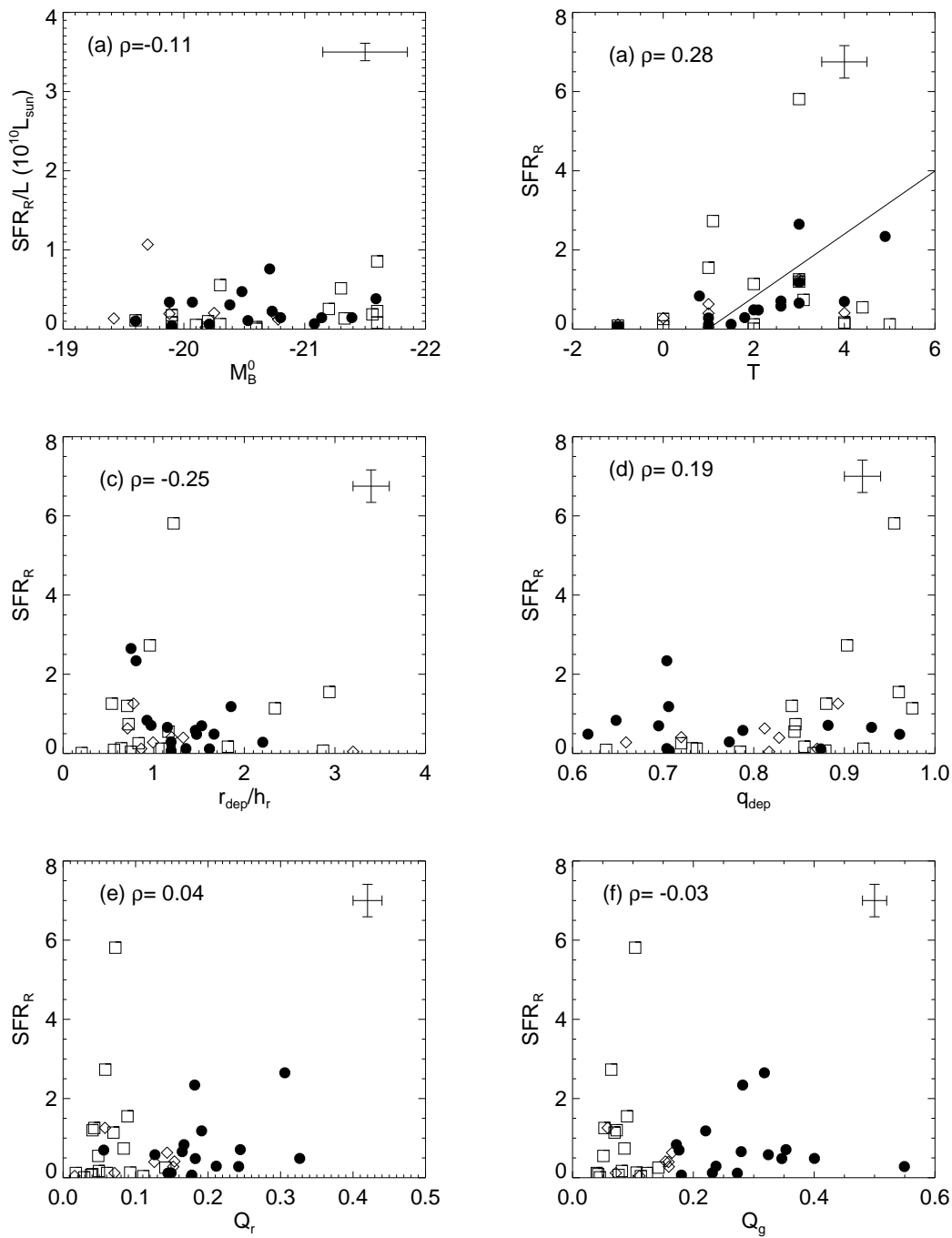}
\vspace{0cm} \caption{Plots of inner ring SFR in units of solar masses 
per year vs. (a) the absolute blue magnitude of the parent galaxy; (b) 
the de Vaucouleurs type of the parent galaxy; (c) the ratio of the deprojected 
radius of the main ring to the radial scale length of the parent galaxy, 
which is unitless; (d) the deprojected axis ratio of the inner  
ring in arcseconds; (e)  $Q_{r}$; and (f) $Q_{g}$.  For all 
panels, the open squares refer to SA galaxies, open diamonds refer 
to SAB galaxies, and solid points refer to SB galaxies.
In the upper left-hand corner of each panel is a representative 
errorbar of median parameters.  The OLS bisector line is included 
in the plots in which the correlation was found to be significant 
as discussed in the text.}\label{sfrplots}
\end{figure}

The data in Tables $1,2,6-9$ provide a sample of 44 
unique galaxies: 26 nonbarred or slightly barred (SA or 
S\underline{A}B) and 18 more strongly barred (SAB, SA\underline{B}, or SB) 
galaxies.  Figure~\ref{sfrplots} examines correlations between the 
SFRs of the inner rings and various galaxy parameters.  Three of the 
galaxies listed in Table 1 are also included in the CBB96 sample of 
Table 2, so we use the average values for these cases.  The open 
squares in these plots correspond to the SA galaxies; asterisks 
represent the SAB galaxies; and the solid points 
correspond to the SB galaxies.  The solid line included in each graph 
represents the "Ordinary Least Squares bisector" (OLS bisector) linear 
fit to the data as derived by the program SLOPES (Isobe et al. 1990).  
Using the OLS bisector was recommended by Isobe et al. when 
the errors of the data on both axes are either unknown or only vaguely 
known.  

To test for any correlations, we used the Spearman's rank correlation 
coefficient which measures the strength of the relationship between 
two parameters by assigning a number ranging between $\rho=-1$ and 
$+1$.  A rank correlation coefficient of $-1$ is considered a perfect 
anti-correlation; a value close to $0$ means no correlation is found; 
and a value close to $+1$ is considered a strong positive correlation.  
A non-parametric test quantity (i.e. rank correlation rather than Pearson's 
linear correlation coefficient) was chosen, in order to avoid making any 
assumptions about the unknown distributions of the compared quantities, 
or about the exact functional form of the correlation: note that the rank 
correlation coefficient between $log(y)$ and $log(x)$ is exactly the same as 
that between $x$ and $y$.

In addition to the rank correlation coefficient, the significance 
was calculated to establish the reliability of the relationship.  For 
this paper, we adopted the following convention: a significance less 
than $0.05$ is the maximum value to be considered significant, and 
a value less than $0.01$ is classified as very significant.

Figure~\ref{sfrplots}(a) compares the absolute blue magnitude normalized 
to the Galaxy's luminosity to the SFR of the inner ring.  The barred and 
nonbarred galaxies cover approximately the same 
range in $M_B^0$, and no correlation was found between the two 
parameters.

Figure~\ref{sfrplots}(b) shows the relationship of the SFR of the 
galaxy's inner ring to the de Vaucouleurs type index.  The plot shows a
larger spread in the SFR values toward higher index values, and the 
parameters have a correlation of $\rho=0.38$ with a significance of 0.02.

In Figure~\ref{sfrplots}(c), we search for any dependence between the 
SFR and the relative size of the inner ring. The parameter 
$r_{dep}/h_r$ is the deprojected semimajor axis radius of the ring
relative to the fitted radial scale length. There is a suggestion in
the plot that relatively smaller rings have higher SFRs; however, no
significant correlation was found between the two parameters.

In Figure~\ref{sfrplots}(d), the inner ring SFRs are compared with the
deprojected H$\alpha$ ring axis ratios, $q_{dep}$. The deprojections
for the present paper's sample used orientation parameters from
Grouchy (2008), while those for the CBB96 sample are from Table 4 of 
CBB96.  The issue of ring axis ratio is significant because CBB96 
showed that the distribution of HII regions around inner rings is 
sensitive to ring shape but not necessarily to $Q_g$. The rounder
nonbarred galaxy rings in our sample have a higher average SFR
compared to the barred galaxy rings of similar shape. For example,
the relatively circular inner ring of NGC 6935 from CBB96 has a
higher SFR than all of the CBB96 barred galaxies. In our present
sample, the extremely luminous galaxy ESO 231$-$1 has the highest
estimated ring SFR and is also a case where the inner ring is
relatively circular. In general, however, no significant correlation
between the inner ring SFRs and the ring shape was found.

In Figure~\ref{sfrplots}(e) and ~\ref{sfrplots}(f), the inner ring SFRs 
are compared with $Q_r$ and $Q_g$, respectively, and no significant 
correlation is found in either case. It is generally accepted that 
bars are efficient at collecting gas near major resonances, such as 
outer Lindblad resonance (OLR), ILR, and the inner 4:1 resonance (UHR), 
based on the early sticky-particle models of Schwarz (1981; 1984a; 1984b) 
and subsequent follow-up papers (e.g., Byrd et al.  1994; Rautiainen \& Salo 2000). 
More recent studies have suggested that resonances are less of an issue, 
and that ``orbit transition regions" (Regan \& Teuben 2004) or ``invariant 
manifolds" (Romero-G\'{o}mez et al. 2006, 2007) are better physical 
ways of describing rings. In all
of these interpretations, however, it is expected that stronger bars
will have a greater impact on gas flow and collecting material into
rings than would weaker bars.

In the classical resonance interpretation of inner rings, as material
is funneled inward to the ILR, some of the gas will collect and remain
in the UHR (see, e.g., Simkin et al. 1980). An inner ring will
form in this location if the bar strength near the UHR is strong enough
to trigger shocks in the area (see review of Buta \& Combes 1996). Compared 
to the ILR, the UHR is considered a relatively weak resonance, and 
some of the gas will leave the inner ring to settle in the nuclear ring. In order
to have continual star formation in an inner ring, the gas must be
replenished either through gas released through stellar evolution, by
the continued effects of the bar, or by external gas accretion
(Bournaud \& Combes 2002). If the bar weakens, we would expect the rate
at which the material moves inward to lessen, resulting in an eventual
drop in a ring's SFR. However, we find that there is not a significant 
correlation between the strength of the bar torque 
and the inner ring SFR of our sample.  The limitations of our sample 
may make it difficult to detect such an effect, but it is also clear 
that other factors besides bar strength must be involved in ring star 
formation, such as the amount of gas in the ring region.  For example, 
it is interesting that NGC 7267, which has the strongest bar in the 
sample, has only a weak, mostly open inner pseudoring.  In this case, 
most of the H$\alpha$ emission is confined to the bar itself.  Therefore, 
without a complete view of the SFR history and gaseous content of our 
sample, it is difficult to draw any strong conclusions from the lack 
of correlation between the ring SFR and the bar torque strength.

\subsection{Ring Axis Ratios}
\begin{figure}
\includegraphics[width=1.0\textwidth]{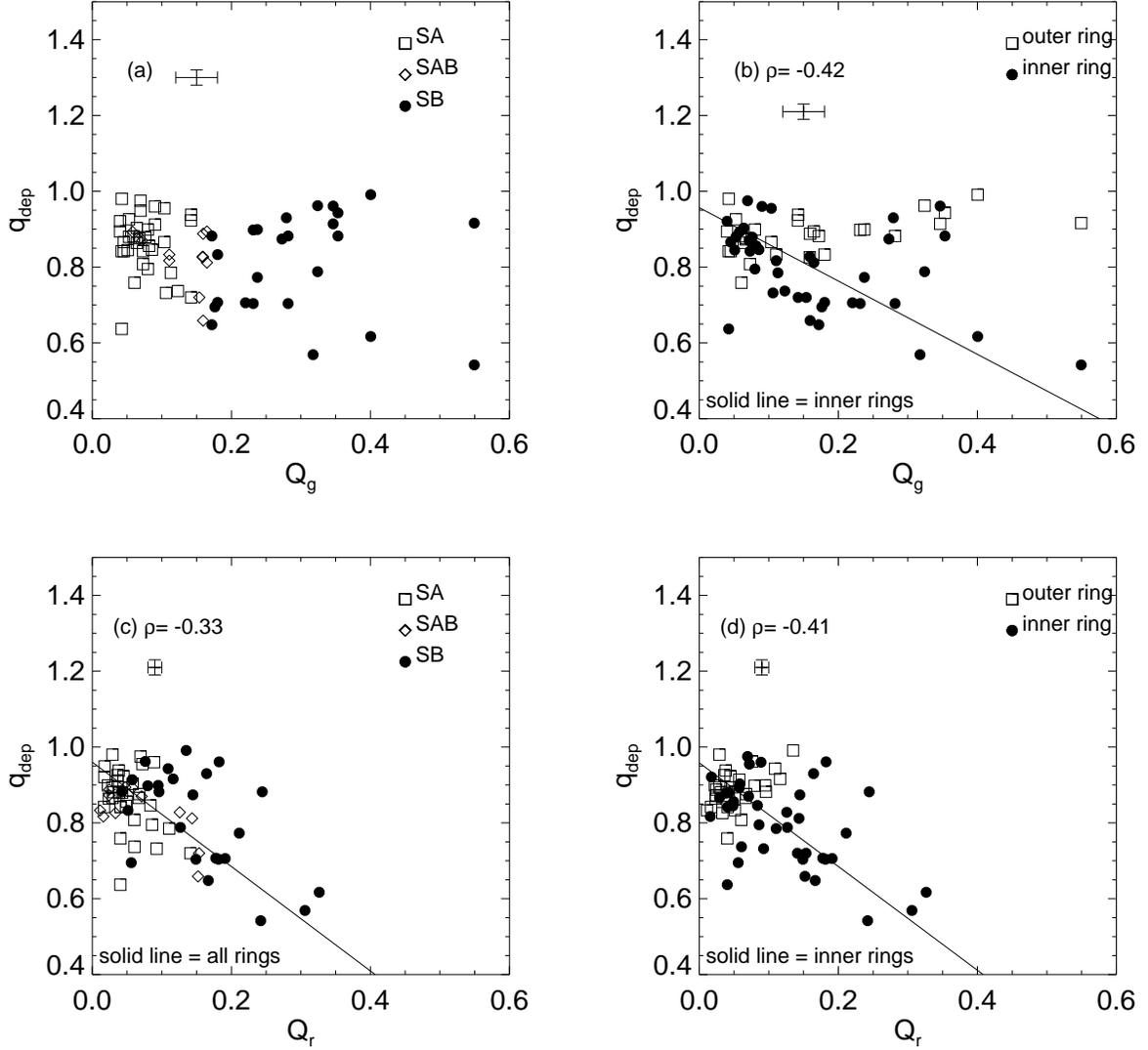}
\caption{Deprojected ring shape vs. (a) and (b) $Q_g$,
and (c) and (d) $Q_{r}$.  The shape of the outer ring is not
correlated to $Q_g$ or $Q_{r}$.  However, inner rings are 
related to $Q_{g}$ and $Q_{r}$.  For plots (a) and (c) open 
squares refer to SA galaxies, open diamonds refer to SAB 
galaxies, and solid points refer to SB galaxies.  For 
plots (b) and (d) open squares refer to outer rings, and 
solid points refer to inner rings.  The OLS bisector line is 
included in the plots in which the correlation was found to be 
significant as discussed in the text with the corresponding 
rank correlation coefficient displayed in the upper left 
corner.}\label{qgrings}
\end{figure}

From current theories of galaxy dynamics, we would expect {\it inner 
rings} to be more elongated in strongly barred galaxies than in 
weakly-barred ones (e.g, Salo et al. 1999; Schwarz 1985). However, 
Buta (2002) used the CBB96 sample to show that intrinsic inner ring 
shape does not depend strongly on bar strength as defined by the 
parameter $Q_g$. This was based on preliminary derivations of $Q_{g}$ 
and a few of the more exceptional examples, such as UGC 12646 and NGC 
6782 as barred galaxies with exceptionally elongated rings, and NGC 53 
and NGC 7329 as similar cases with much rounder rings.  Here we 
approach the issue differently by using a larger sample as well as a 
newly defined parameter $Q_r$.  Figure~\ref{qgrings} shows plots of 
deprojected ring axis ratio $q_{dep}$ versus $Q_g$ and $Q_r$, 
separating galaxies by family classification (SA, SAB, SB) and by 
ring type (inner, outer).  

When the deprojected axis ratio for all rings in the sample is plotted 
against $Q_g$, no correlation is found (see Figure~\ref{qgrings}(a)).  
However, when we compare the deprojected ring shape to $Q_r$, we 
find a correlation of $\rho=-0.33$ and a significance of 0.005 
(see Figure~\ref{qgrings}(c)). When focussing on only the {\it inner} 
rings, we find that both $Q_g$ and $Q_r$ correlate significantly 
with the ring shape (see Figure~\ref{qgrings}(b) and Figure~\ref{qgrings}(d), 
respectively).  For $Q_g$, we find a correlation of $\rho=-0.42$ with a 
significance of 0.007, and for $Q_r$, we find a correlation of $\rho=-0.41$ 
with a significance of 0.008.  Therefore, we are able to say that while 
inner rings appear to be affected by $Q_g$, outer rings are not.  
This is obvious by looking at the correlation of ring shape with $Q_r$: 
the bar induced tangential force is signifcant at the position of the 
inner ring, but not at the distance of the outer ring.

\begin{figure}
\includegraphics[width=0.85\textwidth]{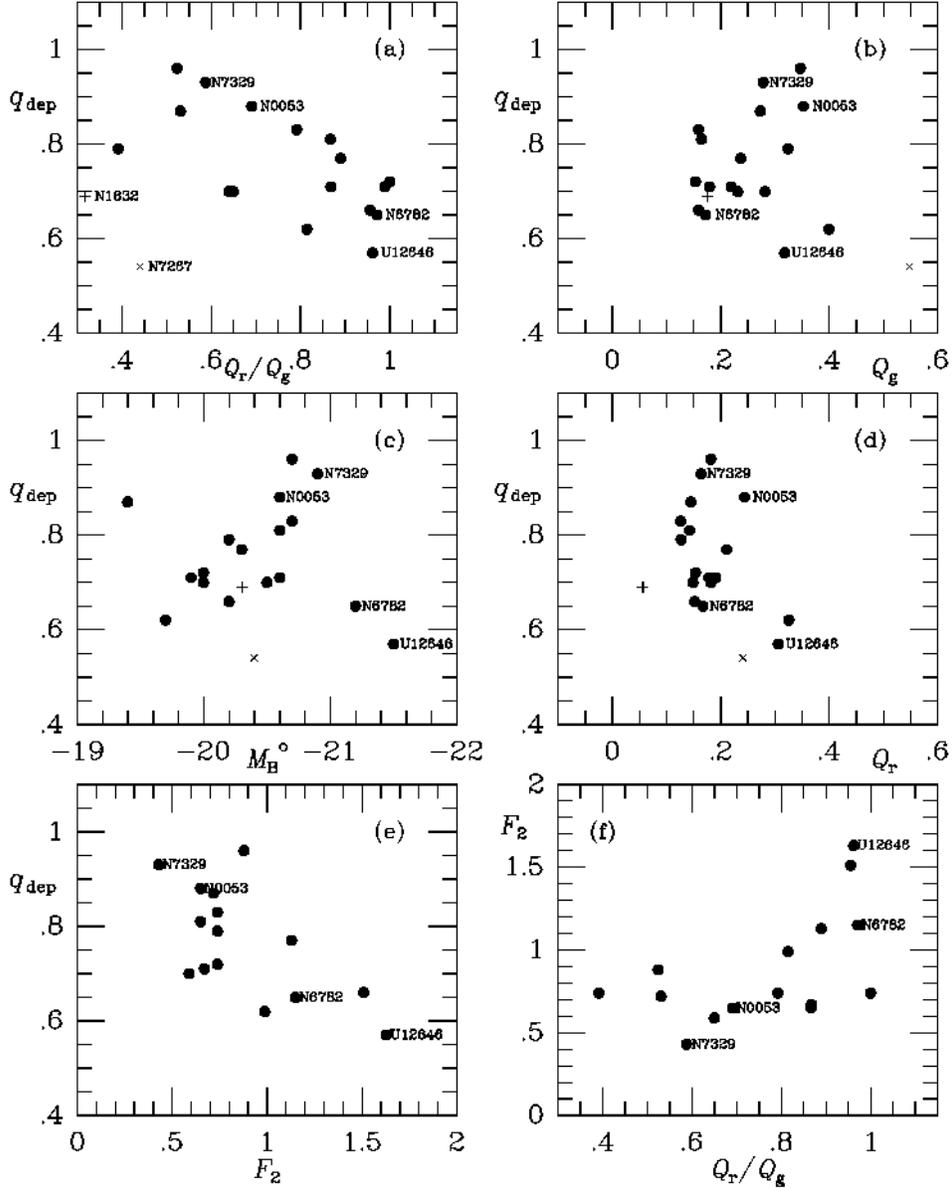}
\caption{Plots of 15 CBB96 galaxies having $Q_g$ $\geq$ 0.15.  
Constraining $Q_{g}$ in this way ensures that barred galaxies 
are selected.  The first five panels show the deprojected ring 
shape ($q_{dep}$) vs. (a) the relative forcing parameter 
($Q_{r}$/$Q_g$), (b) $Q_g$,  (c) absolute blue magnitude $M_B^o$,
(d) $Q_{r}$, and (e) the relative amplitude of the H$\alpha$ 
flux around the ring, $F_2$. Panel (f) shows $F_2$ vs. 
$Q_{r}$/$Q_g$} 
\label{qg015}
\end{figure}

\begin{figure}
\includegraphics[width=0.85\textwidth]{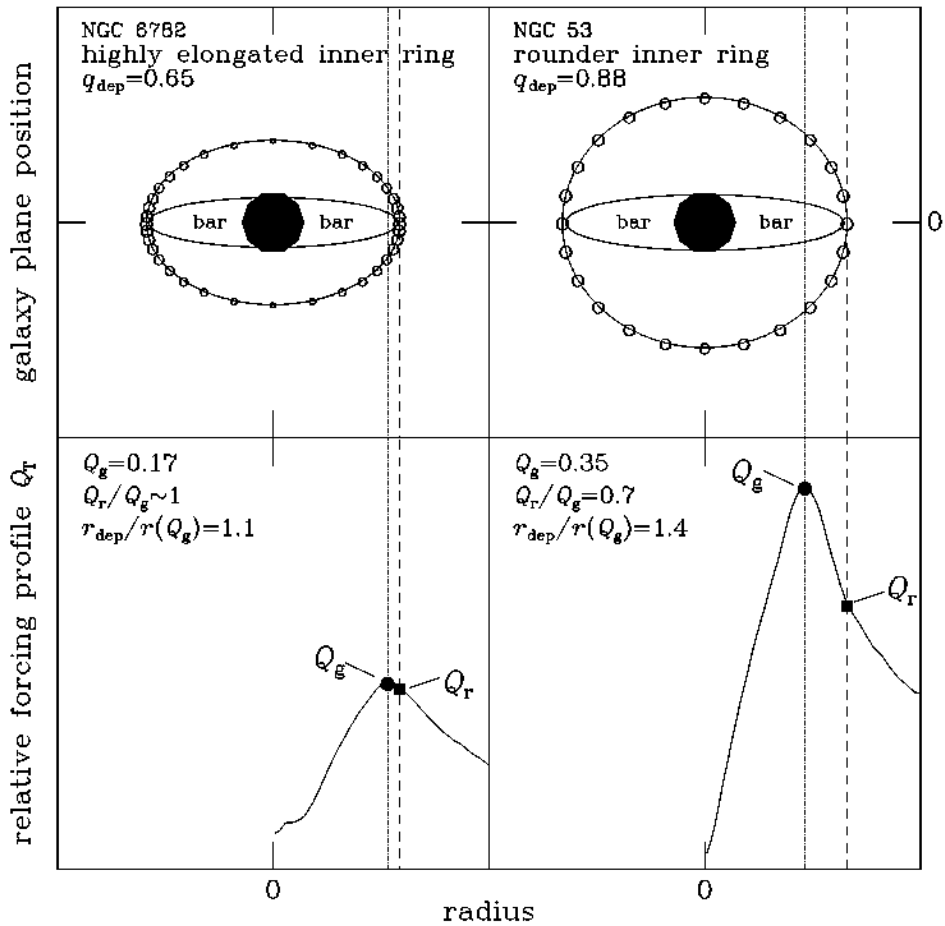}
\caption{Schematics of the deprojected inner rings of NGC 6782
and NGC 53 to highlight how the shape of the ring may be tied
to the relative forcing parameter $Q_{r}/Q_g$, rather than to
just $Q_g$ or $Q_{r}$. NGC 53 has a rounder inner ring than
does NGC 6782, even though $Q_{r}$ for NGC 53 exceeds $Q_g$
for NGC 6782. The small open circles lining the inner rings
are intended to highlight how the distribution and even luminosities
of HII regions vary around the rings in a manner dependent on
the inner ring axis ratio.}
\label{schematic}
\end{figure}

A greater insight into what is driving ring shape can be gained by
restricting the analysis to the barred galaxies. The reason for doing 
this is that we see round and highly elongated inner rings in strongly 
barred galaxies, and including nonbarred galaxies could confuse the 
possible reason for this. Figure~\ref{qg015} shows $q_{dep}$ versus 
$Q_g$ and $Q_r$ for the inner rings of 19 CBB96 galaxies having $Q_g$ 
$\geq$ 0.15. Several galaxies are highlighted, including the four 
best cases (NGC 53, 6782, 7329, and UGC 12646) of round versus 
elliptical SB inner rings from Buta (2002). In addition, 
Figure~\ref{qg015}(a) shows $q_{dep}$ versus the {\it relative forcing}
$Q_r$/$Q_g$. While there is little correlation between $q_{dep}$
and $Q_g$ or $Q_r$ (Figures~\ref{qg015}(b) and ~\ref{qg015}(d)) 
individually for this subsample, Figure~\ref{qg015}(a) shows a strong 
correlation of $\rho=-0.63$ with a significance of 0.002 between 
$q_{dep}$ and $Q_r/Q_g$. The two discrepant points, NGC 1832 and 7267, 
may depart the correlation for different reasons. NGC 1832 has a 
dominant spiral and a relatively weak bar, unlike most of the other 
galaxies, and although NGC 7267 has the strongest bar in the sample, 
it also has the weakest inner ring, a highly elongated, open pseudoring 
with an ill-defined major axis radius.  For the remaining 17 galaxies, 
Figure~\ref{qg015}(a) shows that the most elongated inner rings are 
found in galaxies having $Q_r/Q_g \approx 1$, while the roundest inner 
rings are found in galaxies having $Q_r/Q_g \approx 0.5$. This suggests 
that {\it the key factor determining ring shape in the presence of a 
significant bar is the location of the ring relative to the bar maximum}. 
This is schematically illustrated in Figure~\ref{schematic} using NGC 53 
and NGC 6782 as examples. When $r_{dep}$ is close to $r(Q_g)$, the ring 
lies essentially on top of the bar and can be highly elongated, even 
if $Q_g$ is relatively weak (as in NGC 6782 and IC 1438). But when 
$r_{dep}$ is well outside $r(Q_g)$, the ring has formed far enough from 
the bar maximum that it can be quite round, even if $Q_g$ is relatively 
strong (as in NGC 53 and 7329). These effects are manifested in the 
forcing ratio $Q_r/Q_g$.

Figures~\ref{qg015}(c),~\ref{qg015}(e), and ~\ref{qg015}(f) show other 
correlations for the same subset of galaxies. Figure~\ref{qg015}(c) 
shows that of the four labeled galaxies, the two with the most elliptical 
inner rings have the highest blue luminosity. The two with the most circular inner rings 
are 0.5$-$1 mag fainter than these. There is no significant correlation 
between $q_{dep}$ and the blue luminosity. Figure~\ref{qg015}(e)
shows $q_{dep}$ versus the CBB96 ring star formation distribution 
parameter $F_2$, defined as the relative amplitude of the $m=2$
Fourier component of the H$\alpha$ distribution around the rings.
$F_2$ is taken from Figure 15 of CBB96 and shows a similar correlation
with $q_{dep}$ for the H$\alpha$ distribution as for the continuum
ring light (Figure 16 of CBB96). As expected from Figure~\ref{qg015}(a),
$F_2$ roughly increases with increasing $Q_r$/$Q_g$.

\subsection{Other Correlations}

\begin{figure}
\includegraphics[width=1.0\textwidth]{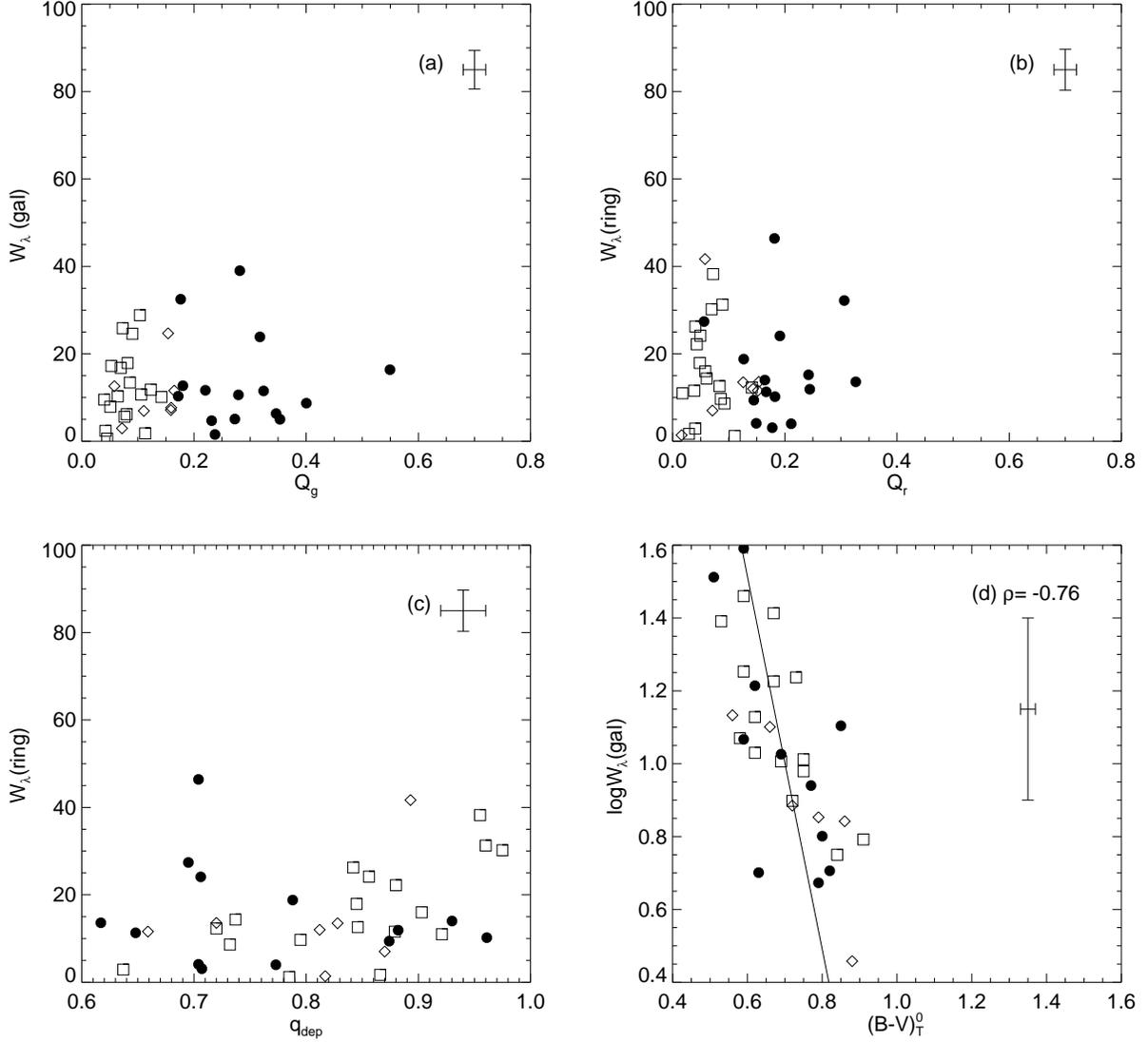}
\caption{Plots of (a) the global galaxy equivalent width of 
H$\alpha$+[NII] emission vs. $Q_{g}$; (b) the equivalent 
width of inner ring H$\alpha$+[NII] emission vs. $Q_{r}$; 
(c) the equivalent width of inner ring H$\alpha$+[NII] emission 
vs. the inner ring's deprojected axis ratio; (d) the 
log of the global equivalent width of H$\alpha$+[NII] vs. 
the corrected total color index, following Kennicutt (1983). 
In each plot, open squares refer to SA galaxies, open diamonds 
refer to SAB galaxies, and solid points refer to SB galaxies. 
In the upper right hand corner of each panel is a representative 
errorbar based on median parameters. The OLS bisector line and 
the Spearman rank coefficient are included in the plots in which 
the correlation was found to be significant as discussed in the 
text.} \label{ewplots}
\end{figure}

Figure~\ref{ewplots} shows several plots using the equivalent width,
$W_{\lambda}$, of the H$\alpha$+[NII] emission instead of the SFR.
$W_{\lambda}$ conceivably could give a different view of ring star
formation. Figure~\ref{ewplots}(a) shows the relationship between the
galaxy-wide H$\alpha$+[NII] equivalent width and $Q_g$. This plot 
includes the Table 1 sample and did not result in any correlation.
Our values of the galaxy-wide equivalent width are comparable to
those found by Hameed \& Devereux (1999) for another sample of
early-type galaxies. Figures~\ref{ewplots}(b) and ~\ref{ewplots}(c)
show the relationship between the equivalent width of H$\alpha$+[NII] 
emission from the inner ring to $Q_{r}$ and $q_{dep}$, respectively. 
No significant correlation was found in either case. 

Kennicutt (1983) showed that there is a correlation between $log
W_{\lambda}$ and the corrected total color index, ($B-V)_T^o$ of normal
galaxies. The correlation can be tied to properties of the initial mass
function, or IMF. Figure~\ref{ewplots}(d) shows the same plot for our
combined barred/nonbarred galaxy sample, using RC3 colors for the CBB96
sample and measured colors from Grouchy (2008) for the Table 1 sample.
The latter were corrected for Galactic extinction, tilt, and redshift 
using RC3 prescriptions. As Kennicutt found, the bluer galaxies in our 
sample have larger H$\alpha$+[NII] equivalent widths with a strong 
correlation of $\rho=-0.77$ and a significance $\le 0.0001$.

\begin{figure}
\includegraphics[width=1.0\textwidth]{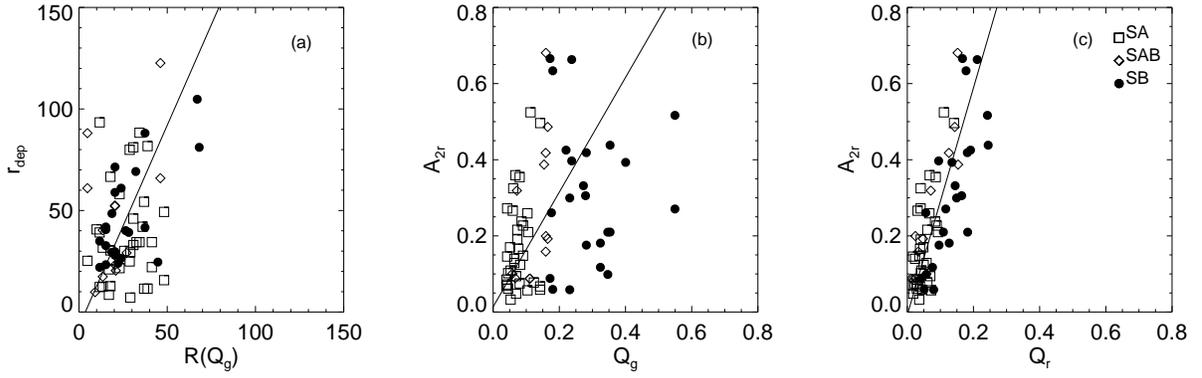}
\vspace{0cm} \caption{Plots of (a) the semimajor axis of the
deprojected ring, in arcseconds, vs. the location of $Q_g$; 
(b) the $m=2$ Fourier amplitude at the location of the ring 
vs. $Q_g$; and (c) the $m=2$ Fourier amplitude at the position 
of the ring vs. $Q_g$. The OLS bisector line is included in 
the plots in which the correlation was found to be significant 
as discussed in the text.}\label{A2plots}
\end{figure}

Figure~\ref{A2plots}(a) looks at the relationship between the semimajor
axis of the deprojected inner ring, $r_{dep}$, and the radius at which 
$Q_g$ is measured, both in arcseconds. There is a slight spread in the 
values of $r_{dep}$ and $r(Q_g)$, but a correlation was detected 
with $\rho=0.27$ and a significance 0.02.  Further investigation 
indicates that this relationship is due to the strongly barred SB galaxies 
which have a correlation of $\rho=0.57$ and a significance 0.003. 
Typically, the length of a bar is greater than or equal to the location 
of $Q_g$ (see Laurikainen et al. 2004 Figure 5; also Buta et al. 2009, 
Section 7). The location of an inner ring is typically thought to be 
related to the UHR (Rautiainen \& Salo 2000) whereas a 
self-sustained bar does not extend beyond the radius of corotation 
(Contopoulos 1980; but see Zhang \& Buta 2007 and Buta \& Zhang 2009). 
Our data show that the ratio of $r_{dep}$ to $r(Q_g)$ is greater than 
unity for the barred galaxies in the sample (see Figure~\ref{A2plots}(a)). 
Therefore, as expected, the rings in these barred galaxies lie at a 
radius larger than the radius of $Q_g$ and the radius of the bar. 
For nonbarred galaxies the correlation disappears, and the relationship 
becomes mostly random.

The last two panels of Figures~\ref{A2plots} compare the $m=2$ Fourier 
amplitude at the position of the ring with $Q_g$ (see Figure~\ref{A2plots}(b)) 
and $Q_{r}$ (see Figure~\ref{A2plots}(c)).  While the amplitude range is the 
same in both plots, the values of $Q_g$ have a wider spread than those of 
$Q_{r}$.  This spread appears to be attributed to the strongly barred 
galaxies of the sample which have $Q_g$ ranging from 0.2 to 0.6 and 
$Q_{r}$ ranging from 0.05 to 0.26.  At the same time, for nonbarred 
galaxies, $Q_{r}$ and $Q_g$ both range between 0 and 0.2 due to the lack
of a strong bar.  Since rings are usually located just outside the ends 
of a strong bar, we would expect $Q_{r}$ to be, on average, lower than 
$Q_g$ for barred galaxies.  Even with these differences, both $Q_g$ and 
$Q_{r}$ are strongly correlated with the $m=2$ amplitude.

\section{Discussion}

\subsection{Evolving Ringed Barred Galaxies}

The dynamical theory of ring formation in the context of bar resonances
is reviewed by Buta \& Combes (1996). In simulations, rings form by gas
accumulation at specific resonances, owing to gravity torques from the
bar pattern. The gas settles into a main periodic orbit near the
resonance that does not cusp or cross another orbit. There is no net
torque on the orbit, so that any further evolution of the ring will
involve mainly how it forms stars, how its sharpness will change over
time as ring stars age, and, in principle, how the bar itself (pattern
speed, strength) may change over time. In an equilibrium situation, a
ring will be aligned either parallel or perpendicular to the bar. Rings
should form stars as long as gas is available and presumably above a
certain threshold of surface density, and gas should be gathered
smoothly as long as the bar exists. A recent simulation of a
classical ringed barred galaxy is given by Treuthardt et al. (2008).

Given this basic scenario, a few questions arise: how does the time lag
between the gathering of gas into a resonance region and the onset of
star formation compare with the evolutionary timescale of the bar?
Once the bar has dissolved, how long does it take for the star
formation in the ring to cease? And, finally, once the bar has dissolved
and star formation in the ring has ceased, how long can the ring
persist as a visible ring-like feature? These questions are relevant
because bar dissolution is one possible explanation for rings in
nonbarred galaxies. We can make some judgments based on published
models.

The dissolution of a bar by the buildup of a CMC was considered by 
Hasan \& Norman (1990), who showed that such an
object would cause chaos in bar-supporting orbits near the center. If
the mass of this object increased such that the ILR reached the ends of
the bar, then the bar would be destroyed. Shen \& Sellwood (2004)
reconsidered the problem by determining not only how massive 
but also how dense a CMC needs to be to dissolve a bar. Their
conclusion was that a CMC would not, in most galaxies, have the
mass or central density needed to dissolve bars.

Bournaud \& Combes (2002) developed models where the flow of gas to the
center, driven by a bar, builds up the central mass sufficiently to
destroy the bar. But they also considered how external gas accretion
might prevent total bar destruction and even lead to bar regeneration
after the end of a previous major bar episode. In the presence of gas
accretion, a galaxy could be seen as barred for most of its lifetime.
Depending on model details, a galaxy may have one or more bar episodes
during a Hubble time. Even in the absence of accretion, a galaxy can
appear as barred for up to 2Gyr, before dissolving into an oval lens.
We can argue based on such models that in a worst-case scenario, a bar
may last only for 1Gyr. Bar development depends on model parameters.
For example, Rautiainen \& Salo (2000) show two-dimensional $n$-body 
models with variable Toomre stability parameter, $Q$. Considering $Q$ 
ranging from 1.25 to 2.5, they found that bars formed quickly for 
low $Q$ values and much more slowly or not at all for the high $Q$ values.

Test-particle simulations with an analytical bar and disk (e.g.,
Simkin et al. 1980; Schwarz 1981, 1984a; Byrd et al. 1994)
have provided insights into the timescales of ring formation, usually
in units of the bar rotation period. For a galaxy of absolute blue
magnitude $-$20.4, a characteristic rotation speed is 250 km s$^{-1}$,
based on the $B$-band Tully-Fisher relation of Yasuda et al. (1997). A
typical inner ring in our sample has a radius of 6 kpc, and if we
assume a flat rotation curve and that the inner ring lies near the
inner 4:1 ultraharmonic resonance, this would give a typical bar
rotation period of 2$\times$10$^8$ yr. The cited papers show that
gaseous inner rings may form in $3-7$ bar rotations ($0.6-1.4$ Gyr), 
and gaseous outer rings in $10-20$ bar rotations ($2-4$ Gyr), for a bar 
turned on fully in about two bar rotations.

The Rautiainen \& Salo (2000) models also used test particles for gas,
but for the stellar component, an $n$-body treatment was used,
therefore the bar formed from a natural instability. The timescales for
the rings were found to depend on the timescales for the bar, but once
the bar formed, ring formation timescales thereafter were similar to
those from the analytical models.

Given these approximate timescales, is it possible that a bar could
gather gas into a resonance region over some period of time, and then
dissolve {\it before} the ring is seen as an active star-forming zone,
giving us a nonbarred ringed galaxy with a ring SFR comparable to that
of a barred ringed galaxy? Tamburro et al. (2008) have considered the
time lag between the densest HI gaseous phase and the initial massive
star formation phase in spiral arms based on a comparison between HI
and 24 $\mu$m emission peaks in a sample of spiral galaxies. They
concluded that this time lag is fairly rapid, on the order of $1-4$ Myr.
If we consider that, in simulations, rings usually begin as pseudorings
made of wrapped spiral arms, then a similar timescale may be relevant
to the HI $\rightarrow$ 24 $\mu$m conversion in rings. From this, we
suggest that, after the bar has dissolved, it is unlikely that
significant star formation will suddenly turn on in a ring because of a
long time lag, which would have to be on the order of a Gyr for such a
scenario to work. It is most likely that star formation in a ring will
begin when the bar is strong, because that is when the bar will be most
effective at gathering gas into resonant regions. How long the star
formation proceeds in the absence of the bar will depend on the star
formation efficiency.

The case of NGC 7702 suggests that a ring may experience a rapid burst
of star formation, and then evolve in a quiescent manner for a long
time afterward. Buta (1991) isolated the bright inner ring of NGC 7702
from its surrounding background old disk light, and found that the net
colors are consistent with a ``star formation cutoff" model where the
star formation proceeds uniformly for 10$^7$ yr, and then cuts off
(Larson \& Tinsley 1978). The observed net colors suggest that star
formation ended in the ring $1-2$ Gyr ago, which may be long enough for a
primary bar to dissolve in that case. Our H$\alpha$ imaging confirms
the lack of any recent star formation in either of the rings of NGC
7702. In contrast to the ring colors, the colors of the background
star light at the ring position are consistent with those of a normal
S0$^+$ galaxy having a monotonically declining SFR. NGC 7702 may be the
kind of ``nonbarred" ringed galaxy we might expect from bar
dissolution. Even so, the galaxy is not completely unbarred. As we have
noted, it has a small nuclear bar and its bright inner ring is an
intrinsically oval, bar-like feature.

Athanassoula (1996) has shown that the morphology of a galaxy like NGC
7217 can be explained in terms of a dissolved bar model. In her model,
a bar is turned on after two bar rotations, and then is completely turned
off by 12 bar rotations. While the bar is present, the galaxy forms an
$R_2^{\prime}$ outer pseudoring that evolves to a more detached
feature. The ring is initially slightly elongated, but becomes more
circular once the bar is gone. Interestingly, the ring persists for
more than 20 additional bar rotations, only becoming more diffuse with
time. The model does not tell us how long star formation in the ring
might proceed, but as long as the ring remains a region of enhanced gas
density (which it is in the model), it will probably form stars. This
model suggests that star formation may proceed in a ring for as long as
$2-4$ Gyr after the bar has dissolved. Our H$\alpha$ image of NGC 7217
shows that the outer ring is where most of the HII regions are found,
and Buta et al. (1995) also found the ring to be a concentration of HI
gas. The authors derived the net colors of the outer ring of NGC 7217 
and found that they lie essentially on the normal galaxy
sequence, consistent with a roughly uniform SFR for most of the
lifetime of the galaxy. Although this result is very different from 
NGC 7702, NGC 7217 could still be another example of a nonbarred ringed
galaxy that evolved from a former barred ringed galaxy.

\subsection{Star Formation Triggering in Rings}
One of our main goals in this paper was to examine how important a bar
might be in triggering the star formation seen in inner rings. That is,
does the presence of a strong bar enhance star formation in a normal
inner ring? For our limited sample, the range of SFRs for barred and
nonbarred galaxy inner rings is similar. For a similar range in
absolute blue magnitudes, the average inner ring SFR for rounder
nonbarred galaxy rings is actually higher than for barred galaxy rings
of similar shape (Figure~\ref{sfrplots}(d)). This suggests that while a
bar may be capable of concentrating star formation in an inner ring, it
does not increase the overall efficiency of star formation in such a
ring. This may not be true for nuclear rings, where the star formation
efficiency can be comparatively high (e.g., Schinnerer et al.1997) and
enhanced nuclear star formation can cause correlations between IRAS
flux ratios and bar strength and length (Martinet \& Friedli 1997). Our
result for bars and inner rings is very similar to what McCall \&
Schmidt (1986) and Elmegreen \& Elmegreen (1986) concluded concerning
the impact of spiral density waves on global SFRs (Kennicutt 1998b).
Both studies concluded that density waves are not the main triggers of
star formation in galaxies.

Although our combined sample cannot be considered statistical, this
result is consistent with other previous studies. Moss \& Whittle
(1993) found no difference in detected H$\alpha$ emission between
barred and nonbarred galaxies in cluster and field samples.  Tomita 
et al. (1996) found no significant differences in the far-infrared to 
blue flux ratio, an indicator of the current SFR, between samples of 
barred and nonbarred spirals of
types Sa, Sb, and Sc. They concluded that an optical bar structure has
no impact on present-day star formation activity in galaxies. In
contrast to this, Devereux (1987) showed that 10$\mu$m luminosities are
definitely higher for early-type barred galaxies than in nonbarred
early types or late types.

Hameed \& Devereux (1999) also studied the H$\alpha$ emission for a
sample of early-type galaxies, including several with strong rings.
In addition to showing enhanced star formation in the rings, these
authors found that some early-type galaxies have a rate of massive
star formation comparable to Sc galaxies. Hameed \& Young (2003)
also showed that galactic interactions can have a significant effect
on the current SFR in many early-type galaxies.

An alternative interpretation of the lack of correlation between bar
strength and ring SFR in our sample could simply be sample
inhomogeneity. Some galaxy rings have formed in response to a bar,
while others have formed via a different mechanism, so that the ring
SFRs may depend on different factors. (See for example, Struck (2009), 
for an interpretation of our sample galaxy ESO 236$-$29 in terms of 
analytical collisional ring galaxy theory.)  Also, the sample may 
include a mix of rings in different evolutionary states. Martinet 
\& Friedli (1997) interpreted four classes of star formation activity 
among late-type barred spirals in terms of possible evolutionary paths
including the age of the bar.

\subsection{Ring Star Formation and Intrinsic Ring Shape}

The intrinsic shape ($q_{dep}$) of the ring appears to affect the star
formation process in rings more than does the maximum strength ($Q_g$)
of the bar. Nonbarred galaxy rings tend to be round, and HII regions
distribute fairly randomly around them. The same is true for a round
barred galaxy inner ring. But significant intrinsic elongation causes
HII regions to bunch up around the bar axis. If, in an equilibrium
situation, gas circulates around the ring as if it is a periodic orbit,
then clouds will spend more time near the ring major axis points
(around the bar ends) than they will near the ring minor axis points
(Contopoulos 1979). Byrd et al. (2006) argue that velocity
perturbations in the inner ring of NGC 3081 due to its extreme
elongated shape can explain the concentrated star formation in the ring
major axis regions. In this case, crowding, rather than spiral shock
fronts, might explain the HII region bisymmetry. Not only is the HII
region distribution in NGC 3081 affected by the ring shape, but also
the luminosity function of young clusters is affected in that more
luminous clusters are found near the bar major axis (e.g., Buta et al.
2004). Byrd et al. (2006) argue that clusters in the inner ring of NGC
3081 show a position-age sequence suggesting that they form around the
bar ends and move further upstream along the ring away from the bar
ends. In spite of the clear shape effects, when total ring SFR are 
considered, there is little dependence on $q_{dep}$ in our small sample.

Our new sample has allowed us to look into the relationship between
ring shape, bar strength, and star formation properties. Buta (2002)
had shown that for barred galaxies, there is little correlation between
$q_{dep}$ and $Q_g$, which acts as a good approximation to the bar
strength. By adding in our new sample of nonbarred galaxy inner rings,
some correlation between $q_{dep}$ and $Q_g$ is now found.  However,
for the whole sample, a better correlation is found between $q_{dep}$
and the maximum relative forcing at the ring major axis radius, $Q_r$.
Thus, at some level, $Q_r$ is a more important controlling parameter
for ring shape than is $Q_g$. Nevertheless, we have shown that for
inner rings in barred galaxies having $Q_g$ $\geq$ 0.15, the more
likely controlling parameter for ring shape is $Q_r/Q_g$. When this
ratio is close to unity, SB inner rings tend to be highly elongated,
while when significantly less than unity, SB inner rings tend to be
rounder. Weaker correlations are found between $q_{dep}$ and $Q_g$ and
$Q_r$ for this more restricted subsample. As bar strength decreases,
the ratio $Q_r/Q_g$ becomes ill-defined and subject to noise or other
nonaxisymmetric features, such as spirals.  The correlation between
$q_{dep}$ and $Q_r/Q_g$ indicates that indeed, on some level, a bar
does play a role in ring star formation. This is because $Q_r/Q_g$
determines $q_{dep}$, which in turn affects how ionized gas is
distributed in the inner ring. Yet, with our small samples, we have not
been able to uncover a connection between the rate of star formation in
inner rings and either the nonaxisymmetric torque strength or the ring
shape.

\section{Conclusions}

Our study, based on a combined sample of 26 nonbarred or weakly barred 
galaxies and 18 strongly barred galaxies, has led to the following conclusions:

\noindent
1. Rings in nonbarred galaxies are often well-defined, narrow zones of
HII regions and H$\alpha$ emission. This is consistent with the
enhanced blue colors seen in broadband images, indicating that the
rings are well-organized, active zones of star formation, a trait which
is very much in common with the more abundant rings seen in barred
galaxies (CBB96). The best examples we illustrate here are found in ESO
111$-$22, ESO 198$-$13, ESO 231$-$1,ESO 236$-$29, ESO 526$-$7, IC 1993,
NGC 7020, NGC 7187, NGC 7217, and NGC 7742. 

\noindent
2. For the 20 ringed galaxies illustrated in this paper,
azimuthally averaged H$\alpha$ surface brightness profiles follow the
approximate shape of $B$-band azimuthally averaged profiles. This is in
contrast to the finding of Ryder \& Dopita (1994) that H$\alpha$
profiles have a longer radial scale length than optical $V$ or $I$-band
profiles.

\noindent
3. The organized nature of the H$\alpha$ emission from rings allows us
to define the rings well enough to selectively integrate the fluxes and
estimate SFRs {\it for the rings alone}.  Combining our sample of 20 
nonbarred ringed galaxies with a larger sample of ringed
and mostly barred galaxies from CBB96, we were able to investigate
possible correlations between ring SFRs and bar strength. The analysis
showed that for a typical ringed galaxy having an absolute blue
magnitude of $\approx$$-$20, inner ring SFRs show little or no
dependence on the strength of the nonaxisymmetric perturbation. There 
are galaxies showing exceptional star-forming rings with little or no 
trace of a bar. The few outer rings in the sample are consistent with 
the results from the inner rings.

\noindent
4. Our combined sample allowed us to further investigate the
correlation, if any, between bar strength and intrinsic inner ring
shape. A previous study (Buta 2002) had suggested that inner ring shape
did not depend on bar strength as defined by $Q_g$. Galaxies having 
similar values of $Q_g$ can have very different values of $q_{dep}$ 
(see, for example, Buta et al. 2007). When our sample of nonbarred 
ringed galaxies is considered, some correlation with $Q_g$ is found. 
However, a better correlation is found when a newly defined parameter, 
the maximum relative torque at the position of the ring, $Q_r$, is used. 
This is in agreement with the results from numerical simulations, and 
implies that the metric properties of the rings are dictated by the local 
bar strength.  

\noindent
5. For barred galaxies having $Q_g$ $\geq$ 0.15, a better correlation
is found between $q_{dep}$ and $Q_r/Q_g$ than between $q_{dep}$ and
$Q_r$. For barred galaxies, this suggests that the controlling
parameter for inner ring shape is the location of the ring relative to
the bar maximum. If the ring major axis radius is at
$\approx$1.4$r(Q_g)$, an inner ring can be nearly circular, while if
the ring major axis radius is at $\approx$1.1$r(Q_g)$, the ring is
practically on top of the bar and is highly elongated. $Q_r/Q_g$ loses
its usefulness when a bar is weak or absent.

\noindent
6. The combined barred/nonbarred sample includes some intrinsically
very large rings. A comparison of the deprojected linear diameters
of inner rings in our sample with an earlier analysis of nearby
galaxies by Buta \& de Vaucouleurs (1982) shows that the average
nonbarred ring diameters are comparable to those of the barred
galaxies. In the Buta \& de Vaucouleurs analysis, SA inner rings
averaged a factor of 2 smaller than SB inner rings. The emphasis
on large SA rings in our sample is not unexpected because of the way
the sample was selected.

\noindent
7. We verify previous analyses which have shown that nonbarred ringed
galaxies are a diverse and relatively inhomogeneous class of objects
whose ring formation mechanisms are likely varied. Evolved bar
resonance rings, spiral density wave resonance rings, interaction-produced
rings, and even the limitations of $B$-band images for recognizing bars
could account for most of the features observed in our combined sample.

R.G. and R.B. have been supported by NSF grant AST 050-7140 to the University 
of Alabama. R.G. is now supported by an NSF International Research Fellowship 
(OISE-0852959).  E.L., H.S., and R.G. acknowledge the support of the Academy
of Finland. This research has made use of the NED, which is operated by the 
Jet Propulsion Laboratory, California Institute of Technology, under contract 
with NASA. We thank Drs$.$ Fran\c{c}oise Combes, Lia Athanassoula, Patrick 
Treuthardt, and the anonymous referee for their helpful comments. We also thank NOAO 
and CTIO for use of the 1.5 m telescope.

\clearpage

\clearpage

\clearpage
\begin{deluxetable}{llccrr}
\tablewidth{0pc} 
%%\tabletypesize{\scriptsize} 
\tablecaption{Basic Parameters of Observed
Galaxies\tablenotemark{a}} \tablehead{\colhead{Name} &\colhead{Type}
&\colhead{Type} &\colhead{$B_{T}^{0}$} &\colhead{$V_{sun}$}
&\colhead{Distance}\\
\colhead{} & \colhead{} & \colhead{Source} &\colhead{} &\colhead{(km
s$^{-1}$)} &\colhead{(Mpc)}} \startdata
ESO 111-22   & (R)SA(r)b                            &          deVA  & $14.38 \pm 0.15$  &$ 9796 \pm 26$ & $133.0 \pm 9.3$ \\
ESO 198-13   & (R)SA(r)ab                           &          deVA  & $14.02 \pm 0.21$  &$ 5452 \pm 27$ & $ 73.1 \pm 5.1$ \\
ESO 231-1    & (R$^{\prime}$)SA(r)b                 &          CSRG  & $13.46 \pm0.16$   &$ 7575 \pm 13$ & $103.1 \pm 7.2$ \\
ESO 234-11   & (R)SA(r)0$^+$                        &          CSRG  & $14.35 \pm0.04$   &$ 5604 \pm 26$ & $ 76.4 \pm 5.4$ \\
ESO 236-29   & (R)SA(r)a                            &          deVA  & $15.12 \pm1.05$   &$15990 \pm 52$ & $218.6 \pm 15.3$ \\
ESO 286-10   & (R$^{\prime}$)SA(r)a                 &          deVA  & $13.07 \pm0.11$   &$ 5368 \pm 25$ & $ 73.4 \pm 5.2$  \\
ESO 297-27   & SA($\underline{\rm r}$s)bc           &          GBSLS & $14.51 \pm0.03$   &$ 6362 \pm 10$ & $ 86.1 \pm 6.0$ \\
ESO 399-25   & (R$^{\prime}$)SA(s)0/a               &          CSRG  & $12.92 \pm0.06$   &$ 2542 \pm  \phantom{0}6$ & $ 35.1 \pm 2.5$ \\
ESO 409-3    & (R$^{\prime}$)SA(s)ab                &          CSRG  & $14.00 \pm0.13$   &$ 8507 \pm 15$ & $116.7 \pm 8.2$ \\
ESO 526-7    & SA(r)b                               &          deVA  & $12.98 \pm0.07$   &$ 5959 \pm  \phantom{0}3$ & $ 82.4 \pm 5.8$ \\
ESO 576-57   & (R)SA(r)ab p                         &          CSRG  & $14.32 \pm0.11$   &$ 4627 \pm 10$\tablenotemark{b} & $ 61.7 \pm 4.3$  \\
IC 1993      & (R$^{\prime}$)SA(s)b                 &          deVA  & $12.43 \pm0.14$   &$ 1080 \pm  \phantom{0}3$ & $ 13.1 \pm 0.9$ \\
IC 5267      & (R)SA(rs)0/a                         &          CSRG  & $11.20 \pm0.02$   &$ 1712 \pm  \phantom{0}4$ & $ 23.2 \pm 1.6$\\
NGC 5364     & SA(r)bc                              &          deVA  & $10.81 \pm0.09$   &$ 1241 \pm  \phantom{0}4$ & $ 16.7 \pm 1.2$\\
NGC 5530     & SA($\underline{\rm r}$s)c            &          CSRG  & $10.91 \pm0.01$   &$ 1193 \pm  \phantom{0}2$ & $ 14.5 \pm 1.0$ \\
NGC 7020     & (R)SAB(r?)0/a                        &          deVA  & $12.88 \pm0.13$   &$ 3201 \pm 26$  & $ 42.7 \pm 3.0$ \\
NGC 7187     & (R)S$\underline{\rm A}$B(r)0$^+$     &          deVA  & $13.46 \pm0.16$   &$ 2670 \pm 21$ & $ 37.0 \pm 2.6$ \\
NGC 7217     & (R)SA(r)ab                           &          deVA  & $10.47 \pm0.13$   &$  952 \pm  \phantom{0}2$ & $ 16.0 \pm 1.1$ \\
NGC 7702     & (R)S$\underline{\rm A}$B(r)0$^{+}$                      &          CSRG  & $12.98 \pm0.06$   &$ 3231 \pm 24$ & $ 43.2 \pm 3.0$ \\
NGC 7742     & S$\underline{\rm A}$B(r)ab           &          deVA  & $12.30 \pm0.05$   &$ 1663 \pm  \phantom{0}1$ & $ 24.8 \pm 1.7$ \\
\enddata
\tablenotetext{a}{Column 1: galaxy name; Column 2: galaxy classification; Column 3: source of galaxy classification where
deVA is from the de Vaucouleurs Atlas of Galaxies (Buta et al. 2007) and CSRG is from the Catalogue of Southern Ringed Galaxies (Buta
1995); Column 4: the corrected total apparent magnitude of the galaxy; Column 5: the recessional velocity; Column 6: the galactocentric 
distance from NASA Extragalactic Database (NED).} \tablenotetext{b}{This is the current
value listed on NED; however, when we observed the galaxy in 2002, the listed recessional velocity was $8883 \pm 53$ km s$^{-1}$.}
\end{deluxetable}

\begin{deluxetable}{llcrrcrccr}
\tablewidth{0pc} 
%%\tabletypesize{\scriptsize} 
\tablecaption{Basic 
Parameters of CBB96 Sample\tablenotemark{a}} \tablehead{
\colhead{Name} &\colhead{Type} &\colhead{Type} &\colhead{$B_{T}^{0}$}
&\colhead{$V_{sun}$} &\colhead{Distance}\\
\colhead{} &\colhead{} &\colhead{Source} &\colhead{} 
&\colhead{(km s$^{-1}$)} &\colhead{(Mpc)} } 
\startdata
NGC   53  &     (R$^{\prime}$)SB(r)ab                  &CSRG &13.20  &$4568 \pm 26$ &$61.2 \pm 4.3$\\
NGC 1317  &     (R$^{\prime}$)SA\underline{B}(rl)0/a   &CSRG &11.81  &$1941 \pm 14$ &$25.0 \pm 1.8$\\
NGC 1326  &     (R$_{1}$)SB(rl)0/a                     &CSRG &11.25  &$1360 \pm 1$  &$17.0 \pm 1.2$\\
NGC 1350  &     (R$^{\prime}_{1}$)SB(r)ab              &CSRG &10.87  &$1905 \pm 3$  &$24.5 \pm 1.7$\\
NGC 1433  &     (R$^{\prime}_{1}$)SB(\underline{r}s)ab &CSRG &10.64  &$1075 \pm 3$  &$12.7 \pm 0.9$\\
NGC 1832  &     SB(r)bc                                &RC3  &11.59  &$1939 \pm 5$  &$24.8 \pm 1.7$\\
NGC 6300  &     SB(\underline{r}s)b                    &CSRG &10.20  &$1109 \pm 3$  &$13.7 \pm 1.0$\\
NGC 6753  &     (R$^{\prime}$)SA(r)b                   &CSRG &11.58  &$3169 \pm 8$  &$42.5 \pm 3.0$\\
NGC 6761  &     (R$^{\prime}$)SB(r)0/a                 &CSRG &13.63  &$5650 \pm 51$ &$76.8 \pm 5.4$\\
NGC 6782  &     (R$^{\prime}_{1}$)SB(r)0/a             &CSRG &12.21  &$3920 \pm 14$ &$52.6 \pm 3.7$\\
NGC 6902  &     SA(r)b                                 &CSRG &11.58  &$2796 \pm 4$  &$38.2 \pm 2.7$\\
NGC 6935  &     (R)SA(r)a                              &CSRG &12.65  &$4543 \pm 18$ &$61.7 \pm 4.3$\\
NGC 6937  &     (R$^{\prime}$)SB(r)c                   &CSRG &13.35  &$4777 \pm 6$  &$64.9 \pm 4.5$\\
NGC 7020\tablenotemark{b}  &     (R)SAB(r?)0/a         &CSRG &12.67  &$3201 \pm 26$ &$42.7 \pm 3.0$\\
NGC 7098  &     (R)SA\underline{B}(r)a                 &CSRG &11.63  &$2381 \pm 6$  &$30.9 \pm 2.2$\\
NGC 7187  &     (R)S\underline{A}B(r)0$^+$             &CSRG &13.42  &$2670 \pm 21$ &$37.0 \pm 2.6$\\
NGC 7219  &     (R$^{\prime}_{2}$)SA(r)b               &CSRG &13.07  &$2956 \pm 10$ &$39.2 \pm 2.7$\\
NGC 7267  &     (R$^{\prime}_{1}$)SB(rs)a              &CSRG &12.80  &$3355 \pm 12$ &$46.3 \pm 3.2$\\
NGC 7329  &     SB(r)b                                 &CSRG &12.03  &$3252 \pm 4$  &$43.1 \pm 3.0$\\
NGC 7417  &     (R$^{\prime}$)SB(r)a                   &CSRG &12.93  &$3196 \pm 22$ &$42.1 \pm 3.0$\\
NGC 7531  &     S\underline{A}B(r)bc                   &CSRG &11.41  &$1596 \pm 3$  &$21.5 \pm 1.5$\\
NGC 7702  &     (R)S$\underline{\rm A}$B(r)0$^+$                          &CSRG &13.03  &$3231 \pm 24$ &$43.2 \pm 3.0$\\
IC  1438  &     (R$_{1}$R$^{\prime}_{2}$)SA\underline{B}(s)ab  &CSRG &12.46  &$2596 \pm 3$ &$36.6 \pm 2.6$\\
IC  4754  &     (R$^{\prime}$)SB(r)ab                  &CSRG &13.79  &$4406 \pm 50$ &$59.1 \pm 4.2$\\
IC  5240  &     SB(r)ab                                &CSRG &12.29  &$1765 \pm 7$  &$23.9 \pm 1.7$\\
UGC 12646 &     (R$^{\prime}_{1}$)SB(r)b               &deVA &13.67  &$8034 \pm 7$  &$112.6 \pm 7.9$\\
ESO 152$-$26 &  (R$_{1}$R$^{\prime}_{2}$)S\underline{A}B(r)ab &CSRG  &13.95  &$6157 \pm 27$ &$82.7 \pm 5.8$\\
\enddata
\tablenotetext{a}{Column 1: galaxy name; Column 2: galaxy classification; Column 3: source of the galaxy 
classification where deVA is from the de Vaucouleurs Atlas of Galaxies (Buta et al. 2007) 
and CSRG is from the Catalogue of Southern Ringed Galaxies (Buta 1995); Column 4: the corrected total 
apparent magnitude of the galaxy; Column 5: heliocentric velocity of the galaxy; Column 6: the galactocentric 
distance from NED.}
\tablenotetext{b}{All ring parameters are for inner rings except for
NGC 7020 whose measured values are for the outer ring.}
\end{deluxetable}

\begin{deluxetable}{lcccc}
\tablewidth{0pc} 
%%\tabletypesize{\scriptsize} 
\tablecaption{Planetary Nebulae H$\alpha$
Emission\tablenotemark{a}} \tablehead{ \colhead{Name} &
\colhead{Aperture} &\colhead{This Paper (GB)}
& \colhead{Kohoutek (KM)} &\colhead{$log(\frac{F_{GB}}{F_{KM}})$} \\
\colhead{} & \colhead{($\arcsec$)} &\colhead{ (erg s$^{-1}$
cm$^{-2}$)} &\colhead{ (erg s$^{-1}$ cm$^{-2}$)} &\colhead{}}
\startdata
NGC 6326 &  34.8 &$3.96\times 10^{-11}$ & $3.43\times 10^{-11}$ & 0.06\\
NGC 6326 &  34.8 &$3.84\times 10^{-11}$ & $3.43\times 10^{-11}$ & 0.05\\
NGC 6818 &  43.4 &$1.43\times 10^{-10}$ & $1.30\times 10^{-10}$ & 0.04\\
NGC 6818 &  43.4 &$1.50\times 10^{-10}$ & $1.30\times 10^{-10}$ & 0.06\\
\enddata
\tablenotetext{a}{Column 1: name of planetary nebula; Column 2: size 
of the aperture used to measure H$\alpha$ + [NII] flux emitted from 
the nebula; Column 3: H$\alpha$ + [NII] flux measured
for this paper by Grouchy et al.(GB); Column 4: H$\alpha$ + [NII] flux
measured by Kohoutek \& Martin (KM); Column 5: The log of the ratio of the
fluxes measured by Grouchy et al. (GB) to that measured by Kohoutek
\& Martin (KM).}
\end{deluxetable}

\begin{deluxetable}{lccccccccl}
\tabletypesize{\scriptsize}  
\tablewidth{0pc} \tablecaption{Galaxy 
H$\alpha$ emission\tablenotemark{a}} \tablehead{ \colhead{Name} &
\colhead{ID}& \colhead{Radius} & \colhead{$F_{H\alpha + [NII]}$} & 
\colhead{\% Error} & \colhead{log$L_{H\alpha + [NII]}$} &
\colhead{$L_{H\alpha}^{0}$}&\colhead{SFR} &\colhead{$W_{\lambda}$} \\
\colhead{} &\colhead{} & \colhead{($\arcsec$)} &
\colhead{($ \frac{10^{-14} ergs}{s * cm^{2}} $)} &\colhead{Cont Sub} &
\colhead{} &\colhead{($10^{41}$ ergs s$^{-1}$)} &
\colhead{(M$_\bigodot$ yr$^{-1}$)} &\colhead{}}
\startdata
ESO 111-22 & s&  65.16  & $\phantom{0}21.5 \pm \phantom{0}7.4$ & 2.8& $41.66 \pm 0.37$  & $8.6 \pm 3.2$ & $\phantom{0}6.8 \pm 2.5$ & $ 25.9 \pm 4.4$\\
ESO 198-13 & l&  43.44  & $\phantom{0}23.0 \pm \phantom{0}7.9$ & 0.1&$41.17 \pm 0.37$  & $2.8 \pm 1.1$ & $\phantom{0}2.3 \pm 0.8$& $16.8 \pm 4.7$\\
ESO 231-1  & p&  34.75  & $\phantom{0}59.9 \pm 20.7$ & 2.0&$41.88 \pm 0.37$ & $16.4 \pm 6.1$ & $13.1 \pm 4.9$ & $28.9 \pm 5.0$\\
ESO 234-11 & m&  47.78  & $\phantom{0}\phantom{0}3.3  \pm \phantom{0}1.1$ & 4.3 &$40.36 \pm 0.37$  & $0.5 \pm 0.2$ & $\phantom{0}0.4 \pm 0.1$ & $\phantom{0}2.4 \pm 4.6$\\
ESO 236-29 & t&  43.44  & $\phantom{0}\phantom{0}8.2  \pm \phantom{0}2.8$ & 5.5 &$41.71 \pm 0.35$ & $10.1 \pm 3.6$ & $\phantom{0}8.0 \pm 2.8$ & $24.6 \pm 4.6$\\
ESO 286-10 & k&  86.88  & $\phantom{0}19.0 \pm \phantom{0}6.6$ & 2.1 &$41.09 \pm 0.37$  & $2.5 \pm 0.9$ & $\phantom{0}2.0 \pm 0.7$& $\phantom{0}4.3 \pm 4.6$\\
ESO 297-27 & o&  80.36  & $\phantom{0}12.5 \pm \phantom{0}4.3$ & 7.1 &$41.04 \pm 0.37$  & $2.1 \pm 0.8$ & $\phantom{0}1.7 \pm 0.6$ & $ 17.9 \pm 4.8$\\
ESO 399-25 & g&  43.44  & $\phantom{0}\phantom{0}8.9  \pm \phantom{0}3.1$ & 3.2 &$40.12 \pm 0.37$  & $0.3 \pm 0.1$ & $\phantom{0}0.2 \pm 0.1$ & $\phantom{0}1.0 \pm 4.3$\\
ESO 409-3  & q&  45.61  & $\phantom{0}\phantom{0}3.8  \pm \phantom{0}1.3$ & 8.6 &$40.80 \pm 0.37$  & $1.2 \pm 0.5$ & $\phantom{0}1.0 \pm 0.4$  & $\phantom{0}2.0 \pm 4.7$\\
ESO 526-7  & n& 106.43  & $\phantom{0}25.0 \pm \phantom{0}8.6$ & 7.2 &$41.31 \pm 0.37$  & $5.0 \pm 1.9$ & $\phantom{0}4.0 \pm 1.5$ & $\phantom{0}7.9 \pm 4.7$\\
ESO 576-57 & r&  45.61  & $\phantom{0}11.4 \pm \phantom{0}3.9$ & 9.1 &$40.72 \pm 0.37$ & $1.2 \pm 0.4$ & $\phantom{0}0.9 \pm 0.3$& $9.5 \pm 4.7$\\
IC 1993    & c&  65.59  & $\phantom{0}\phantom{0}9.0  \pm \phantom{0}3.1$ & 3.7 &$39.27 \pm 0.37$  & $0.04 \pm 0.01$ & $\phantom{0}0.03 \pm 0.01$ & $\phantom{0}1.6 \pm 4.1$\\
IC 5267    & f& 193.31  & $141.0  \pm 48.8$  & 4.5 &$40.96 \pm 0.37$   & $1.7 \pm 0.7$ & $\phantom{0}1.4 \pm 0.5$ & $\phantom{0}5.6 \pm 3.8$\\
NGC 5364   & d& 217     & $265.0  \pm 91.5$  & 5.1 &$40.95 \pm 0.37$   & $1.7 \pm 0.7$ & $\phantom{0}1.4\pm0.5$ &  $10.7\pm4.0$\\
NGC 5530   & a& 136.84 & $217.0  \pm 75.1$  & 0.7 &$40.74 \pm 0.37$  & $1.3 \pm 0.5$ & $\phantom{0}1.0 \pm 0.4$ &  $11.8 \pm 4.0$\\
NGC 7020   & i&  86.9   & $\phantom{0}16.5 \pm \phantom{0}5.7$ & 0.4 &$40.56 \pm 0.37$  & $0.7 \pm 0.3$ & $\phantom{0}0.6 \pm 0.2$ & $\phantom{0}2.7 \pm 4.2$\\
NGC 7187   & h&  22.59  & $\phantom{0}\phantom{0}9.2  \pm \phantom{0}3.2$ & 2.5 &$40.18 \pm 0.37$  & $0.3 \pm 0.1$ & $\phantom{0}0.2 \pm 0.1$  & $\phantom{0}3.7 \pm 3.8$\\
NGC 7217   & b&  152    & $\phantom{0}22.7 \pm \phantom{0}7.8$ & 1.1 &$39.84 \pm 0.37$  & $0.2 \pm 0.1$ & $\phantom{0}0.1\pm0.1$ & $\phantom{0}0.6 \pm 4.3$\\
NGC 7702   & j&  100    & $\phantom{0}\phantom{0}10.9  \pm \phantom{0}3.8$ & 3.7 &$40.39 \pm 0.37$  & $0.5 \pm 0.2$ & $\phantom{0}0.4 \pm 0.1$   & $2.5 \pm 4.2$\\
NGC 7742   & e&  86.9   & $119.0  \pm 41.2$  & 2.4 & $40.94 \pm 0.37$   & $1.8 \pm 0.7$ & $\phantom{0}1.5\pm 0.5$  & $12.6 \pm 4.4$\\
\enddata
\tablenotetext{a}{Column 1: galaxy name; Column 2: ID assigned to the galaxy which corresponds to the labels in Figure 1;
Column 3: aperture radius used to measure the H$\alpha$ + [NII] flux emitted from each galaxy; Column 4: the H$\alpha$ + [NII] 
flux; Column 5: estimated error of the continuum subtraction calculated from the residuals of foreground stars; Column 6: the 
log of the H$\alpha$ + [NII] luminosity emitted from the galaxy; Column 7: the H$\alpha$ luminosity corrected for internal 
extinction, galactic extinction, and [NII] contamination; Column 8: the current SFR assuming a constant SFR through time, 
a solar metallicity, and a Salpeter (1955) IMF; Column 9: equivalent width of H$\alpha$ + [NII] emission.}
\end{deluxetable}

\begin{deluxetable}{crrrrrrrr}
%%\tabletypesize{\scriptsize} 
\tablewidth{0pc} \tablecaption{Calculated Sensitivity ($S$($\lambda$)) of Our
Filter System\tablenotemark{a,b}} \tablehead{ \colhead{} &\colhead{6477\AA\
} &\colhead{6563\AA\ } &\colhead{6606\AA\ } &\colhead{6649\AA\ }
&\colhead{6693\AA\ } &\colhead{6737\AA\ } &\colhead{6781\AA\ }
&\colhead{6916\AA\ }} \startdata
LTT 7987 & 11.62 & 10.05 & 10.57 & 10.82 & 11.75 & 10.07 & 10.06 & 11.42\\
LTT 7987 & 11.68 & 10.19 & 10.53 & 10.93 & 11.72 & 10.09 & 10.14 & 11.32\\
LTT 9239 & 11.65 &  9.98 & 10.50 & 10.88 & 11.78 & 10.08 & 10.14 & 11.41\\
LTT 9239 & 11.27 &  9.80 & 10.35 & 10.65 & 11.65 &  9.85 &  9.96 & 11.45\\
LTT 7987 & n/a\tablenotemark{c} & 10.12 & 10.52 & 10.77 & 11.81 & 10.03 & 10.12 & 11.35\\
LTT 7987 & n/a\tablenotemark{c} &  9.79 & 10.46 & 10.96 & 11.40 & 10.05 & 10.11 & 11.29\\
LTT 7987 & n/a\tablenotemark{c} &  9.96 & 10.27 & 10.80 & 11.68 &  9.92 &  9.90 & 11.41\\
LTT 9239 & n/a\tablenotemark{c} &  9.75 & 10.31 & 10.65 & 11.57 &  9.93 &  9.98 & 11.44\\
EG 274   & n/a\tablenotemark{c} & n/a\tablenotemark{d} & 10.36 & 10.78 & 11.72 & 10.10 & n/a\tablenotemark{c} &  n/a\tablenotemark{c}\\
\hline
$\langle S$($\lambda$)$\rangle$  & 11.56 &  9.96 & 10.43 & 10.80 & 11.68 & 10.01 & 10.05 & 11.39\\
$\sigma$  &  0.19 &  0.16 &  0.11 &  0.11 &  0.13 &  0.09 &  0.09 &  0.06\\
$< \sigma >$ & 0.10 &  0.06 &  0.04 &  0.04 &  0.04 &  0.03 &  0.03 & 0.02\\
\enddata
\tablenotetext{a}{Rows 1 and 2: night four values of star LTT 7987; Rows 3 and 4: night 
four values of star LTT 9239; Rows $5-7$: night five values for star LTT 7987; Row 8: night 
five value for star LTT 9239; Row 9: night five value for star EG 274; Row 10: the average 
value of $S$($\lambda$) for all stars; Row 11: the standard deviation of the mean; Row 12: the 
standard error of the mean.} \tablenotetext{b}{All values listed in units of $10^{-16}$ erg 
s$^{-1}$ cm$^{-2}$.} \tablenotetext{c}{No observation made.} 
\tablenotetext{d}{Image not used due to tracking problems.} 
\end{deluxetable}

\clearpage
\begin{deluxetable}{llcrrcrccr}
\tabletypesize{\small} 
\tablewidth{0pc} 
\tablecaption{Ring SFRs
in the CBB96 Sample\tablenotemark{a,b}} \tablehead{
\colhead{Name} &\colhead{$logF$} &\colhead{$r_{min}$} &\colhead{Width} 
&\colhead{$q$} &\colhead{P.A.} &\colhead{Flux} &\colhead{SFR} &\colhead{$W_{\lambda}$}
\\
\colhead{} &\colhead{H$\alpha$+[NII]}
&\colhead{($\arcsec$)} &\colhead{($\arcsec$)} &\colhead{}
&\colhead{($\degr$)} &\colhead{Fraction} &\colhead{Ring}
&\colhead{Ring} } \startdata
NGC   53  &   $-12.99 \pm 0.35$ &     17.0 &    17.4 &   0.795 &      164.0 &    0.906 &    $0.71 \pm 0.21$ & $11.9 \pm 3.8$\\
NGC 1317  &   $-12.11 \pm 0.33$ &   39.1 &    26.1 &   1.000 &       90.0 &    0.045 &    $0.05 \pm 0.02$ &  $1.4 \pm 0.5$\\
NGC 1326  &   $-11.62 \pm 0.31$ &    21.8 &    17.4 &   0.822 &       30.5 &    0.046 &    $0.07 \pm 0.02$ &  $\phantom{0}3.1 \pm 1.0$\\
NGC 1350  &   $-12.42 \pm 0.34$ &    53.9 &    37.0 &   0.499 &       13.0 &    0.630 &    $0.29 \pm 0.08$ &  $4.0 \pm 1.3$\\
NGC 1433  &   $-11.57 \pm 0.31$ &    56.5 &    65.2 &   0.721 &       97.6 &    0.564 &    $0.49 \pm 0.14$ & $13.6 \pm 4.4$\\
NGC 1832  &   $-11.59 \pm 0.31$ &    13.1 &    10.9 &   0.768 &      147.7 &    0.191 &    $0.70 \pm 0.20$ & $27.4 \pm 8.8$\\
NGC 6300  &   $-11.50 \pm 0.31$ &    39.1 &    43.5 &   0.594 &      125.2 &    0.814 &    $1.18 \pm 0.34$ & $24.1 \pm 7.7$\\
NGC 6753  &   $-11.61 \pm 0.31$ &    8.7 &     4.3 &   0.840 &       57.1 &    0.124 &     $1.26 \pm 0.37$ & $22.2 \pm 7.1$\\
NGC 6761  &   $-12.89 \pm 0.35$ &    19.1 &     7.0 &   0.951 &       95.7 &    0.272 &    $0.48 \pm 0.14$ & $10.2 \pm 3.3$\\
NGC 6782  &   $-12.25 \pm 0.33$ &    17.4 &    15.7 &   0.643 &      175.6 &    0.240 &    $0.84 \pm 0.24$ & $11.3 \pm 3.6$\\
NGC 6902  &   $-11.95 \pm 0.32$ &    13.1 &    14.4 &   0.883 &      154.7 &    0.209 &    $0.74 \pm 0.22$ & $12.6 \pm 4.0$\\
NGC 6935  &   $-12.28 \pm 0.33$ &    11.4 &    17.4 &   0.860 &        0.8 &    0.647 &    $2.73 \pm 0.79$ & $16.0 \pm 5.1$\\
NGC 6937  &   $-12.05 \pm 0.33$ &    16.5 &    11.3 &   0.868 &       91.0 &    0.294 &    $2.34 \pm 0.68$ & $46.4 \pm 14.9$\\
NGC 7020\tablenotemark{b}  & $-12.45 \pm 0.34$ &    52.6 &    52.2 &   0.484 &      162.0 &    0.448 &    $0.62 \pm 0.18$ & $ 9.7 \pm 3.1$\\
NGC 7098  &   $-12.18 \pm 0.33$ &    50.0 &    26.1 &   0.523 &       63.8 &    0.275 &    $0.40 \pm 0.12$ & $13.5 \pm 4.3$\\
NGC 7187  &   $-13.14 \pm 0.36$ &    13.9 &     8.7 &   0.968 &      123.3 &    0.557 &    $0.12 \pm 0.04$ & $ 6.7 \pm 2.1$\\
NGC 7219  &   $-12.56 \pm 0.34$ &    6.5 &     6.6 &   0.762 &       53.1 &    0.293 &    $0.26 \pm 0.08$ & $12.3 \pm 3.9$\\
NGC 7267  &   $-12.27 \pm 0.33$ &    25.2 &     8.7 &   0.601 &       99.0 &    0.121 &    $0.28 \pm 0.81$ & $15.2 \pm 4.9$\\
NGC 7329  &   $-12.16 \pm 0.33$ &    26.1 &    17.4 &   0.726 &      106.7 &    0.246 &    $0.66 \pm 0.19$ & $14.0 \pm 4.5$\\
NGC 7417  &   $-12.80 \pm 0.35$ &    17.4 &    13.1 &   0.701 &       28.2 &    0.203 &    $0.12 \pm 0.04$ & $ 4.1 \pm 1.3$\\
NGC 7531  &   $-11.90 \pm 0.32$ &    21.8 &    13.1 &   0.347 &       14.8 &    0.349 &    $0.41 \pm 0.12$ & $24.7 \pm 7.9$\\
NGC 7702  &   $-13.35 \pm 0.36$ &    28.3 &     8.7 &   0.463 &      113.7 &    0.000 &    $0.00 \pm 0.00$ & $ 0.0 \pm 0.0$\\
IC  1438  &   $-12.49 \pm 0.34$ &    14.4 &    14.4 &   0.641 &      124.0 &    0.300 &    $0.28 \pm 0.08$ & $11.6 \pm 3.7$\\
IC  4754  &   $-12.77 \pm 0.35$ &    15.2 &     7.8 &   0.869 &       47.2 &    0.393 &    $0.58 \pm 0.17$ & $18.8 \pm 6.0$\\
IC  5240  &   $-12.42 \pm 0.34$ &    33.9 &    10.4 &   0.632 &      106.8 &    0.262 &    $0.11 \pm 0.03$ & $ 9.4 \pm 3.0$\\
UGC 12646 &   $-12.53 \pm 0.34$ &    15.2 &    13.1 &   0.629 &      138.0 &    0.325 &    $2.65 \pm 0.77$ & $32.2 \pm 10.3$\\
ESO 152$-$26 &$-12.75 \pm 0.35$ &     9.1 &    10.4 &   0.757 &        8.4 &    0.248 &    $0.63 \pm 0.18$ & $12.0 \pm 3.8$\\
\enddata
\tablenotetext{a}{Column 1: galaxy name; Column 2: log of total H$\alpha$+[NII] flux based on improved 
continuum subtraction, to be compared with values in Column 2 of Table 2 of CBB96; Column 3: inner 
radius of elliptical annulus defining the inner ring (outer ring in the case of NGC 7020); Column 4: 
width of same elliptical annulus; Column 5: shape of elliptical annulus; Column 6: P.A. of elliptical 
annulus; Column 7: fraction of total H$\alpha$ flux included within this annulus; Column 8: ring SFR 
($M_{\odot}$ yr$^{-1}$); Column 9: equivalent width of ring H$\alpha$+[NII] ($\AA$), derived as in 
Romanishin (1990).}
\tablenotetext{b}{All ring parameters are for inner rings except for NGC 7020 whose measured values are for the outer ring.}
\end{deluxetable}

\begin{deluxetable}{lccccccc}
\tablewidth{0pc} \tabletypesize{\small} \tablecaption{Galaxy Ring
Properties\tablenotemark{a}} \tablehead{ \colhead{Galaxy} &
\colhead{Ring} &\colhead{$a$} & \colhead{Width} & \colhead{P.A.}
& \colhead{$q$} &\colhead{SFR$_{ring}$}
&\colhead{$\frac{SFR_{ring}}{SFR_{total}}$} \\
\colhead{Name} &\colhead{} &\colhead{($\arcsec$)} &\colhead{($\arcsec$)}
&\colhead{($\degr$)} &\colhead{} &\colhead{(M$_{\bigodot}$
yr$^{-1}$)} &\colhead{}} \startdata
ESO 111$-$22 &r & $15.6 \pm 0.1$  &  5.0 &   $\phantom{0}\phantom{0}8.8\pm0.2$ & $0.41 \pm0.03$ & $1.2\pm0.5$ & 0.18\\
           &R &  $48.9\pm 0.3$  &  4.2 &   $\phantom{0}\phantom{0}9.2 \pm0.1$ & $0.45 \pm 0.03$ & $0.16\pm0.06$ & 0.02\\
ESO 198$-$13 & nr & $\phantom{0}6.9 \pm 0.1$  &  1.4 & $-55.6 \pm 0.6$ &$0.75 \pm 0.05$ & $0.4\pm 0.1$ & 0.17\\
           & r & $24.4 \pm 0.3$ &  2.0 &  $\phantom{0}64 \pm 2$ &$0.88 \pm 0.05$ & $1.1\pm0.4$ & 0.51\\
ESO 231$-$1  &r &  $12.2 \pm 0.1$  &  4.6 &  $\phantom{0}66.4 \pm 0.5$ & $0.84 \pm 0.05$ & $5.8\pm2.1$ & 0.44\\
           &R$^{\prime}$  &$29.5 \pm 0.2$  &  4.2 & $-55.9 \pm 0.7$ & $0.83 \pm 0.05$ & $2.1\pm0.7$ & 0.16\\
ESO 234$-$11 &r &  $10.8 \pm 0.1$  &  2.5 &  $-62.5 \pm 0.5$ & $0.53 \pm 0.03$ & $0.09 \pm 0.03$ & 0.22\\
           &R &  $31.7 \pm 0.6$  &  1.5 &  $\phantom{0}57.5 \pm 0.3$ & $0.40 \pm 0.03$ & $0.05\pm0.02$ & 0.13\\
ESO 236$-$29 &nr &  $\phantom{0}5.7 \pm 0.1$  &  2.3 &  $\phantom{0}73 \pm 1$ & $0.85 \pm 0.05$ & $0.6\pm0.2$ & 0.07\\
           &r &  $19.3 \pm 0.3$ &  4.6 &  $\phantom{0}76 \pm 2$ & $0.62 \pm 0.04$ & $1.6\pm0.5$ & 0.19\\
ESO 286$-$10 & R & $60.5 \pm 1.2$ &  1.5 & $-46 \pm 3$ & $0.84 \pm 0.03$ & $0.08\pm0.03$ & 0.04\\
ESO 297$-$27 & $\underline{\rm r}$s & $18.0 \pm 0.6$  &  0.7 &  $\phantom{0}68 \pm 2$ & $0.65 \pm 0.05$ & $0.2\pm0.1$ & 0.10\\
ESO 399$-$25 & R$^{\prime}$  &$92.9 \pm 2\phantom{0}$  &  0.4 & $-25.1 \pm 0.3$ & $0.31 \pm 0.02$ & $0.008\pm 0.003$ & 0.04\\
ESO 409$-$3  & R$^{\prime}$  &$32.2 \pm 0.2$ &  3.2 &  $-\phantom{0}5 \pm1$ & $0.88 \pm 0.05$ & $0.5\pm0.2$ & 0.53\\
ESO 526$-$7  & r &  $27.0 \pm 0.3$  &  2.3 & $-74.3 \pm 0.5$ & $0.58 \pm 0.04$ & $0.6\pm0.2$ & 0.14\\
ESO 576$-$57 &r &  $22.1 \pm 0.4$  &  1.4 &  $\phantom{0}14 \pm 2$ & $0.71 \pm 0.05$ & $0.12\pm0.04$ & 0.13\\
           & R &  $34.4 \pm 0.8$  &  1.3 &  $\phantom{0}16 \pm 3$ & $0.81 \pm 0.05$ & $0.09\pm0.03$ & 0.10\\
IC 1993    &R$^{\prime}$   &$\phantom{0}45.2 \pm 0.3$ &  0.7 &  $\phantom{0}57 \pm 2$ & $0.94 \pm 0.02$ & $0.03\pm0.01$ & 0.94\\
IC 5267    &rs &  $ 78 \pm 1$  &  0.6 & $-50 \pm 2$ & $0.78 \pm 0.02$ & $0.07\pm0.03$ & 0.05\\
           &R  &  $147 \pm 1$  &  0.7 & $\phantom{0}34.3 \pm 0.2$ & $0.68 \pm 0.04$ & $0.05\pm0.02$ & 0.04\\
NGC 5364   &r  &  $41.1 \pm 0.6$ &  1.4 & $\phantom{0}44.5 \pm 0.2$ & $0.49 \pm 0.03$ & $0.13\pm0.05$ & 0.09\\
NGC 5530   &$\underline{\rm r}$s  & $32.6 \pm 0.7$ &  1.0 & $-65.6 \pm 0.7$ & $0.54 \pm 0.03$ & $0.12\pm0.04$ & 0.12\\
NGC 7020   &R  &  $81.2 \pm 0.7$  &  2.7 & $-19.7 \pm 0.2$ & $0.49 \pm 0.03$ & $0.16\pm0.06$ & 0.27\\
NGC 7187   &r  &  $19.6 \pm 0.2$ &  1.6 &  $\phantom{0}84 \pm 2$ & $0.88 \pm 0.05$ & $0.13\pm0.05$ & 0.54\\
NGC 7217   &r &  $11.1 \pm 0.1$ &  0.2 & $-86.4 \pm 0.6$ & $0.77 \pm 0.05$ & $0.01\pm0.002$ & 0.04\\
           &R &  $76.7 \pm 0.5$  &  1.4 & $-86.1 \pm 0.5$ & $0.85 \pm 0.05$ & $0.2\pm0.1$ & 1.46\\
NGC 7702   & r & $32.6 \pm 0.3$ &  1.8 & $-61.7 \pm 0.3$ & $0.41 \pm 0.03$ & $0.07\pm0.03$ & 0.18\\
NGC 7742   & r & $\phantom{0}9.4 \pm 0.1$  &  1.3 & $-50 \pm 7$ & $0.97 \pm 0.06$ & $1.3\pm0.5$ & 0.86\\
\enddata
\tablenotetext{a}{Column 1: galaxy name; Column 2: ring feature; Column 3: average diameter of the H$\alpha$ 
ring feature; Column 4: width of aperture used to measure H$\alpha$ flux in the ring; Column 5: P.A. of 
elliptical aperture used.  Zero is north with positive angles moving toward the east and negative to the 
west; Column 6: axis ratio of aperture; Column 7: SFR of ring; Column 8: fraction of the ring SFR to the 
galaxy's total SFR.}
\end{deluxetable}

\begin{deluxetable}{lccrccccc}
\tablewidth{0pc} 
\tabletypesize{\small} \tablecaption{Non-axisymmetric Torque Strengths,
Radial Scale Lengths and Deprojected Ring Parameters for 20
Nonbarred Ringed Galaxies\tablenotemark{a}} \tablehead{
\colhead{Name} &\colhead{$r_{dep}$ ($\arcsec$)} &\colhead{$q_{dep}$}
&\colhead{$h_r$ ($\arcsec$)} &\colhead{$Q_{r}$}
&\colhead{$A_2(ring)$} &\colhead{$Q_{g}$} &\colhead{$r_{max}$
($\arcsec$)}} \startdata
ESO 111$-$22 &  $15.7 \pm 0.8$ &     $0.85 \pm 0.04$ &     22.6\tablenotemark{b} &     0.041    &0.191    &0.073      &48.2\\
ESO 198$-$13 &  $ 7.1 \pm 0.4$ &     $0.97 \pm 0.05$ &     10.5\tablenotemark{b} &     0.018    &0.048    &0.070      &29.1\\
ESO 231$-$1 &   $12.7 \pm 0.7$ &     $0.86 \pm 0.05$ &      9.4 &     0.072    &0.057    &0.104      &17.8\\
ESO 234$-$11\tablenotemark{c} &  $32 \pm 2$ &     $0.85 \pm 0.04$ &     23.8 &     0.017    &0.146    &0.042      &13.5\\
ESO 236$-$29 &  $ 8.5 \pm 0.4$ &     $0.67 \pm 0.04$ &     12.0\tablenotemark{b} &     0.037    &0.148    &0.090      &16.9\\
ESO 286$-$10 &  $67 \pm 4$ &     $0.88 \pm 0.05$ &     38.2 &     0.031    &0.266  &0.059 &17.8\\
ESO 297$-$27 &  $25 \pm 1$ &     $0.59 \pm 0.05$ &     11.8 &     0.049    &0.124    &0.082      &4.8\\
ESO 399$-$25\tablenotemark{c} &   $93 \pm 5$  &  $0.67 \pm 0.04$  &     41.0\tablenotemark{b} &    0.041    &0.325   &0.061      &11.7\\
ESO 409$-$3  &  $34 \pm 2$ &     $0.90 \pm 0.05$ &      9.7\tablenotemark{b} &    0.068    &0.359    &0.068      &32.6\\
ESO 526$-$7  &  $30 \pm 2$ &     $0.75 \pm 0.04$ &     24.7\tablenotemark{b} &    0.049    &0.170    &0.051      &25.6\\
ESO 576$-$57 &  $22 \pm 1$ &     $0.85 \pm 0.05$ &     18.4 &    0.018    &0.069    &0.040      &41.3\\
IC  1993  &     $46 \pm 2$ &     $0.95 \pm 0.05$ &     27.7\tablenotemark{b} &    0.029    &0.072    &0.043      &30.8\\
IC  5267  &     $80 \pm 4$ &     $0.92 \pm 0.05$ &     28.4\tablenotemark{b} &    0.039    &0.166    &0.076      &28.7\\
NGC 5364  &     $42 \pm 2$ &     $0.59 \pm 0.03$ &     59.5 &    0.093    &0.210    &0.106      &36.1\\
NGC 5530\tablenotemark{c}  &     $41 \pm 2$ &     $0.73 \pm 0.04$ &     41.0 &    0.061    &0.077    &0.123      &10.0\\
NGC 7020  &     $90 \pm 5$ &     $0.89 \pm 0.05$ &     26.3\tablenotemark{b} &    0.021    &0.050    &0.074      &34.3\\
NGC 7187  &     $21 \pm 1$ &     $0.82 \pm 0.05$ &     16.6\tablenotemark{b} &    0.077    &0.338    &0.077      &21.3\\
NGC 7217  &     $11.5 \pm 0.6$ &     $0.85 \pm 0.04$ &     56.0\tablenotemark{b} &    0.029    &0.061    &0.045      &38.9\\
NGC 7702\tablenotemark{c}  &     $33 \pm 2$ &     $0.80 \pm 0.04$ &     54.0 &    0.081    &0.395    &0.084      &30.8\\
NGC 7742  &     $ 9.8 \pm 0.5$ &     $0.96 \pm 0.05$ &     13.1 &    0.058    &0.101    &0.058      &9.1\\
\enddata
\tablenotetext{a}{Column 1: galaxy name; Columns 2 and 3: deprojected radius and axis ratio of the main ring, typically 
the inner ring, as seen in the H$\alpha$ image. Values are based on an ellipse fit to the H$\alpha$ distribution 
in the ring. If the ring was not visible in the H$\alpha$ then the parameters are based on the $B$-band image and the 
galaxy name is marked with the letter ``c''; Column 4: radial scale length based on one-dimensional (denoted with the 
letter ``b'') and two-dimensional decompositions (see the text and Grouchy (2008)); Column 5: gravitational torque strength at 
the location of the ring's major axis; Column 6: the $m=2$ Fourier amplitude at the position of the ring's major axis; 
Column 7: maximum relative non-axisymmetric torque strength based on Grouchy (2008) $I$-band images; Column 8: the radius of 
the $Q_{g}$. } 
\tablenotetext{b}{One-dimensional fits to the disk of the surface brightness profile were used for these galaxies in 
which the two-dimensional decomposition had difficulty fitting the disk.} \tablenotetext{c}{These rings had 
no detectable H$\alpha$ emission; therefore, the ring parameters were taken from the broadband stellar ring.}
\end{deluxetable}

\begin{deluxetable}{lcrrrccc}
\tablewidth{-3pc} 
\tabletypesize{\small} \tablecaption{Bar Strengths, Radial Scale Lengths 
and Deprojected Ring Parameters for the CBB96 
Sample\tablenotemark{a}} \tablehead{ \colhead{Name}
&\colhead{$r_{dep}$ ($\arcsec$)} &\colhead{$q_{dep}$}
&\colhead{$h_r$ ($\arcsec$)} &\colhead{$Q_{r}$}
&\colhead{$A_2(ring)$} &\colhead{$Q_{g}$} &\colhead{$r_{max}$
($\arcsec$)}} \startdata
NGC    53   &     $30 \pm 2$ &     $0.88 \pm 0.04$ &     30.6 &    0.244 &0.438 &0.353 &20.4\\
NGC  1317   &     $61 \pm 3$ &     $0.82 \pm 0.04$ &     19.1 &    0.016 &0.087 &0.111 &$\phantom{0}4.8$\\
NGC  1326   &     $42 \pm 2$ &     $0.71 \pm 0.04$ &     34.9 &    0.178 &0.634 &0.180 &37.4\\
NGC  1350   &     $81 \pm 4$ &     $0.77 \pm 0.04$ &     68.1 &    0.211 &0.663 &0.237 &68.1\\
NGC  1433   &    $105 \pm 5$ &     $0.62 \pm 0.03$ &     62.9 &    0.326 &0.912 &0.400 &67.0\\
NGC  1832   &     $25 \pm 1$ &     $0.69 \pm 0.03$ &     16.1 &    0.056 &0.261 &0.176 &44.7\\
NGC  6300   &     $69 \pm 4$ &     $0.71 \pm 0.04$ &     37.3 &    0.191 &0.425 &0.220 &32.2\\
NGC  6753   &     $11.5 \pm 0.6$ &     $0.88 \pm 0.04$ &     21.4 &    0.043 &0.109 &0.053 &36.9\\
NGC  6761   &     $23 \pm 1$ &     $0.96 \pm 0.05$ &     15.8 &    0.182 &0.210 &0.347 &15.2\\
NGC  6782   &     $26 \pm 1$ &     $0.65 \pm 0.03$ &     28.5 &    0.167 &0.666 &0.172 &23.9\\
NGC  6902   &     $25 \pm 1$ &     $0.85 \pm 0.04$ &     34.5 &    0.084 &0.238 &0.086 &28.7\\
NGC  6935   &     $22 \pm 1$ &     $0.90 \pm 0.05$ &     22.6 &    0.058 &0.129 &0.064 &23.0\\
NGC  6937   &     $28 \pm 1$ &     $0.70 \pm 0.04$ &     34.7 &    0.182 &0.419 &0.282 &20.4\\
NGC  7020   &     $87 \pm 4$ &     $0.90 \pm 0.05$ &     40.5 &    0.086 &0.354 &0.086 &34.3\\
NGC  7098   &     $66 \pm 3$ &     $0.83 \pm 0.04$ &     49.7 &    0.126 &0.418 &0.159 &46.1\\
NGC  7187   &     $20 \pm 1$ &     $0.92 \pm 0.05$ &     30.7 &    0.065 &0.299 &0.067 &20.4\\
NGC  7219   &     $12.3 \pm 0.6$ &     $0.72 \pm 0.04$ &     14.8 &    0.141 &0.496 &0.142 &11.7\\
NGC  7267   &     $33 \pm 2$ &     $0.54 \pm 0.03$ &     14.8 &    0.242 &0.517 &0.549 &15.2\\
NGC  7329   &     $39 \pm 2$ &     $0.93 \pm 0.05$ &     34.1 &    0.164 &0.306 &0.279 &28.2\\
NGC  7417   &     $29 \pm 2$ &     $0.70 \pm 0.04$ &     21.7 &    0.149 &0.300 &0.232 &18.7\\
NGC  7531   &     $29 \pm 2$ &     $0.72 \pm 0.04$ &     24.3 &    0.154 &0.387 &0.154 &27.0\\
NGC  7702   &     $33 \pm 2$ &     $0.77 \pm 0.04$ &     34.2 &    0.140 &0.654 &0.142 &30.8\\
IC   1438   &     $23 \pm 1$ &     $0.66 \pm 0.03$ &     23.4 &    0.152 &0.681 &0.159 &20.4\\
IC   4754   &     $22 \pm 1$ &     $0.79 \pm 0.04$ &     15.1 &    0.127 &0.181 &0.324 &11.7\\
IC   5240   &     $40 \pm 2$ &     $0.87 \pm 0.04$ &     24.8 &    0.145 &0.332 &0.273 &26.5\\
UGC 12646   &     $24 \pm 1$ &     $0.57 \pm 0.03$ &     32.0 &    0.306 &1.090 &0.318 &22.1\\
ESO 152-26  &     $17.4 \pm 0.9$ &     $0.81 \pm 0.04$ &     24.5 &    0.143 &0.486 &0.165 &13.5\\
\enddata
\tablenotetext{a}{Column 1: galaxy name; Columns 2 and 3: deprojected radius and axis ratio of 
the main ring, typically the inner ring, as seen in the H$\alpha$ image. These values are
based on an ellipse fit to the H$\alpha$ distribution in the ring; Column 4: radial scale length 
based on two-dimensional decompositions (see the text and Grouchy(2008)); Column 5: gravitational 
torque strength at the location of the ring's major axis; Column 6: the $m=2$ Fourier amplitude 
at the position of the ring's major axis; Column 7: maximum relative non-axisymmetric torque 
strength based on Grouchy (2008) $I$-band images; (8) the radius of the $Q_{g}$. }
\end{deluxetable}

\end{document}